\documentclass[twocolumn]{aastex631}

\usepackage{makecell}
\usepackage{multirow}
\usepackage{amsmath}

\begin{document}

\title{Changing-look Active Galactic Nuclei from the Dark Energy Spectroscopic Instrument. V.\\ Dramatic Variability in High-Ionization Broad Emission Lines}

\author[0009-0008-3338-393X]{Zhi-Qiang Chen} 
    \affiliation{School of Physics and Technology, Nanjing Normal University, No. 1,
	Wenyuan Road, Nanjing, 210023, P. R. China\\ Email:\href{mailto:yuanqirong@njnu.edu.cn}{yuanqirong@njnu.edu.cn}}

    \author[0000-0002-8402-3722]{Jun-Jie Jin}
    \affiliation{Key Laboratory of Optical Astronomy, National Astronomical Observatories, Chinese Academy of Sciences, Beijing 100012, P.R. China\\ Email:\href{mailto:guowj@bao.ac.cn，jjjin@bao.ac.cn}{guowj@bao.ac.cn, jjjin@bao.ac.cn}}

    \author[0000-0001-9457-0589]{Wei-Jian Guo}
    \affiliation{Key Laboratory of Optical Astronomy, National Astronomical Observatories, Chinese Academy of Sciences, Beijing 100012, P.R. China\\ Email:\href{mailto:guowj@bao.ac.cn，jjjin@bao.ac.cn}{guowj@bao.ac.cn, jjjin@bao.ac.cn}}


    \author[0000-0002-1234-552X]{Sheng-Xiu Sun}
    \affiliation{Kavli Institute for Astronomy and Astrophysics at Peking University, Yiheyuan Road, Haidian District, Beijing 100871, P.R. China}

    \author[0000-0003-0230-6436]{Zhi-Wei Pan}
    \affiliation{Kavli Institute for Astronomy and Astrophysics at Peking University, Yiheyuan Road, Haidian District, Beijing 100871, P.R. China}

    \author[0000-0001-5561-2010]{Chen-Xu Liu}
    \affiliation{South-Western Institute for Astronomy Research, Yunnan University, Kunming 650500, P.R. China}

    \author[0000-0003-4200-9954]{Hua-Qing Cheng}
    \affiliation{Key Laboratory of Optical Astronomy, National Astronomical Observatories, Chinese Academy of Sciences, Beijing 100012, P.R. China\\ Email:\href{mailto:guowj@bao.ac.cn，jjjin@bao.ac.cn}{guowj@bao.ac.cn, jjjin@bao.ac.cn}}

    \author[0000-0002-0779-1947]{Jing-Wei Hu}
    \affiliation{Key Laboratory of Optical Astronomy, National Astronomical Observatories, Chinese Academy of Sciences, Beijing 100012, P.R. China\\ Email:\href{mailto:guowj@bao.ac.cn，jjjin@bao.ac.cn}{guowj@bao.ac.cn, jjjin@bao.ac.cn}}

    \author[0000-0001-6938-8670]{Zhen-Feng Sheng}
    \affiliation{Institute of Deep Space Sciences, Deep Space Exploration Laboratory, Hefei 230026, P.R. China}

    \author[0000-0002-6684-3997]{Hu Zou}
    \affiliation{Key Laboratory of Optical Astronomy, National Astronomical Observatories, Chinese Academy of Sciences, Beijing 100012, P.R. China\\ Email:\href{mailto:guowj@bao.ac.cn，jjjin@bao.ac.cn}{guowj@bao.ac.cn, jjjin@bao.ac.cn}}
    
    \author[0009-0006-9628-7318]{Zhao-Bin Chen}
    \affiliation{School of Physics and Technology, Nanjing Normal University, No. 1,
	Wenyuan Road, Nanjing, 210023, P. R. China\\ Email:\href{mailto:yuanqirong@njnu.edu.cn}{yuanqirong@njnu.edu.cn}}

    \author[ 0000-0003-4583-3383]{Qi Zheng}
    \affiliation{School of Physics and Technology, Nanjing Normal University, No. 1,
	Wenyuan Road, Nanjing, 210023, P. R. China\\ Email:\href{mailto:yuanqirong@njnu.edu.cn}{yuanqirong@njnu.edu.cn}} 
     
    \author[0000-0002-9244-3938]{Qi-Rong Yuan}
    \affiliation{School of Physics and Technology, Nanjing Normal University, No. 1,
	Wenyuan Road, Nanjing, 210023, P. R. China\\ Email:\href{mailto:yuanqirong@njnu.edu.cn}{yuanqirong@njnu.edu.cn}}
    \affiliation{University of Chinese Academy of Sciences, Nanjing 211135, P. R. China}

\begin{abstract}
We present a systematic search for changing-look (CL) quasars at high redshift $z > 0.9$ by cross-matching the spectroscopic datasets from the Dark Energy Spectroscopic Instrument Data Release 1 and Sloan Digital Sky Survey Data Release 18.  
We identify 97 CL quasars showing significant variability in high-ionization broad emission lines (BELs), including 45 turn-on and 52 turn-off events, corresponding to a detection rate of $\sim$0.042\%. The low rate relative to low-ionization CL quasars searches, likely due to selection and physical effects in high-ionization lines.
Based on the CL quasar sample, we find that CL quasars generally exhibit lower accretion rates compared to typical quasars, with average Eddington ratios of $\log \lambda_{\rm Edd} \sim -1.14$ in the bright state and $\sim -1.39$ in the dim state, compared to $\sim -0.65$ for typical quasars.
Furthermore, while high-ionization lines in CL quasars follow the Baldwin effect on a population level, some sources can display inverse Baldwin trends. In addition, we find a positive correlation between the variability in high-ionization lines (e.g., $\rm Mg\,\textsc{ii}$, $\rm C\,\textsc{iii}]$) and the change in bolometric luminosity. We also estimate a characteristic rest-frame timescale of $\sim$3 years for CL transitions, with no significant difference between turn-on and turn-off cases. Taken as a whole, these findings support an accretion-driven origin of the CL phenomenon, and provide new insights into the variability of high-ionization emission lines.

\end{abstract}


\section{Introduction} \label{sec:intro}

Changing-look (CL) active galactic nuclei (AGNs) represent a rare subclass of AGNs characterized by the appearance (turn-on) or disappearance (turn-off) of broad emission lines (BELs) (\citealt{2015ApJ...800..144L, 2016ApJ...826..188R, 2018ApJ...862..109Y, 2019ApJ...883...94T, 2022ApJ...939L..16Z}). This behavior poses a direct challenge of the standard unified model (UM) of AGNs, which attributes the observed diversity of AGN types primarily to orientation effects (\citealt{1993ARA&A..31..473A, 1995PASP..107..803U}). 
While changes in the accretion rate have been proposed as an intrinsic mechanism, where a significant drop could alter the central engine's structure, leading to the dimming or disappearance of the broad-line region (BLR) and the transformation of the optically thick accretion disk into a radiatively inefficient accretion flow (\citealt{2008ARA&A..46..475H}).
However, the observed transition timescales of CL AGNs, ranging from months to years, are often too rapid to be fully explained by existing theoretical models. Therefore, the nature of CL AGNs underscores the need for intrinsic physical changes within the AGN itself, the exact mechanisms of which remain a key open question in the field.

The physical mechanisms responsible for the optical CL phenomenon remain an active area of research. In the optical domain, where CL behavior is defined by the appearance or disappearance of BELs, two main categories of intrinsic explanations have been proposed:
(1) High-energy phenomena such as tidal disruption events (TDEs) that temporarily enhance accretion (\citealt{2015MNRAS.452...69M, 2017ApJ...843..106B, 2022ApJ...933...70L}). However, the long-term light curves of most CL quasars are stochastic rather than exhibiting the rapid rise and slow decay trend typical of TDEs.
(2) Abrupt changes in the accretion rate, potentially associated with state transitions in the accretion disk (\citealt{2019ApJ...874....8M, 2020A&A...641A.167S, 2022ApJ...926..184J, 2023ApJ...953...61Y}), which represent the most widely accepted scenario. A key challenge here is that the observed transition timescales are much shorter than the theoretical viscous timescales of the disk, prompting the development of alternative models such as disk winds or an unstable disk (\citealt{2020MNRAS.492.5540M, 2020A&A...641A.167S}).
Although a few extreme cases have suggested that obscuration of the inner BLR by moving clouds or intervening material could be a potential mechanism (\citealt{2022ApJ...939L..16Z}), this scenario is generally disfavored for most optical CL AGNs. Characteristic CL timescales, combined with infrared and polarimetric observations, often fail to support this interpretation (e.g., \citealt{2017ApJ...846L...7S, 2019A&A...625A..54H}).

Early CL quasars were identified through visual inspection (e.g., \citealt{2015ApJ...800..144L, 2016ApJ...826..188R, 2018ApJ...862..109Y}). With the increasing availability of survey data, current searches for CL quasars mainly rely on two approaches: (1) direct comparison of spectroscopic emission lines (e.g., \citealt{2024ApJ...966...85Z,2024ApJS..270...26G, 2025ApJS..278...28G, 2024arXiv240807335D}), and (2) photometric variability searches in the optical or infrared to select candidates, followed by spectroscopic confirmation (e.g., \citealt{2020ApJ...889...46S, 2022ApJ...933..180G,2023MNRAS.524..188L,2024MNRAS.530.3538Z,2025ApJ...980...91Y,2025RAA....25i5012C}).
While CL quasars have been increasingly identified through variability in low- and intermediate-ionization BELs such as $\rm H\alpha$, $\rm H\beta$, and Mg\,\textsc{ii} (Ionization Potential, IP $\approx 7.6$ eV), systematic studies of high-ionization BELs in CL transitions remain in their infancy. The few existing investigations underscore this gap: the identification of merely three CL quasars through C\,\textsc{iv} variability with $z > 2$ by \cite{2020MNRAS.498.2339R}, a sample of 23 at $z > 1.5$ by \cite{2020ApJ...905...52G}, and the first-ever reported $\rm Ly\alpha$ CL quasar by \cite{2025ApJ...981L...8G} collectively highlight the rarity and underexplored nature of the high-ionization CL phenomenon. These high-ionization lines such as C\,\textsc{iii}] (IP $\approx 24.4$ eV), C\,\textsc{iv} (IP $\approx 47.9$ eV), and Si\,\textsc{iv} (IP $\approx 33.5$ eV) arise from regions of the BLR closer to the central engine, where the ionization parameter is higher and the radiation pressure is more intense (e.g., \citealt{1991ApJ...366...64C, 2013peag.book.....N}). In addition, strong recombination lines like $\rm Ly\alpha$ (IP = 13.6 eV), although not a high-ionization line in the strict sense, also primarily originate in these inner BLR regions and can respond sensitively to rapid changes in the ionizing continuum.
These lines are more responsive to changes in the ionizing continuum, therefore, systematic investigations of CL quasars involving high-ionization lines are crucial for probing rapid inner disk or accretion structure variations (\citealt{1992MNRAS.255..502P, 2016ApJ...824...11G}).

However, detecting CL behavior in these lines is challenging due to observational limitations at high redshift, such as low signal-to-noise ratios (SNR) and strong associated absorption features (\citealt{2013ApJ...777..168F}), as well as intrinsic physical effects such as the Baldwin effect, which suppresses equivalent widths as luminosity increases (\citealt{1977ApJ...214..679B, 2011ApJS..194...45S}). 
With the advent of large spectroscopic surveys such as the Dark Energy Spectroscopic Instrument (DESI; \citealt{2016arXiv161100036D}) and the latest release of the Sloan Digital Sky Survey (SDSS; \citealt{2023ApJS..267...44A}), it is now possible to explore high-redshift quasars with sufficient spectral depth and temporal baselines. 
In this work, we present a systematic search for high-redshift ($z > 0.9$) CL quasars based on spectral variability in high-ionization BELs. By cross-matching DESI Data Release 1 (DR1) and SDSS Data Release 18 (DR18) spectroscopic catalogs, and incorporating multi-epoch photometric data from the optical surveys, we build a robust selection pipeline that mitigates observational artifacts such as fiber drop and miscalibration.

The paper is structured as follows. In Section \ref{sec:data}, we describe the datasets employed in this study, including both spectroscopic and photometric observations, along with the data preprocessing procedures. Section \ref{Sec: result} describes the selection strategy for high-redshift CL quasars, and the criteria applied to ensure spectral reliability. In Section \ref{Sec: spec fitting}, the spectral decomposition approach is described. A comprehensive analysis of the CL quasar sample is carried out in Section \ref{Sec: Discussion}, followed by a brief summary. Throughout this study, a concordance $\rm \Lambda CDM$ cosmology is adopted, with parameters $\Omega_{m}=0.3$,  $\Omega_{\Lambda}=0.7$, and $H_{0}=70\, \rm km\, s^{-1}\, Mpc^{-1}$ (\citealt{2003ApJS..148..175S}).

\section{Data} \label{sec:data}

\subsection{Spectroscopic Survey} \label{subsec:spec survey}

The DESI is a Stage IV ground-based dark energy experiment conducted with the 4-meter Mayall Telescope at Kitt Peak National Observatory (\citealt{2013arXiv1308.0847L, 2016arXiv161100036D, 2016arXiv161100037D, 2022AJ....164..207D, 2023AJ....165....9S, 2024AJ....168...95M}).
It represents the most extensive multi-object spectroscopic survey to date, capable of simultaneously observing 5000 targets with robotic fibers positioners on its focal plane and reaching an $r$-band limiting magnitude of $\sim 23$.  (\citealt{2016arXiv161100036D, 2024AJ....168...95M}).
In addition, the DESI spectrographs split the light into three channels: blue (3600-5900\,\AA), green (5660-7220\,\AA), and red (7470-9800\,\AA), with spectral resolutions of $R \sim$2100, 3200, and 4100, respectively (\citealt{2016arXiv161100037D, 2022AJ....164..207D}).

We also use the spectra from the SDSS. SDSS-V uses the Sloan Foundation 2.5m telescope located at the Apache Point Observatory in New Mexico, USA, and the du Pont 2.5m telescope at the Las Campanas Observatory in Chile (\citealt{2006AJ....131.2332G}). 
The SDSS DR18 is the first data release of $\rm \text{SDSS-} \textsc{V}$, encompassing all spectroscopic data from the previous four phases of the survey (\citealt{2023ApJS..267...44A}).
The SDSS spectra in the later three stages span a wavelength range of 3600\,\AA\, to 10600~\AA, with a spectral resolution ranging from $R \sim1300$ to $\sim3000$(\citealt{2004AJ....128..502A, 2009ApJS..182..543A, 2011AJ....142...72E, 2013AJ....146...32S}).

Both the DESI and the SDSS spectroscopic pipelines compare the target spectra with galaxy, quasar, and star templates and classify the object as {\tt GALAXY}, {\tt QSO}, and {\tt STAR}, automatically (\citealt{2012AJ....144..144B, 2016AJ....152..205H, 2023AJ....165..144G, 2023AJ....166..259S, 2024AJ....167...62D, 2024AJ....168...58D}).
During this work, we focus on objects that are classified as {\tt GALAXY} or {\tt QSO}.
To ensure that the classification is reliable, we require the average SNR of the spectra to be larger than 2.

To facilitate the subsequent analysis, we perform preprocessing on the DESI and SDSS spectral data before their use.  We corrected all spectra for Galactic extinction using the \cite{1999PASP..111...63F} extinction law with $\rm R_{v}=3.1$, and subsequently shifted them to the rest frame.
To directly subtract the spectra, we rebin the spectra onto a common wavelength grid with 2 \AA~per pixel. 

\subsection{Light Curve}

The photometric light curves are essential for investigating the luminosity and color variations associated with CL behavior. Critically, they also enable checks for fiber drops and flux calibration(the detail is shown in Sec \ref{sec: selection}). 

Accordingly, we incorporate long-baseline optical light curves from multiple surveys, including the Catalina Real-time Transient Survey (CRTS; $V \text{-band}$ from 2005 to 2013; \citealt{2009ApJ...696..870D, 2012IAUS..285..306D}), the Pan-STARRS1 (PS1; $g\text{-}$ and $r\text{-band}$ from 2010 to 2014; \citealt{2016arXiv161205560C}), and the Zwicky Transient Facility (ZTF; $g\text{-}$ and $r\text{-band}$; \citealt{2019PASP..131a8002B, 2019PASP..131a8003M}) for the optical band.
We specifically utilize the ZTF DR23 for data from 2018 to the present, which provides forced-photometry light curves from the ZTF Public Sky Survey.
We performed several data quality cuts to ensure the reliability of the light curves. First, we discarded data points with photometric errors larger than twice the average error of the long-term light curve. Second, we applied a $3\sigma$ clipping algorithm twice to remove statistical outliers. Finally, to mitigate the impact of correlated noise and systematic errors, we binned the data into 3-day intervals, following the methodology of \cite{2019PASP..131a8003M} and \cite{2022ApJ...933..180G}.
We caution that analyses near the survey limits require care because the limiting magnitudes of these surveys vary slightly. For example, the CRTS has a $V \rm \text{-band}$ limiting magnitude of approximately 20.5, while PS1 reaches about 22.0 in the $g$-band and 21.8 in the $r$-band. For ZTF, the limiting magnitudes are around 20.8 and 20.5 in the $g\text{-}$ and $r$-bands, respectively (\citealt{2009ApJ...696..870D, 2012IAUS..285..306D, 2012ApJ...750...99T, 2019PASP..131a8002B, 2019PASP..131a8003M}).

\section{Sample Selection and Result}\label{Sec: result}

\subsection{Sample Selection \label{sec: selection}}

Based on previous works (\citealt{2024ApJS..270...26G, 2025ApJS..278...28G}), we select targets with redshifts $z>0.9$ to ensure that key high-ionization emission lines are in the observed wavelength range and sufficiently separated from the spectral edge of the detector, where the SNR is critically low. 
We then perform a positional cross-match between the pre-selected DESI DR1 and SDSS DR18 spectroscopic catalogs. To ensure the matches are astrophysically relevant and minimize false positives, we adopt a matching radius of 1 arcsecond and further require a redshift difference of $\lvert \Delta z \rvert < 0.01$. From these robust matches, we keep sources classified as {\tt QSO} in one epoch and {\tt GALAXY} or {\tt QSO} in the other.
In our final sample, we keep 826 pairs of non-repeated {\tt GALAXY-QSO} and 228,623 pairs of {\tt QSO-QSO}. SDSS-V has already conducted systematic comparisons of repeat spectra to search for CL quasars (e.g., \citealt{2024ApJ...966...85Z}), and since the DESI time baselines are relatively short, we retain only the latest spectrum per source in each survey to avoid redundancy.

To efficiently identify a robust sample of CL candidates from the parent sample, we employ a set of spectral variability metrics that probe both the statistical significance and the physical magnitude of BEL changes. Our methodology follows and extends the work of \cite{2016MNRAS.457..389M} and \cite{2024ApJS..270...26G}, focusing on three key parameters.

First, we quantify the statistical significance of the spectral change at each wavelength pixel using an SNR metric:
\begin{equation}
    N_{\sigma}(\lambda) = (f_{\rm bright} - f_{\rm dim})/\sqrt{\sigma_{\rm bright}^{2} + \sigma_{\rm dim}^{2}} ,
\end{equation}
here, $N_{\sigma}(\lambda)$ represents the significance of the flux difference between the bright- and dim-state spectra, where the $f$ and $\sigma$ are the flux and uncertainty at $\lambda$ in $\rm erg\,cm^{-2}\,s^{-1}\,$\AA$^{-1}$. This parameter allows us to identify regions of the spectrum where the change is robust against observational noise.

Second, to measure the physical magnitude of the change in the BELs, we define a continuum-subtracted integrated flux ratio, $R$:
\begin{equation}
    R = \frac{F_{\rm bright} - F_{\rm dim}}{F_{\dim}}, 
\end{equation}
where $F$ represents the total integrated flux of a specific BEL, obtained by summing the continuum-subtracted flux within the rest-frame windows listed in Table \ref{tab:inte_windows}. The uncertainty $\sigma_{R}$ is calculated using the propagation of the standard error. 
However, a large $R$ value alone could be misleading if the line is barely detected in the dim state. To ensure that a disappearing line is genuinely absent and not merely diluted, we introduce a third parameter, $F_{\rm \sigma, dim}$, to quantify the prominence of the emission line in the dim state:
\begin{equation}
    F_{\rm \sigma, dim} = \sum f_{\rm dim} / \sqrt{\sum \sigma_{\rm dim}^{2}}.
\end{equation} 
This is effectively the integrated SNR of the line in the dim-state spectrum.

In summary, our candidate selection leverages these complementary metrics: $N_{\sigma}(\lambda)$ maps significant spectral changes, $R$ quantifies the amplitude of the BEL flux change, and $F_{\rm \sigma, dim}$ confirms the line's non-detection in the dim state. This multi-faceted approach minimizes false positives caused by poor data quality while effectively capturing genuine CL events.

\begin{deluxetable*}{ccccc}[!ht]
    \setlength{\tabcolsep}{5pt}
    \renewcommand{\arraystretch}{1.2}
    \tablenum{1}
    \centering
    \tablecaption{The integration windows for BELs \label{tab:inte_windows}}
    \tablehead{\colhead{Line Name} & \colhead{Wavelength (\AA)} & \colhead{Window (\AA)} & \colhead{Left Continuum (\AA)} & \colhead{Right Continuum (\AA)} \\
    \colhead{(1)} & \colhead{(2)} & \colhead{(3)} & \colhead{(4)} & \colhead{(5)} }
    \startdata
    $\rm Ly \alpha$ & 1215.67 & 1190-1250 & 1150–1190 & 1250–1290 \\
    $\rm Si\, \textsc{iv}$ & 1396.76, 1402.77 & 1360-1430 & 1310-1360 & 1430, 1480 \\
    $\rm C\, \textsc{iv}$ & 1548.19, 1550.77 & 1510–1590 & 1460–1510 & 1590–1640\\
    $\rm C\,\textsc{iii}]$ & 1906.68, 1908.73 & 1850–1970 & 1820–1850 & 1970–2000  \\
    $\rm Mg\,\textsc{ii}$ & 2795.53, 2802.71 & 2750–2850 & 2680–2720 & 2880–2920
    \enddata
    \tablecomments{Column (1): BEL name; Column (2): Rest wavelength of the BEL; Column (3): Integration window of the BEL; Column (4): Left continuum window of the BEL; Column (5): Right continuum window of the BEL.}
\end{deluxetable*}

We adopt stricter selection criteria than previous works (e.g., \citealt{2024ApJS..270...26G}), requiring $N_{\sigma}(\lambda) > 3$ and $R>2$ to ensure significant variability in the BELs, and we imposed $R>2\sigma_{R}$ to guarantee the reliability of our total broad-line flux measurements. Finally, we required $F_{\rm \sigma, dim}<2R$, which ensures that the emission lines are relatively weak in the dim state. At this step, a total of two {\tt GALAXY-QSO} pairs and 572 {\tt QSO-QSO} pairs satisfy the criteria.

As mentioned in previous studies, the flux calibration or fiber drop issues may affect the assessment of variability of BELs (e.g., \citealt{2020ApJ...905...52G, 2024ApJS..270...26G}). The most common method for calibrating the relative flux scaling of two spectra assumes that the narrow emission line $\rm [O\,\textsc{iii}]$ (5007\,\AA) remains constant over timescales of a few years (e.g., \citealt{2016ApJ...831..157M,2024ApJ...966..128W}), as the narrow-line region typically extends over kiloparsec scales, so significant variations in narrow lines would only occur over millennial timescales (\citealt{2002ApJ...574L.105B, 2018MNRAS.477.4615D}). However, this approach is unsuitable for high-redshift sources, as the prominent narrow lines are shifted beyond the observational windows of SDSS and DESI. 
To mitigate contamination from flux calibration errors or fiber drop effects, we compute pseudo-magnitude by convolving each spectrum with the $g\text{-}$, $r\text{-}$ and $\rm V\text{-}$band filter transmission curves and compare them to the photometric magnitudes measured at the corresponding epochs in the light curves. Spectra with $>0.5$ mag discrepancy are excluded unless verified visually (see Figure \ref{fig: flux_calibratio} and Appendix \ref{fiber_drop}).
In total, we excluded 326 spurious sources attributed to fiber drop issues.

Since the limiting magnitudes of ZTF and PS1 in the $g$-band are 20.8 mag and 23.0 mag, respectively, therefore, for sources with magnitudes fainter than 21 mag, we take extra caution. Instead of directly comparing them with the observational magnitude, we scale the two spectra to the same photometric level before examining the differences in their emission lines and do a visual check for them to confirm that the CL phenomenon is not due to the fiber drop in Section \ref{sec: visual}.

\subsection{Visual Check}\label{sec: visual}

For high-redshift CL quasars, visual inspection is a key step for identification, as it avoids the need for complex spectral decomposition.
This is particularly important because the high-redshift spectra typically have relatively low SNR. In addition, the impact of interstellar medium and AGN outflows on BELs cannot be ignored, variable AGN outflows on BELs cannot be ignored, especially for C\,\textsc{iv} and Ly$\alpha$ (e.g., \citealt{1986ApJS...61..249W, 2013ApJ...777..168F}). Strong and variable absorption from outflows, such as those in broad absorption line quasars, can not only affect continuum subtraction, leading to incorrect flux estimates, but can also produce observed spectral changes that might be mistaken for intrinsic CL behavior if the outflow structure varies between epochs (e.g., \citealt{2011MNRAS.413..908C, 2022MNRAS.514.1975A}).

\begin{figure*}
    \centering    \includegraphics[width=1\textwidth]{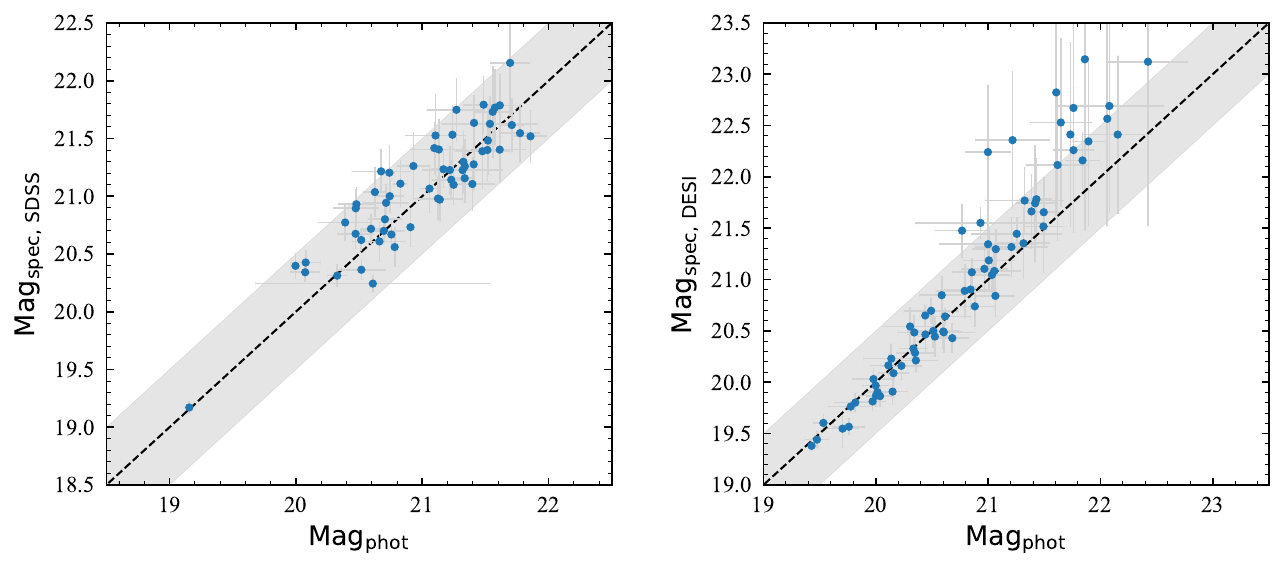}
    \caption{The pseudo-magnitude of our CL quasars versus the actual brightness from several photometric surveys simultaneously. The left and right panels show the magnitude of the targets at the SDSS and DESI observational epochs, respectively. The shaded region indicates an uncertainty range of 0.5 mag.}
    \label{fig: magnitude}
\end{figure*}

Here, we perform visual inspections on all sources that meet the criteria outlined in Section \ref{sec: selection}. First, we check whether strong absorption lines are present within the integration windows of the emission lines of interest. Sources with such absorption lines are discarded unless they exhibit a transition from emission to absorption lines. Secondly, we exclude sources without clear CL signatures as these may result from inaccurate emission line measurements due to low SNR or improper continuum subtraction. Thirdly, a small fraction of sources with inconsistent redshifts between SDSS and DESI spectra are excluded.
Finally, we check that the sources with pseudo-magnitudes fainter than 21 mag and the magnitude difference larger than 0.5 mag (the outliers in Figure \ref{fig: magnitude}), to determine whether the faintness is intrinsic or caused by fiber drop. This is primarily done by examining changes in the spectral shape, since genuine CL transitions are typically accompanied by variations in the continuum (e.g., \citealt{2018ApJ...862..109Y, 2025ApJS..278...28G}).
On the other hand, we perform visual inspection to distinguish CL quasars, characterized by the appearance or disappearance of BELs, and CL quasar candidates, which show dimmer spectra but still retain weak BELs. Such sources may have been caught in a transitional phase of CL behavior when the dim-state spectrum was taken, and are, therefore, classified as CL quasar candidates (e.g., \citealt{2019ApJ...883L..44G}).

To provide a clearer overview of our sample selection process, we summarize the main selection criteria and corresponding sample sizes in Table~\ref{tab:selection}. Each row represents a major step in the filtering procedure, starting from 229,449 matched spectroscopic pairs from SDSS DR18 and DESI DR1, we sequentially applied our criteria: a significance cut ($N_{\sigma}(\lambda) > 3$) retained 145,171 pairs; a broad line flux ratio cut ($R > 2$) selected 697 pairs; a dim-state line prominence cut ($F_{\rm \sigma, dim} < 2R$) yielded 574 pairs; and a final pseudo-magnitude filter ($< 0.5$ mag) identified 248 high-probability candidates. After visual inspection of these candidates, we secured a final sample of 97 robust CL quasars and 135 additional CL quasar candidates, while flagging 16 targets as false positives.

\begin{deluxetable*}{cc}[!ht]
    \setlength{\tabcolsep}{3pt}
    \renewcommand{\arraystretch}{1}
    \tablenum{2}
    \centering
    \tablecaption{CL quasars selection from matching the SDSS DR18 and DESI DR1\label{tab:selection}}
    \tablehead{
        \colhead{Selection} & \colhead{$N_{\rm object}$} 
    }
    \startdata
    \makecell[c]{class={\tt QSO} and spectype={\tt GALAXY} \\[-2pt] class={\tt GALAXY} and spectype={\tt QSO} \\[-2pt] class={\tt QSO} and spectype={\tt QSO}} & 229,449 pairs \\[15pt] 
    $N_{\sigma}(\lambda) > 3$ & 145,171 pairs \\[5pt]
    $R > 2$ & 697 pairs\\[5pt]
    $F_{\rm \sigma, dim} < 2R$ & 574 pairs\\[5pt]
    pseudo-magnitude $<$ 0.5 mag & 248 pairs \\[10pt]
    visual check & \makecell[c]{16 fake targets \\ [-2pt ]97 CL quasars \\[-2pt] 135 CL quasar candidates \\}
    \enddata
\end{deluxetable*}

\subsection{Result}

\begin{figure*}
    \centering
    \includegraphics[width=1\textwidth]{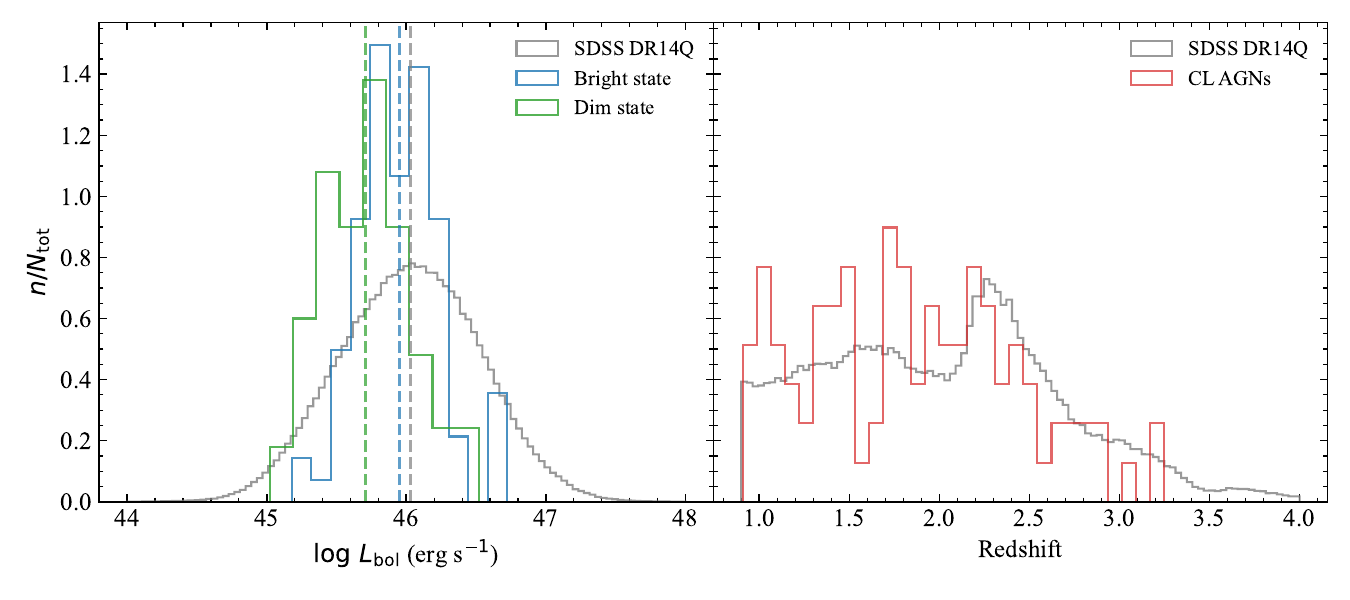}
    \caption{The luminosity and redshift distribution of the high-redshift CL quasars in this work. The grey histogram is the quasar sample at the same redshift range from the SDSS DR14 QSO catalog (\citealt{2020ApJS..249...17R}). The corresponding dashed lines in the left panel represent the average values.}
    \label{fig: luminosity_z}
\end{figure*}

In this work, we identified 97 high-redshift CL quasars from 826 pairs of {\tt GALAXY-QSO} and 228,623 pairs of {\tt QSO-QSO}, comprising 45 turn-on and 52 turn-off events. Figure \ref{fig: luminosity_z} shows their luminosity and redshift distributions, which are broadly consistent with the overall SDSS quasar population at similar redshifts (\citealt{2020ApJS..249...17R}).
In terms of luminosity, even when CL quasars are in their bright state, their median luminosity ($\log L_{\rm bol} = 45.95 \pm 0.043$) is slightly lower than that of typical quasars at the same redshift range (median $\log L_{\rm bol} = 46.03 \pm 0.001$). However, in either bright or dim state (median $\log L_{\rm bol} = 45.71 \pm 0.041$ for the dim state), CL quasars are still more luminous than normal AGNs, reaching quasar-level luminosities ($L \gtrsim 10^{45}\ \mathrm{erg\ s^{-1}}$). This is largely due to selection effects, as fainter objects at $z>0.9$ fall below the SDSS detection limits. Uncertainties on median values are estimated via bootstrap resampling with 1,000 iterations. 

\begin{figure*}
    \centering
    \includegraphics[width=1\textwidth]{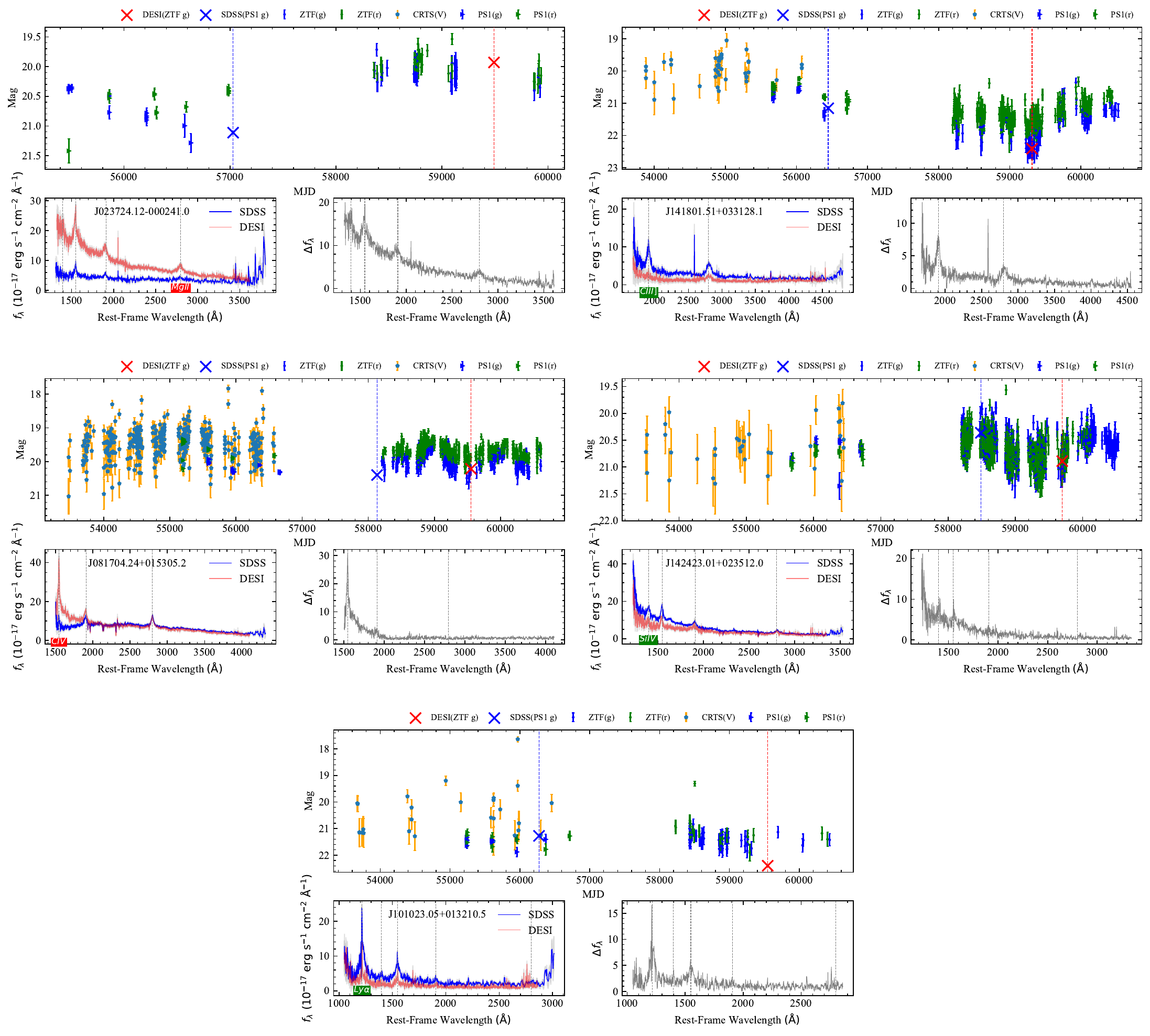}
    \caption{Five examples of CL quasars showing significant transitions in different emission lines. For each example, the top panel shows the nearly 20-year light curve compiled from CRTS $V$-band (dark blue pentagons), PS1 $g$-band (blue triangles), $r$-band (green triangles), ZTF $g$-band (blue circles), and $r$-band (green circles). The blue and red dashed vertical lines indicate the epochs of SDSS and DESI spectroscopic observations, respectively. The corresponding ``X'' markers represent the pseudo-magnitudes derived by convolving the spectra with filter response curves. The bottom-left panel displays the original spectra from SDSS (blue) and DESI (red). Red and green squares at the bottom denote detected turn-on and turn-off transitions, respectively. The lower-right panel shows the difference in flux between the two spectra. Gray dashed lines in the bottom panels mark the expected positions of key emission lines.}
    \label{fig: CLAGN_example}
\end{figure*}

Previous systematic studies have shown that the number of turn-on and turn-off CL quasars is roughly equal (e.g., \citealt{2016MNRAS.457..389M, 2025ApJS..278...28G, 2024arXiv241015587W}). In our high-redshift sample, the ratio is $\rm 1:1.15$, with a slight excess of turn-off sources. This could be attributed to differences in survey depth, as DESI is significantly deeper than SDSS. Some faint sources may only be detectable in DESI, causing high-redshift turn-on CL quasars to remain undetected in the dim state because of the limited sensitivity of SDSS.
In the final CL quasar sample, we identify 28 sources exhibiting CL behavior in the $\rm Mg\,\textsc{ii}$ emission line. The $\rm C\,\textsc{iii}]$ line is the most frequently affected, with 59 CL cases. CL behavior is also observed in 25 $\rm C\textsc{IV}$ and 26 $\rm Si\,\textsc{iv}$ sources. Additionally, we report 8 CLAGNs showing transitions in the $\rm Ly\alpha$ line. Note that some of these cases are overlapping, as a single target may exhibit the CL phenomenon in multiple emission lines.
In Figure \ref{fig: CLAGN_example}, we present representative examples of the CL phenomenon for each emission line, illustrating clear cases of BEL appearance or disappearance. In addition, we cross-match the sample with the Faint Images of the Radio Sky at Twenty-Centimeters (FIRST) catalog\footnote{\url{http://sundog.stsci.edu/}} and find that the vast majority of CL quasars are radio-quiet, with only three sources having radio fluxes above the detection threshold of the FIRST survey (\citealt{1995ApJ...450..559B, 2015ApJ...801...26H}). We will conduct a follow-up analysis on the radio properties of CL quasars in future work. The entire CL quasar sample identified in this work is shown in Table \ref{tab:CLAGN_catalog}.

\begin{figure*}
    \centering
    \includegraphics[width=1\linewidth]{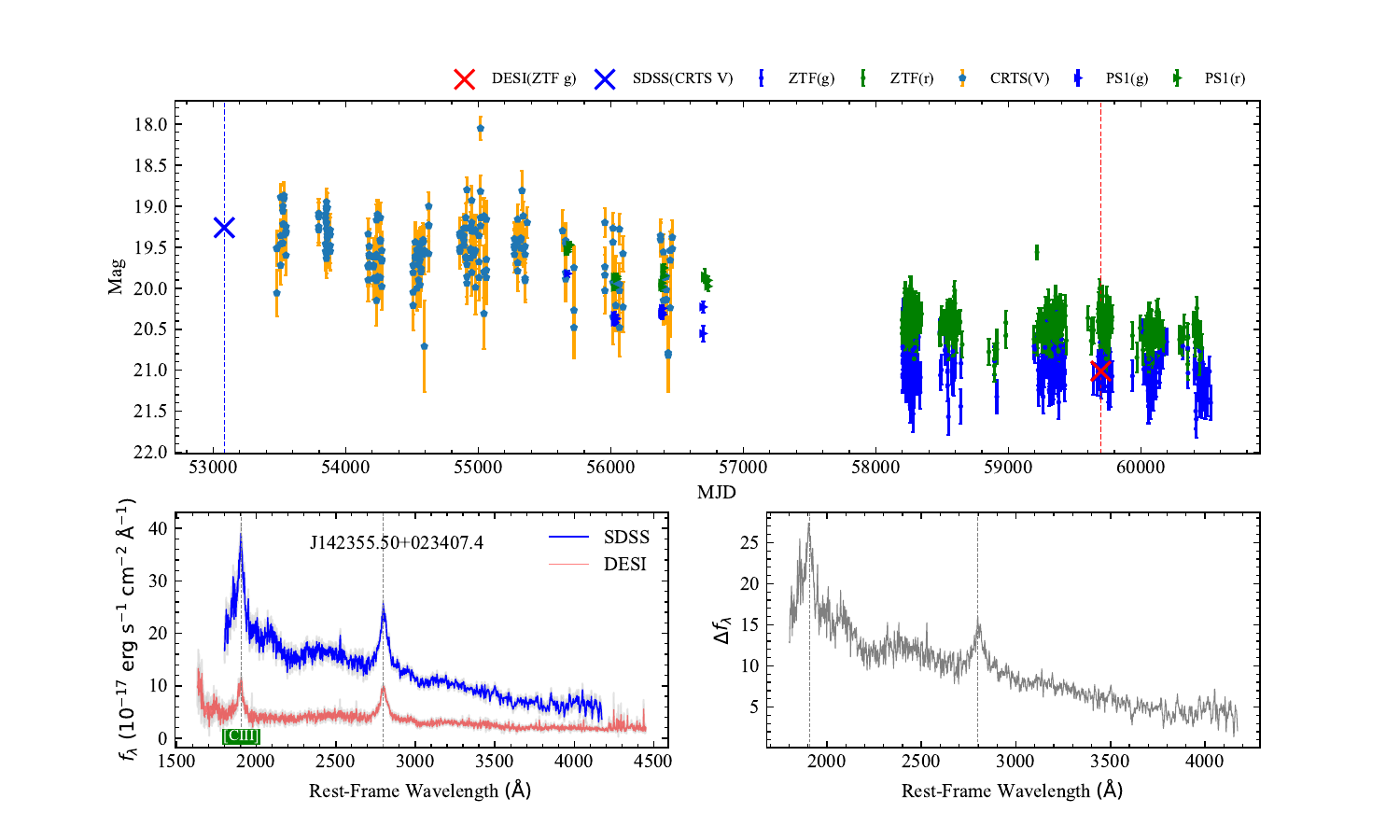}
    \caption{An example of a CL quasar candidate. The panel layout and symbols are the same as in Figure \ref{fig: CLAGN_example}.}
    \label{fig: candidate_example}
\end{figure*}

Moreover, we identified 135 CL quasar candidates that appear to be in intermediate stages of the CL process.
These sources are crucial for probing the physical mechanisms behind CL quasars, such as the continuous evolution of accretion rates and the sequence of emission-line transitions (e.g., \citealt{2011ApJ...733...60T, 2024ApJS..272...13P}). 
Figure \ref{fig: candidate_example} shows a CL quasar candidate identified in the $\rm C\, \textsc{iii}]$ line. The entire CL quasar candidate sample is shown in Appendix \ref{appendix b}. 

\begin{deluxetable*}{lcccccl}[!ht]
    \setlength{\tabcolsep}{5pt}
    \renewcommand{\arraystretch}{1.2}
    \tablenum{3}
    \centering
    \tablecaption{Fits Catalog Description and Column Information of CL quasars in this work. \label{tab:CLAGN_catalog}}
    \tablehead{\colhead{Number} & \colhead{Column Name} & \colhead{Format} & \colhead{Unit} & \colhead{Description} \\
    \colhead{(1)} & \colhead{(2)} & \colhead{(3)} & \colhead{(4)} & \colhead{(5)} }
    \startdata
    0 & SDSS\_NAME & string &  & Unique identifier of SDSS object \\
1 & R.A. & float32 & degree & Right Ascension (J2000) \\
2 & Dec. & float32 & degree & Declination (J2000) \\
3 & Redshift & float32 &  & Redshift \\
4 & MJD\_SDSS & int &  & MJD of the SDSS spectrum \\
5 & MJD\_DESI & float32 &  & Mean MJD of the coadded DESI spectra \\
6 & Transition & string &  & Changing state between SDSS and DESI spectra \\
7 & MgII\_state & string &  & State change in Mg\,\textsc{ii} \\
8 & CIII\_state & string &  & State change in C\,\textsc{iii]} \\
9 & CIV\_state & string &  & State change in C\,\textsc{iv} \\
10 & SiIV\_state & string &  & State change in Si\,\textsc{iv} \\
11 & Lya\_state & string &  & State change in Ly$\alpha$ \\
12 & LOGLBOL\_Bri & float32 & erg\,s$^{-1}$ & Bolometric luminosity in the bright state \\
13 & LOGLBOL\_Bri\_ERR & float32 & erg\,s$^{-1}$ & Error in LOGLBOL\_Bri \\
14 & LOGLBOL\_Dim & float32 & erg\,s$^{-1}$ & Bolometric luminosity in the dim state \\
15 & LOGLBOL\_Dim\_ERR & float32 & erg\,s$^{-1}$ & Error in LOGLBOL\_Dim \\
16 & EW\_MgII\_Bri & float32 & \AA & EW of Mg\,\textsc{ii} (bright) \\
17 & EW\_MgII\_Bri\_ERR & float32 & \AA & Error in EW\_MgII\_Bri \\
18 & EW\_MgII\_Dim & float32 & \AA & EW of Mg\,\textsc{ii} (dim) \\
19 & EW\_MgII\_Dim\_ERR & float32 & \AA & Error in EW\_MgII\_Dim \\
20 & EW\_CIII\_Bri & float32 & \AA & EW of C\,\textsc{iii]} (bright) \\
21 & EW\_CIII\_Bri\_ERR & float32 & \AA & Error in EW\_CIII\_Bri \\
22 & EW\_CIII\_Dim & float32 & \AA & EW of C\,\textsc{iii]} (dim) \\
23 & EW\_CIII\_Dim\_ERR & float32 & \AA & Error in EW\_CIII\_Dim \\
24 & EW\_CIV\_Bri & float32 & \AA & EW of C\,\textsc{iv} (bright) \\
25 & EW\_CIV\_Bri\_ERR & float32 & \AA & Error in EW\_CIV\_Bri \\
26 & EW\_CIV\_Dim & float32 & \AA & EW of C\,\textsc{iv} (dim) \\
27 & EW\_CIV\_Dim\_ERR & float32 & \AA & Error in EW\_CIV\_Dim \\
28 & FWHM\_MgII\_Bri & float32 & km\,s$^{-1}$ & FWHM of Mg\,\textsc{ii} (bright) \\
29 & FWHM\_MgII\_Bri\_ERR & float32 & km\,s$^{-1}$ & Error in FWHM\_MgII\_Bri \\
30 & FWHM\_MgII\_Dim & float32 & km\,s$^{-1}$ & FWHM of Mg\,\textsc{ii} (dim) \\
31 & FWHM\_MgII\_Dim\_ERR & float32 & km\,s$^{-1}$ & Error in FWHM\_MgII\_Dim \\
32 & FWHM\_CIV\_Bri & float32 & km\,s$^{-1}$ & FWHM of C\,\textsc{iv} (bright) \\
33 & FWHM\_CIV\_Bri\_ERR & float32 & km\,s$^{-1}$ & Error in FWHM\_CIV\_Bri \\
    \enddata
\end{deluxetable*}

\begin{deluxetable*}{lcccccl}[!ht]
    \setlength{\tabcolsep}{5pt}
    \renewcommand{\arraystretch}{1.2}
    \tablenum{3}
    \centering
    \tablecaption{(Continued)}
    \tablehead{\colhead{Number} & \colhead{Column Name} & \colhead{Format} & \colhead{Unit} & \colhead{Description} \\
    \colhead{(1)} & \colhead{(2)} & \colhead{(3)} & \colhead{(4)} & \colhead{(5)} }
    \startdata
34 & FWHM\_CIV\_Dim & float32 & km\,s$^{-1}$ & FWHM of C\,\textsc{iv} (dim) \\
35 & FWHM\_CIV\_Dim\_ERR & float32 & km\,s$^{-1}$ & Error in FWHM\_CIV\_Dim \\
36 & Flux\_MgII\_Bri & float32 & $\rm 10^{-17} erg\,s^{-1}\,cm^{2}$ & Flux of Mg\,\textsc{ii} in the bright state \\
37 & Flux\_MgII\_Bri\_ERR & float32 & $\rm 10^{-17} erg\,s^{-1}\,cm^{2}$ & Error in Flux\_MgII\_Bri \\
38 & Flux\_MgII\_Dim & float32 & $\rm 10^{-17} erg\,s^{-1}\,cm^{2}$ & Flux of Mg\,\textsc{ii} in the dim state \\
39 & Flux\_MgII\_Dim\_ERR & float32 & $\rm 10^{-17} erg\,s^{-1}\,cm^{2}$ & Error in Flux\_MgII\_Dim \\
    40 & Flux\_CIII\_Bri & float32 & $\rm 10^{-17} erg\,s^{-1}\,cm^{2}$ & Flux of C\,\textsc{iii}] in the bright state \\
    41 & Flux\_CIII\_Bri\_ERR & float32 & $\rm 10^{-17} erg\,s^{-1}\,cm^{2}$ & Error in Flux\_CIII\_Bri \\
    42 & Flux\_CIII\_Dim & float32 & $\rm 10^{-17} erg\,s^{-1}\,cm^{2}$ & Flux of C\,\textsc{iii}] in the dim state \\
    43 & Flux\_CIII\_Dim\_ERR & float32 & $\rm 10^{-17} erg\,s^{-1}\,cm^{2}$ & Error in Flux\_CIII\_Dim \\
    44 & Flux\_CIV\_Bri & float32 & $\rm 10^{-17} erg\,s^{-1}\,cm^{2}$ & Flux of C\,\textsc{iv} in the bright state \\
    45 & Flux\_CIV\_Bri\_ERR & float32 & $\rm 10^{-17} erg\,s^{-1}\,cm^{2}$ & Error in Flux\_CIV\_Bri \\
    46 & Flux\_CIV\_Dim & float32 & $\rm 10^{-17} erg\,s^{-1}\,cm^{2}$ & Flux of C\,\textsc{iv} in the dim state \\
    47 & Flux\_CIV\_Dim\_ERR & float32 & $\rm 10^{-17} erg\,s^{-1}\,cm^{2}$ & Error in Flux\_CIV\_Dim \\
    48 & LOGMBH & float32 & $M_{\sun}$ & The adopted fiducial black hole mass \\
    49 & LOGMBH\_ERR & float32 & $M_{\sun}$ & Error in LOGMBH \\
    50 & LOGREDD\_Bri & float32 &  & Logarithmic Eddington ratio in the bright state \\
    51 & LOGREDD\_Bri\_ERR & float32 &  &Error in LOGREDD\_Bri \\
    52 & LOGREDD\_Dim & float32 &  & Logarithmic Eddington ratio in the dim state \\
    53 & LOGREDD\_Dim\_ERR & float32 &  &Error in LOGREDD\_Dim \\
    54 & Radio\_Flux & float32 & mJy & Radio flux at 1.4\,GHz in FIRST catalog \\
    \enddata
    \tablecomments{We provide the basic information and the main spectral fitting result of the high-redshift CL quasars identified in this work. The unmeasurable parameters are set to \text{-999}. The errors are obtained from 200 iterations of the Monte Carlo simulation.\\ (This table is available in its entirety in the machine-readable format.)}
\end{deluxetable*}

\section{Spectral Fitting}\label{Sec: spec fitting}

To extract the physical parameters for CL quasars, we adopt the {\tt PyQSOFit}\footnote{\url{https://github.com/legolason/PyQSOFit}} to perform spectral decomposition (\citealt{2019ApJS..241...34S, 2019MNRAS.482.3288G, 2024arXiv240617598R}).
During the fitting procedure, we ignore the host galaxy component, as its contribution is negligible compared to that of the central black hole in high-redshift objects.
This is a standard approach in spectral analyses of high-redshift samples, as demonstrated in previous studies (e.g., \citealt{2011ApJS..194...45S, 2020ApJ...905...52G}).
We model the line-free continuum using a combination of a power-law, a third-order polynomial, and a pseudo-continuum constructed from UV/optical $\rm Fe\,\textsc{ii}$ emission templates. After subtracting the best-fit AGN continuum model, we separate BELs and narrow emission lines using a full width at half maximum (FWHM) threshold of $\rm 1200\, km\, s^{-1}$. 
The narrow lines are modeled using single Gaussian profiles, as they typically originate from a spatially distinct, dynamically cold region with well-defined velocity dispersion. For the broad components, we allow a maximum of three Gaussians to better describe the asymmetry. The specific number of Gaussian components adopted for each broad emission line is fixed for all sources in our sample, as shown in Table \ref{tab:fitting_parameters}. This strategy is chosen to ensure a consistent and homogeneous analysis pipeline crucial for robust statistical comparisons. These numbers are not determined on a source-by-source basis but are instead based on the established practice of previous systematic studies (e.g., \citealt{2011ApJS..194...45S, 2019MNRAS.482.3288G, 2020ApJ...905...52G, 2024ApJ...963....7R}). These studies identified the optimal number of components for various emission lines using the Bayesian Information Criterion, which determines the preferred model by evaluating the trade-off between goodness-of-fit and model complexity. The cited studies systematically compared models with increasing numbers of Gaussians and selected the one that provided a significantly better fit without overfitting.
Consequently, the number of Gaussians varies for different lines, reflecting their distinct intrinsic kinematics and profile shapes. Highly asymmetric lines with prominent broad wings or blue shift, such as $\rm Ly\alpha$ and C\,\textsc{iv}, are typically modeled with three Gaussians. In contrast, lines with generally more symmetric profiles, such as Si\,\textsc{iv}, C\,\textsc{iii}], and Mg\,\textsc{ii}, are adequately fitted with two Gaussian components to prevent overfitting.
Finally, we estimate the uncertainties of the fitted parameters through 200 Monte Carlo realizations, where each pixel is perturbed by Gaussian noise scaled to its measurement error. 

\begin{deluxetable}{cccc}[!ht]
    \setlength{\tabcolsep}{1pt}
    \renewcommand{\arraystretch}{1}
    \tablenum{4}
    \centering
    \tablecaption{The fitting parameters for emission lines\label{tab:fitting_parameters}}
    \tablehead{\colhead{Line Complex} & \colhead{Fitting Window (\AA)} & \colhead{Line Name} & \colhead{$N_{\rm gaussian}$} \\
    \colhead{(1)} & \colhead{(2)} & \colhead{(3)} & \colhead{(4)} }
    \startdata
    $\rm Ly \alpha$ & 1150-1290 & Broad $\rm Ly \alpha$ & 3 \\
     & & $\rm N\,\textsc{v}$ 1240.14 & 1 \\
     $\rm Si\,\textsc{iv}$ & 1290-1450 & $\rm Si\,\textsc{iv}\text{/} O\,\textsc{iv}]$ & 2 \\
      & & $\rm O\,\textsc{iv}$ 1304.35 & 1\\
      & & $\rm C\,\textsc{ii}$ 1335.30 & 1\\
    $\rm C\,\textsc{iv}$ & 1500-1700 & Broad $\rm C\,\textsc{iv}$ & 3\\
     & & Broad $\rm He\,\textsc{ii}$ 1640.42 & 1\\
     & & Narrow $\rm He\,\textsc{ii}$ 1640.42 & 1\\
      & & Broad $\rm O\,\textsc{iii}$ 1663.48 & 1\\
     & & Narrow $\rm O\,\textsc{iii}$ 1663.48 & 1\\
    $\rm C\,\textsc{iii}]$ & 1700-1970 & Broad $\rm C\,\textsc{iii}]$ & 2\\
     & & $\rm [N\,\textsc{iv}]$ 1718.55 & 1\\
     & & $\rm [N\,\textsc{iii}]$ 1750.26 & 1\\
     & & $\rm [Fe\,\textsc{ii}]$ 1786.7 & 1\\
     & & $\rm [Si\,\textsc{ii}]$ 1816.98 & 1\\
     & & $\rm Al\,\textsc{iii}$ 1857.40 & 1\\
     & & $\rm Si\,\textsc{iii}]$ 1892.03 & 1\\
    $\rm Mg\,\textsc{ii}$ & 2700-2900 & Broad $\rm Mg\,\textsc{II}$ & 2\\
     & & Narrow $\rm Mg\,\textsc{ii}$ 2798.75 & 1
    \enddata
\end{deluxetable}

\begin{figure*}
    \centering
    \includegraphics[width=1\linewidth]{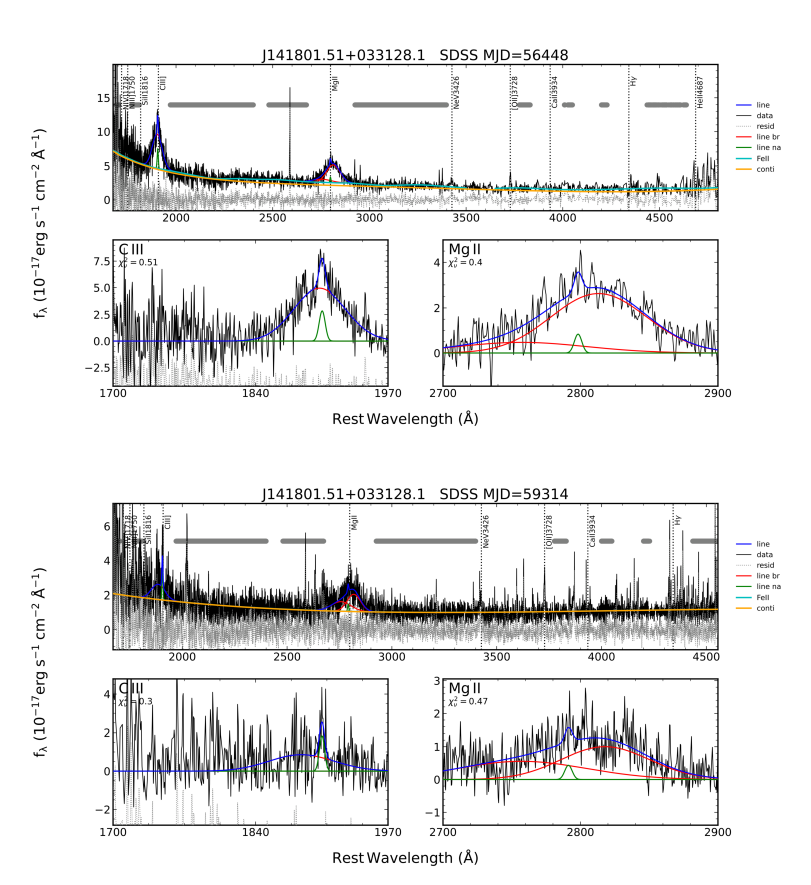}
    \caption{Example of spectral fitting for the bright (SDSS, top panel) and dim (DESI, bottom panel) states of $\rm J141801.51+033128.1$ using {\tt PyQSOFit}.
    The observed spectrum (black) is decomposed into an AGN power-law continuum (orange) and an $\rm Fe\,\textsc{ii}$ emission template (cyan) in the upper panel, under the assumption that the host galaxy contribution is negligible.
    Emission lines are shown in blue and are further decomposed into broad (red) and narrow (green) components. The lower panel provides a zoomed-in view.}
    \label{fig: fitting_example}
\end{figure*}

It is worth noting that high-ionization emission lines are often accompanied by broad or narrow absorption features, which can significantly affect the continuum subtraction and emission-line fitting. To mitigate this, we follow the method proposed by \cite{2020ApJ...905...52G} to mask potential absorption features before the fitting (see Section 3.4 of their paper for details). Here, we show a fitting example in Figure \ref{fig: fitting_example}.

\section{Discussion}\label{Sec: Discussion}

\subsection{Detection Rate}

CL quasars are an extremely rare and peculiar class of objects, with high-redshift CL quasars being even more scarce. Previous studies report CL quasar detection rates of roughly $0.3 \sim 1\%$ from both spectroscopic comparisons and variability-based searches (e.g., \citealt{2023MNRAS.524..188L, 2024ApJ...966..128W, 2024arXiv240807335D, 2025ApJS..278...28G}).
We identify 97 CL quasars from a parent sample of 229,449 matched spectral pairs at $z>0.9$, yielding a detection rate of $\sim$0.042\%.  This rate is approximately an order of magnitude lower than those reported in previous works, which typically focus on low-ionization lines or rely on higher SNR observations (e.g., \citealt{2024arXiv240807335D, 2025ApJS..278...28G}).  This lower detection rate is attributable either to selection effects or to intrinsic differences. On the selection side, a key factor is sample flux limits. The intrinsically higher luminosities of high-redshift CL quasars, though generally bright, can drop below the detection limits of the surveys during their faint phase, causing some targets to be missed. In addition, CL transitions occur on timescales of months to years in the rest frame, with reported median values of about 10 years (the detail is shown in Section \ref{subsec: timescale}). Due to cosmological time dilation, some longer-timescale CL transitions are stretched in the observation frame beyond the temporal baselines of SDSS and DESI ($\sim 20$ yr), and thus cannot be captured within the available observations.  On the intrinsic side, high-ionization emission lines (e.g., $\rm Mg\, \textsc{ii}$, $\rm C\, \textsc{iv}$) generally follow the Baldwin effect (\citealt{1977ApJ...214..679B, 2024ApJ...963....7R}), whereas the definition of CL quasars inherently makes them a special population that tends to exhibit an inverse Baldwin effect (the detail is shown in Section \ref{sec: Baldwin effect}). As a result, CL behavior involving high-ionization lines is expected to be rarer than that involving low-ionization lines.
Moreover, high-redshift quasars are observed near the so-called ``cosmic noon'', when quasar activity is at its peak; the high average accretion rates during this epoch may reduce the fraction of sources undergoing dramatic state changes.

\subsection{Eddington Ratio}

The Eddington ratio is a critical parameter to explore the physical mechanism of the CL quasars. Before measuring the Eddington ratio, we first calculate the mass of the black hole.
Here, we assume a virialized BLR and estimate the black hole mass following \cite{2011ApJS..194...45S}:

\begin{eqnarray}
\log\left(\frac{M_{\rm BH}}{M_{\odot}}\right)
&=& a + b\log\left(\frac{\lambda L_{\lambda}}{10^{44}\ \mathrm{erg\,s^{-1}}}\right) \nonumber \\
& & \quad + 2\log\left(\frac{\mathrm{FWHM}}{\mathrm{km\,s^{-1}}}\right),
\end{eqnarray}
where the values of parameters $a$ and $b$ vary for different emission lines in the virial black hole mass estimation.
For sources with reliable $\rm Mg\,\textsc{ii}$ measurements ($\rm SNR > 2$), we estimate black hole masses using the FWHM of the $\rm Mg\,\textsc{ii}$ broad emission line and the monochromatic luminosity at 3000\,\AA~in the bright state, with coefficients $a = 0.740$ and $b = 0.62$ (\citealt{2011ApJS..194...45S}). 
For the remaining sources, either due to the $\rm Mg\,\textsc{ii}$ line falling outside the spectral coverage or having low SNR, we instead adopt the $\rm C\,\textsc{iv}$ line in the bright spectra, using $L_{1350}$ as the continuum luminosity and coefficients $a = 0.66$ and $b = 0.53$ (\citealt{2006ApJ...641..689V}).

The bolometric luminosity can be derived from the continuum luminosity using bolometric corrections $L_{\rm bol} = 5.15L_{3000}$ or $L_{\rm bol} = 3.81L_{1350}$, following \cite{2006ApJS..166..470R}. Based on the black hole mass estimated from the bright-state spectrum, we then calculate the Eddington ratios for both the bright and dim states:

\begin{equation}
    \lambda_{\rm Edd} = \frac{L_{\rm bol}}{L_{\rm Edd}} = \frac{L_{\rm bol}}{1.26 \times 10^{38} M_{\rm BH}/M_{\rm \sun}}.
\end{equation}

\begin{figure*}
    \centering
    \includegraphics[width=1\linewidth]{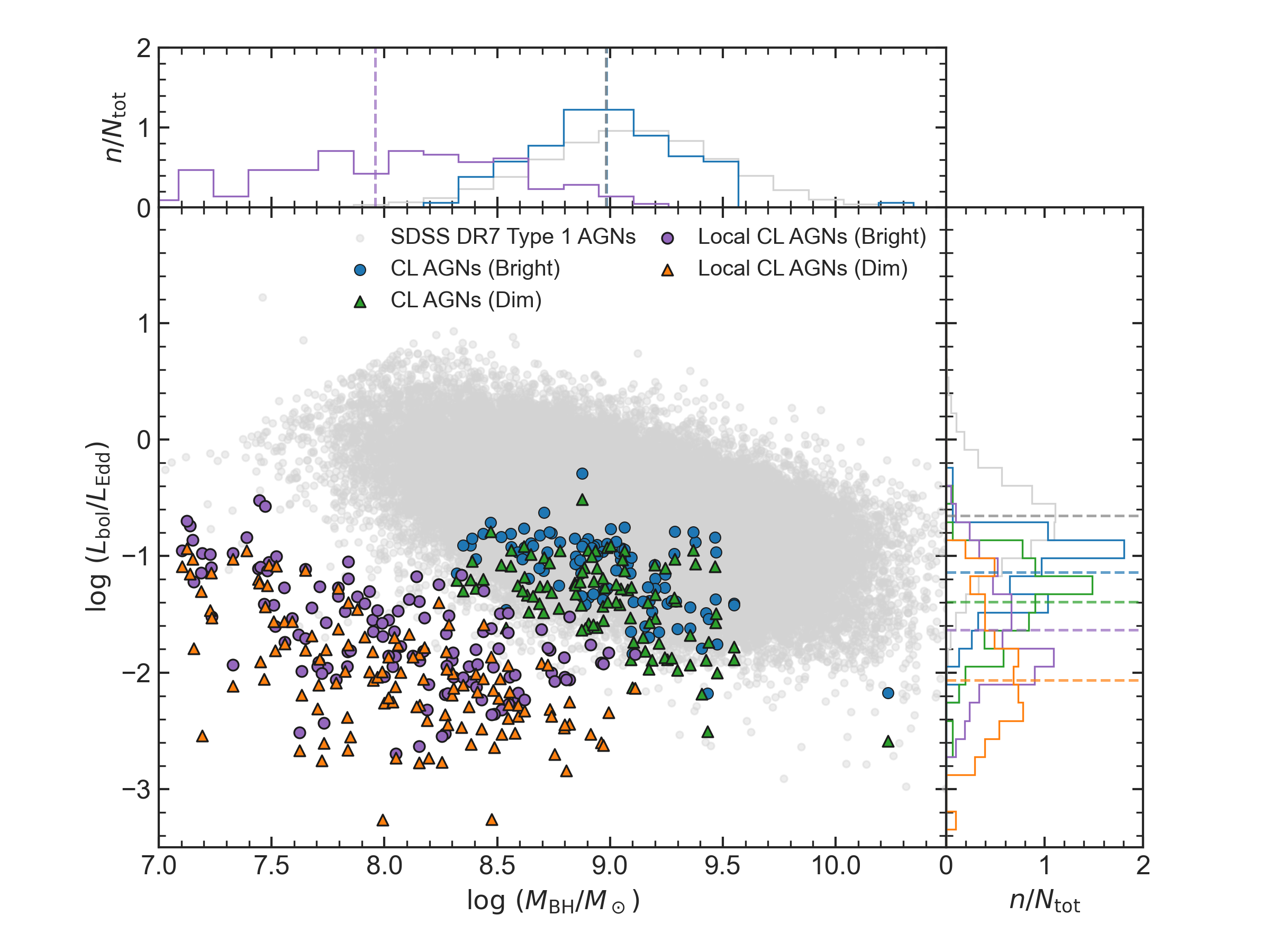}
    \caption{The distribution of CL quasars in this work on the $L_{\rm bol}/L_{\rm Edd}$–$M_{\rm BH}$ plane is shown alongside a comparison quasar sample (grey circles) from the SDSS DR7 QSO catalog at $z > 0.9$ \citep{2011ApJS..194...45S}. Blue circles and green triangles indicate the bright and dim states, respectively. The purple dots and orange triangles are the local CL quasars ($z<0.3$) from \cite{2025ApJS..278...28G} in bright and dim states, respectively. And the dashed lines in the corresponding color represent the linear fitting for the targets in the $L_{\rm bol}/L_{\rm Edd}$–$M_{\rm BH}$ plane. The top and right subpanels display the distributions of $M_{\rm BH}$ and Eddington ratio, with dashed lines marking their respective average values.}
    \label{fig: eddington ratio}
\end{figure*}

Figure~\ref{fig: eddington ratio} presents the distribution of our CL quasar sample in the $L_{\rm bol}/L_{\rm Edd}$–$M_{\rm BH}$ plane, in comparison with SDSS quasars at $z > 0.9$ and a local CL quasar sample ($z < 0.3$) from \citet{2025ApJS..278...28G}.
Compared to typical quasars (median $\log \lambda_{\rm Edd} = -0.65 \pm 0.002$; \citealt{2011ApJS..194...45S}), our CL quasars span a similar range of black hole masses but exhibit a statistically significant shift toward lower Eddington ratios. The median Eddington ratios for our sample are $\log \lambda_{\rm Edd} = -1.14 \pm 0.12$ in the bright state and $-1.39 \pm 0.09$ in the dim state. The uncertainties are estimated by bootstrap resampling with 1000 iterations. We employed the Kolmogorov-Smirnov (KS) test to evaluate the differences in Eddington ratio distributions. The test reveals that both the bright-state ($p < 10^{-4}$) and dim-state ($p < 10^{-4}$) CL quasars exhibit significantly different Eddington ratio distributions compared to the typical SDSS quasar population.

When compared with lower-redshift CL quasars, our high-redshift sample exhibits systematically higher Eddington ratios. Local CL quasars typically have $\log \lambda_{\rm Edd} \sim -2$ (e.g., \citealt{2024ApJ...966..128W, 2025ApJS..278...28G}), significantly lower than our bright-state average of $\log \lambda_{\rm Edd} \sim -1.14 \pm 0.12$ ($p < 10^{-4}$, KS test).
This difference is most likely due to selection bias: at higher redshifts, flux limits preferentially select the most luminous, and thus more highly accreting, CL quasars.

Furthermore, Figure~\ref{fig: eddington ratio} shows the distributions of typical quasars and bright- and dim-state CL quasars in the $L_{\rm bol}/L_{\rm Edd}$–$M_{\rm BH}$ plane. CL quasars are systematically shifted toward lower accretion rates compared to typical quasars, and the bright-state distribution lies consistently above the dim-state distribution, which is also statistically significant ($p < 10^{-4}$, KS test), reinforcing the idea that the CL phenomenon is closely linked to changes in the accretion rate.

\subsection{High-ionization Broad Emission Line Ratio}

The C\, $\rm \textsc{iii}]$/C\, $\rm \textsc{iv}$ emission-line ratio is widely regarded to be sensitive to the ionization parameter ($U$), with a larger C\, $\rm \textsc{iii}]$/C\, $\rm \textsc{iv}$ typically indicating a lower ionization parameter (e.g., \citealt{2005MNRAS.356..778S, 2009A&A...503..721M}). According to the photoionization model (\citealt{1998PASP..110..761F, 2004ApJS..155..675K}):

\begin{equation}
    U = \frac{\int_{\nu_{0}}^{\infty} L_{\nu} \, d\nu/h\nu}{4\pi R^{2} c\, n_{\rm H}},
\end{equation}
where $R$ is the distance from the BLRs to the central ionizing source, and $n_{\rm H}$ is the hydrogen density. A larger $U$ indicates the presence of lower-density gas located closer to the central SMBH.   

\begin{figure}
    \centering
    \includegraphics[width=0.45\textwidth]{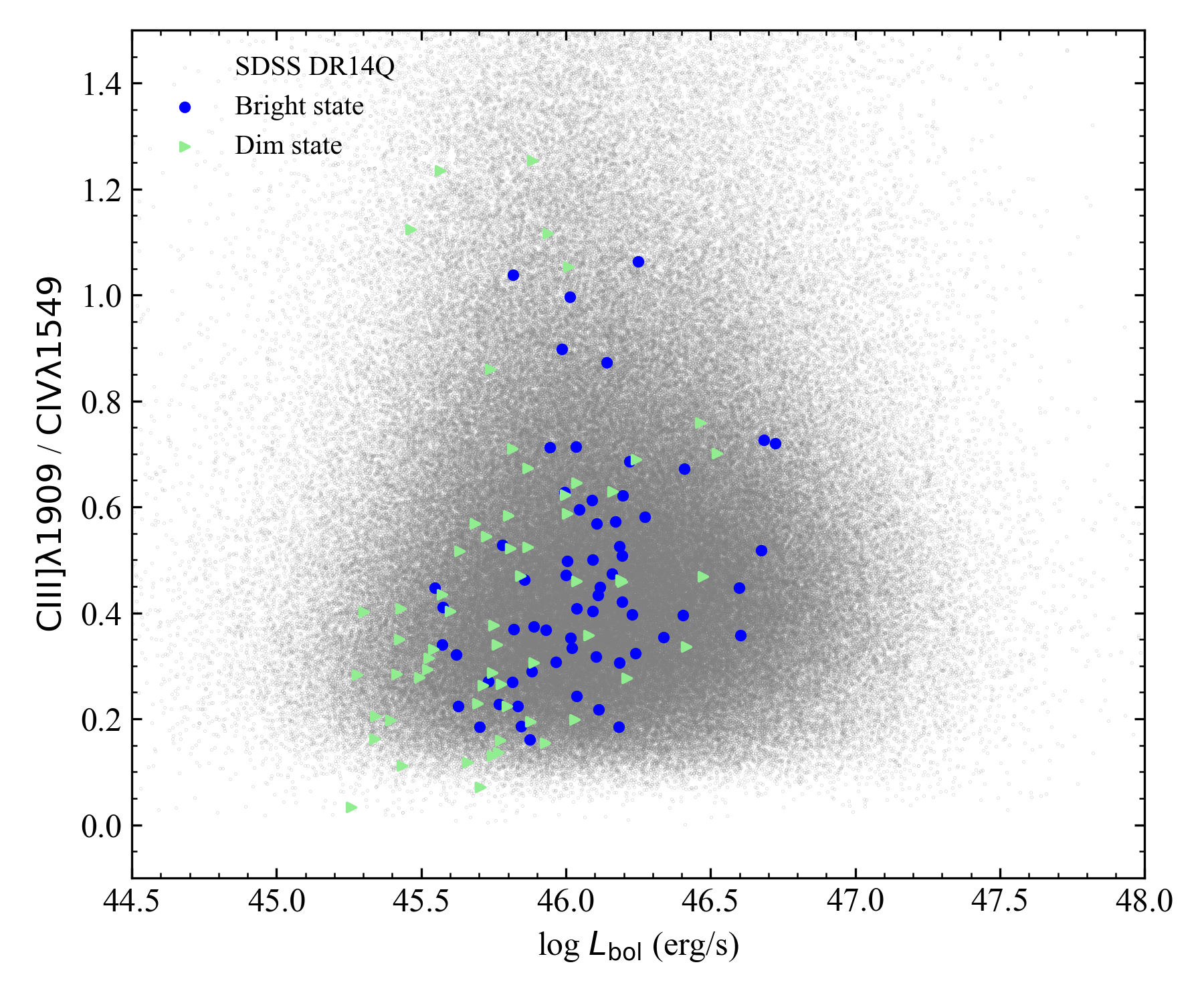}
    \caption{The $\rm C \textsc{iii}] \lambda 1909$/$\rm C \textsc{iv} \lambda 1549$ V.S. $L_{\rm bol}$ diagram for the CL quasars in this work with both $\rm C\, \textsc{iii}]$ and $\rm C\, \textsc{iv}$ emission lines detected. Blue circles and green triangles represent CL quasars in bright and dim states, respectively. Grey dots, shown for comparison, are drawn from the SDSS DR14 QSO catalog (\citealt{2020ApJS..249...17R}).}
    \label{fig: line_ratio}
\end{figure}

In this work, we find no significant correlation between the C\, $\rm \textsc{iii}]$/C\, $\rm \textsc{iv}$ flux ratio and the thermal luminosity, as shown in Figure \ref{fig: line_ratio}, and the distribution for CL quasars is statistically consistent with that of typical quasars. This indicates that the $U$ does not vary significantly between the bright and dim states, suggesting that the distance between the BLR clouds and the central engine remains largely unchanged during the CL transition.

\subsection{Baldwin Effect}\label{sec: Baldwin effect}

The Baldwin effect refers to the phenomenon observed in quasar spectra where the equivalent width (EW) of high-ionization emission lines ($\rm C\, \textsc{iv}$) decreases with increasing continuum luminosity, following the relation $L_{\rm con} \propto 1/EW$ (\citealt{1977ApJ...214..679B}).
However, the underlying physical mechanisms of the Baldwin Effect remain controversial. Several prevailing explanations include: at higher luminosities, the UV-soft X-ray spectrum of the central gravitational source becomes softer, resulting in reduced efficiency for exciting specific emission lines, thereby causing the line intensity to increase less rapidly than the continuum luminosity (\citealt{2004MNRAS.350L..31B, 2009ApJ...702..767W}); Bright AGNs exhibit lower ionization parameters, which similarly lead to relative weakening of line intensity (\citealt{1984ApJ...278..558M}); the time lag of emission-line response leads to a pseudo-Baldwin effect (\citealt{1992AJ....103.1084P}).
Despite the clear Baldwin effect seen in high-ionization lines like $\rm C\, \textsc{iv}$, the behavior of lower-ionization lines, such as $\rm Mg\,\textsc{ii}$, appears to be more complex. Several studies have reported that the equivalent width of $\rm Mg\,\textsc{ii}$ shows a much weaker dependence on continuum luminosity, or even no significant Baldwin effect at all (e.g., \citealt{2011ApJS..194...45S, 2019ApJ...883L..44G}).

\begin{figure*}
    \centering
    \includegraphics[width=1\textwidth]{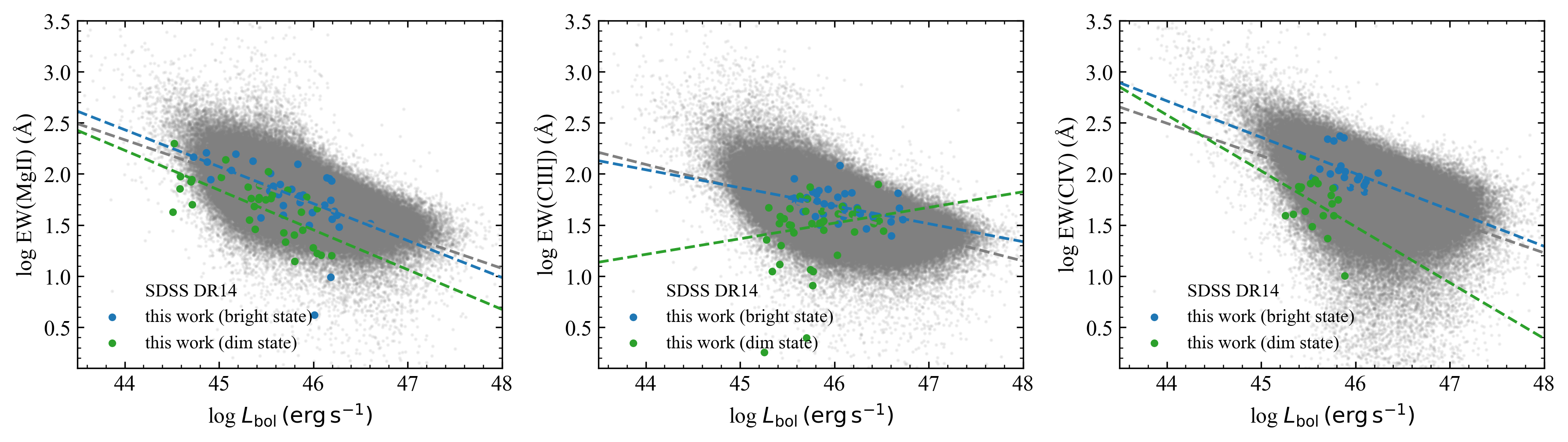}
    \caption{The EW-$L_{\rm bol}$ plane of the CL quasar subsamples in this work. The blue circles and green triangles represent the CL ANGs in the bright states and the dim states, respectively. The grey circles show a comparison quasar sample from the SDSS DR14 QSO catalog (\citealt{2020ApJS..249...17R}). The corresponding color dashed line is the best linear fit for the scatter.}
    \label{fig: Baldwin effect}
\end{figure*}

\begin{figure*}
    \centering
    \includegraphics[width=1\textwidth]{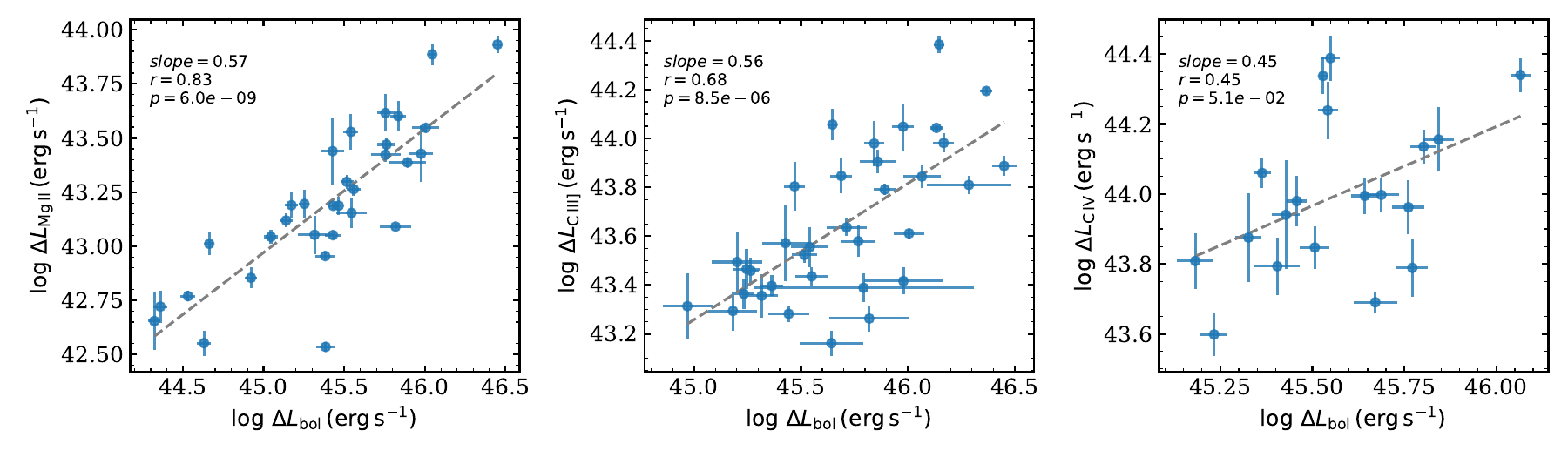}
    \caption{Correlation between the change in BEL luminosity ($\Delta L_{\rm line}$) and the change in bolometric luminosity ($\Delta L_{\rm bol}$) for CL quasars in this work. From left to right, the panels show the results for $\rm Mg\,\textsc{ii}$, $\rm C\,\textsc{iii]}$, and $\rm C\,\textsc{iv}$ lines. Each point represents an individual CL quasar with detected line variability. The dashed lines indicate the best-fit linear relations, with the corresponding slope, Spearman correlation coefficient $r$, and $p$-value labeled in each panel.}
    \label{fig: line_vari}
\end{figure*}

To investigate the Baldwin effect for different BELs, we divide our CL quasars into distinct subsamples based on the emission-line CL behavior. Figure~\ref{fig: Baldwin effect} presents the relationship between the EW and luminosity for several high-ionization emission lines.
At the population level, CL quasars follow the Baldwin effect well for both $\rm Mg\, \textsc{ii}$ and $\rm C\, \textsc{iv}$ lines, with stronger anti-correlations than those observed in the general quasar population, regardless of whether the source is in a bright or dim state. However, the Baldwin effect does not hold for individual CL quasars. We find that the EWs in the bright-state spectra are systematically higher than in the dim-state ones for the same source, which is contrary to the expected Baldwin trend.
This apparent contradiction may result from a selection bias: since CL quasars are defined by the appearance or disappearance of BELs, the bright state is, by definition, more likely to exhibit prominent lines with large EWs.

Notably, the $\rm C\, \textsc{iii}]$ line in the dim state does not exhibit the expected anti-correlation between EW and luminosity. According to the photoionization models of \citet{2020ApJ...888...58G}, the intrinsic line strength of $\rm C\, \textsc{iii}]$ is only about $\sim$10–15\% that of Ly$\alpha$, resulting in larger measurement uncertainties. Thus, the lack of an observable Baldwin effect for $\rm C\, \textsc{iii}$ line in the dim state may be due to its inherently low luminosity.

Furthermore, we examine the correlation between the changes in absolute emission line luminosities and changes in continuum luminosity variations for individual CL quasars, as shown in Figure~\ref{fig: line_vari}. We find a strong positive correlation between the variability of high-ionization lines (e.g., $\rm Mg\, \textsc{ii}$, $\rm C\, \textsc{iii}]$) and the change in bolometric luminosity. The slopes are 0.57 and 0.65, with Spearman correlation coefficients $r$ values are 0.83 and 0.64, respectively, and $p$-values indicating high statistical significance. This result is consistent with a scenario in which the activity of CL quasars is primarily driven by the nuclear region, and the emission-line strength increases with the overall ionizing luminosity, for $\rm C\,\textsc{iv}$.
Interestingly,  the correction is weaker than for $\rm Mg\,\textsc{ii}$ and $\rm C\,\textsc{iii}]$ (Figure~\ref{fig: line_vari}). This is plausibly due to its more compact and dynamic region, likely closer to the accretion disk. As such, it may be more susceptible to complex radiative transfer effects, anisotropic illumination, or perturbations from disk winds and outflows (e.g., \citealt{2011AJ....141..167R, 2012ApJ...759...44D}). These processes can decouple the line response from the overall ionizing continuum, leading to weaker or delayed variability. Exactly, $\rm C\,\textsc{iv}$ is known to be more blueshifted and asymmetric than other broad lines, further suggesting a non-virialized or wind-dominated origin.

\subsection{CL Timescale\label{subsec: timescale}}

The CL timescale refers to the duration over which a Type 1 AGN transforms into a Type 2 AGN (or vice versa), which has later evolved into the timescale for the appearance or disappearance of a particular broad emission line.
The observed CL timescale poses a challenge to the typical accretion disk model of AGNs and is crucial to understanding the physical mechanism behind the CL event.  
Theoretically, the predicted timescale for such transitions is on the order of $10^4$ years (\citealt{1973A&A....24..337S}), which is much longer than the actual observed timescales.
However, even the observed CL timescales represent only upper limits, as they depend on the sampling cadence of spectroscopic observations in the bright and dim states. The typical CL timescale is 1-20 years (e.g., \citealt{2018ApJ...862..109Y, 2023NatAs...7.1282R}).
It is noteworthy that some extreme cases exhibit transitions on timescales of just a few months. For instance, the atypical CL AGN 1ES 1927+654 showcased a rapid turn-on over several weeks (\citealt{2019ApJ...883...94T}), while the CL quasar SDSS J1628+4329 experienced a dramatic dimming and rebrightening within months, potentially driven by variable obscuration (\citealt{2022ApJ...939L..16Z}). Although these specific physical mechanisms may not be representative of the general CL population, they serve to illustrate the observed upper limit of rapid variability that any comprehensive theory must ultimately explain.

\begin{figure*}
    \centering
    \includegraphics[width=1\textwidth]{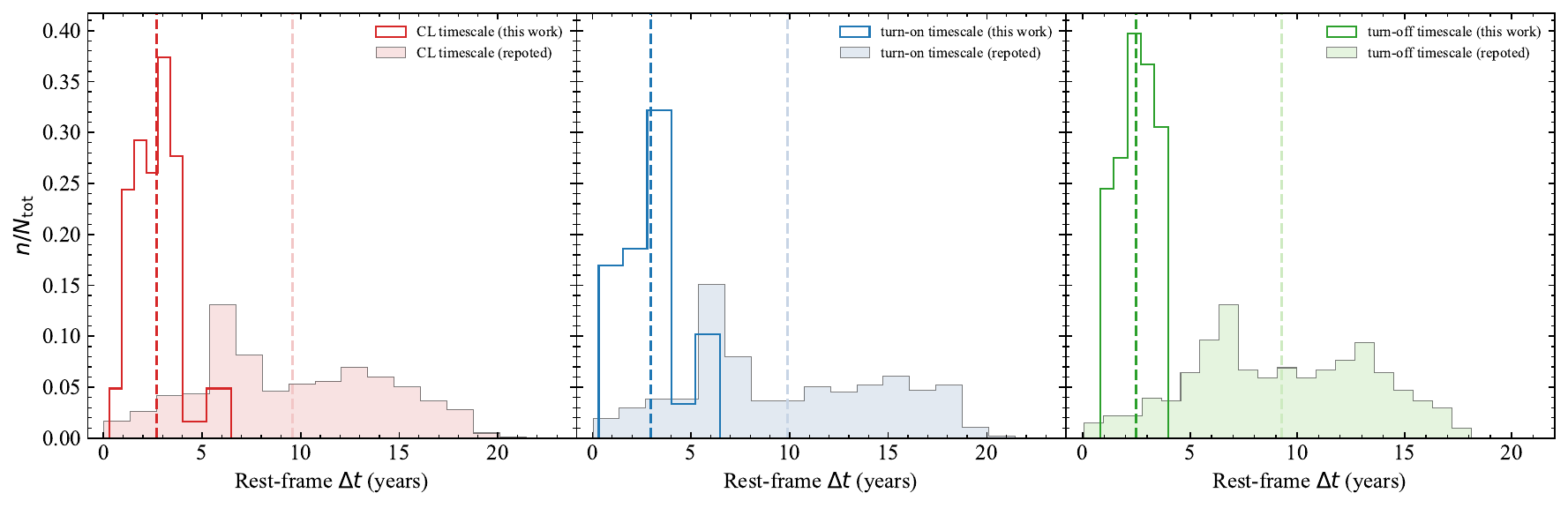}
    \caption{The timescale of the CL phenomenon in this work and reported CL quasars. The left panel is the CL timescale distribution in this work and the previous studies, the middle and the right panels represent the turn-on and turn-off timescale, respectively. The open bars represent the timescale distributions in our sample, and the filled bars correspond to previously reported CL quasars (e.g., \citealt{2016ApJ...831..157M, 2018ApJ...862..109Y, 2023MNRAS.524..188L, 2024ApJ...966..128W, 2024ApJS..270...26G, 2025ApJS..278...28G}). The dashed lines in matching colors indicate the average timescales for each group.}
    \label{fig: CL timescale}
\end{figure*}

In this work, the rest-frame CL timescale for each source is also calculated directly as the time elapsed between the last spectroscopic observation before the transition and the first observation after the transition has occurred. As we lack continuous monitoring, this measured interval represents an upper limit to the true transition timescale.
As illustrated in Figure \ref{fig: CL timescale}, the rest-frame CL timescales span 0.32–6.47 years for turn-on AGNs and 0.82–3.97 years for turn-off AGNs, with one outlier extending to 11.72 years. It is found that there is no significant difference in the timescales between turn-on and turn-off events, with both having an average of around 3 years. However, the CL timescales at high redshift appear to be much shorter than those reported in previous studies, which typically show an average of around 10 years.
We note that the spectroscopic cadence of our data, combining the latest DESI DR1 with the long historical baseline of SDSS, is comparable to, if not longer than, that of these previous works (e.g., \citealt{2018ApJ...862..109Y, 2024ApJ...966..128W, 2024ApJS..270...26G, 2025ApJS..278...28G}). Therefore, the shorter timescales we measure are unlikely to be a cadence artifact.
Also, we do not interpret this as an indication of redshift evolution in CL timescales, but rather as a result of selection effects. Firstly, as mentioned earlier, due to instrumental limitations, the measured CL timescales are strongly influenced by observation cadence. The typical time baselines of existing large-scale spectroscopic surveys are around 10–20 years. Secondly, because of redshift effects, the rest-frame CL timescales inferred for low-redshift sources tend to be $\sim$10 years, while those at higher redshifts are significantly shorter.
Actually, some studies based on light curve analysis suggest that the intrinsic CL timescales may be significantly shorter than the observed ones, possibly even shorter than those estimated in our sample (e.g., \citealt{2017ApJ...846L...7S, 2024ApJ...966..128W}). Therefore, more complete datasets and the discovery of CL quasars at even higher redshifts will be crucial to uncover the true nature of CL timescales.

In the preceding subsections, we investigated the Eddington ratio, the Baldwin effect, and the CL timescale of CL quasars. Taken together, these measurements are consistent with a scenario in which CL activity is primarily triggered by variations in the accretion rate. Specifically, the Eddington ratio of CL quasars in the bright state is systematically larger than that in the dim state, providing direct evidence for the role of accretion-rate changes. The Baldwin effect of the high-ionization lines is consistent with the disk-wind scenario (\citealt{2019A&A...630A..94G}), in which changes in the accretion rate may regulate both the ionization state and the dynamical structure of the BLR. Furthermore, our analysis of the CL timescale constrains the transition period to less than $\sim$3 years, significantly shorter than the $\sim$10 years reported in previous studies. This short timescale cannot be explained by the viscous timescale of a standard disk or obscuration mechanisms, and it places stronger constraints on the accretion-rate variation model.

Taken together, these results outline a coherent picture: CL phenomena are predominantly driven by accretion-rate variations (e.g., disk instability models). When the accretion rate increases, the disk-wind scenario predicts an enhanced supply of BLR material, leading to the emergence of strong BELs. Conversely, when the accretion rate decreases, the BLR gas content diminishes, and BELs weaken or even disappear.

\section{Summary}\label{Sec: Summary}

In this study, we conduct a systematic search for high-redshift CL quasars by cross-matching SDSS DR18 and DESI DR1. We identify 97 pairs of high-redshift CL quasars and 135 CL quasar candidates at redshift $z > 0.9$. Among them, 28 cases exhibit CL transitions in Mg\,\textsc{ii}, 59 in C\,\textsc{iii}], 25 in C\,\textsc{iv}, 26 in Si\,\textsc{iv}, and 8 in Ly$\alpha$. The 97 CL quasars comprise 45 turn-on and 52 turn-off events, with a near-equal occurrence ratio. The resulting detection rate of 0.042\% is significantly lower than the detection rates found for CL transitions from low ionization emission lines. This discrepancy is primarily attributed to four factors: (1) observational limitations, where high-redshift quasars in dim states may fall below the detection thresholds of current surveys; and (2) the baseline between SDSS and DESI is shorter than the CL timescale; (3) physical constraints, as high-ionization lines generally follow the Baldwin effect well, making it more difficult to identify cases with an inverse Baldwin effect; and (4) the high-redshift quasars activity is at its peak that suppression the state change.

Our sample properties support an accretion-rate-driven origin for the CL phenomenon. They have lower Eddington ratios than typical quasars, averaging $\log \lambda_{\rm Edd} \sim -1.14$ and $-1.39$ in the bright and dim states, respectively. This supports the idea that CL behavior preferentially occurs in low accretion states. 
In addition, we find that high-ionization emission lines in CL quasars still follow the Baldwin effect on a population level, though individual sources can exhibit inverse Baldwin trends. Not only that, our results reveal a strong positive correlation between the changes in bolometric luminosity and the flux variability of high-ionization lines such as Mg\,\textsc{ii} and C\,\textsc{iii]}, further supporting that the CL transitions are driven by changes in the accretion rate.  Finally, we examine the timescales of CL behavior and find no significant difference between turn-on and turn-off events. The typical timescale of the CL transition is estimated to be around 3 years in the rest frame, which disfavours the obscuration-based scenario. Together, these results indicate changes in the accretion rate.

\section*{acknowledgements}

We acknowledge the anonymous referee for valuable comments that helped to improve the paper. 
This work is  supported by the National Natural Science Foundation of China (NSFC) under grant No. 12503019 and 12273013. This research is supported by the National Key R\&D Program of China with grant No. 2023YFA1608100. This work is also supported by the National Natural Science Foundation of China (NSFC) under grant No. 12273013. The authors also acknowledge support from the National Key R\&D Program of China (grant Nos. 2023YFA1607804, 2022YFA1602902, 2023YFA1607800), other NSFC projects (grant Nos. 12120101003, 12373010, 12173051, 12233008, 12403022, and 12103048), and the China Manned Space Project (No. CMS-CSST-2025-A06). The authors also acknowledge the Strategic Priority Research Program of the Chinese Academy of Sciences with grant Nos. XDB0550100 and XDB0550000.

We acknowledge the use of DESI data. DESI is a scientific collaboration managed by the U.S. Department of Energy's Lawrence Berkeley National Laboratory, with primary funding provided by the U.S. Department of Energy Office of Science and the National Science Foundation. Additional support is acknowledged from international collaborating institutions and participating universities.
This work also utilizes data from the SDSS. Funding for SDSS-V has been provided by the Alfred P. Sloan Foundation, the Participating Institutions, the National Science Foundation, and the U.S. Department of Energy Office of Science. 

We acknowledge the efforts for public data from CTRS, PS1, and ZTF. The Catalina Sky Survey is funded by the National Aeronautics and Space Administration under grant No. NNG05GF22G, issued through the Science Mission Directorate Near-Earth Objects Observations Program. The CRTS survey is supported by the US National Science Foundation under grants AST-0909182 and AST-1313422.
The CRTS survey is supported by the US National Science Foundation under grants AST-0909182 and AST-1313422. PS1 has been made possible through contributions by the Institute for Astronomy, the University of Hawaii, the Pan-
STARRS Project Ofﬁce, the Max-Planck Society, and its participating institutes, the Max Planck Institute for Astronomy, Heidelberg and the Max Planck Institute for Extraterrestrial Physics, Garching, The Johns Hopkins University, Durham University, the University of Edinburgh, Queen's
University of Belfast, the Harvard–Smithsonian Center for Astrophysics, the Las Cumbres Observatory Global Telescope
Network Incorporated, the National Central University of Taiwan, the Space Telescope Science Institute, the National Aeronautics and Space Administration under grant No.
NNX08AR22G issued through the Planetary Science Division
of the NASA Science Mission Directorate, the National Science Foundation under grant No. AST-1238877, the University of Maryland, and Eotvos Lorand University (ELTE). ZTF is supported by the National Science Foundation under grant No. AST-2034437 and a collaboration including Caltech, IPAC, the Weizmann Institute for Science, the Oskar Klein Center at Stockholm University, the University of Maryland,
Deutsches Elektronen-Synchrotron and Humboldt University, the TANGO Consortium of Taiwan, the University of Wisconsin at Milwaukee, Trinity College Dublin, Lawrence Livermore National Laboratories, and IN2P3, France. Operations are conducted by COO, IPAC, and UW.

\bibliography{refer}{}

@ARTICLE{2013arXiv1308.0847L,
       author = {{Levi}, Michael and {Bebek}, Chris and {Beers}, Timothy and {Blum}, Robert and {Cahn}, Robert and {Eisenstein}, Daniel and {Flaugher}, Brenna and {Honscheid}, Klaus and {Kron}, Richard and {Lahav}, Ofer and {McDonald}, Patrick and {Roe}, Natalie and {Schlegel}, David and {representing the DESI collaboration}},
        title = "{The DESI Experiment, a whitepaper for Snowmass 2013}",
      journal = {arXiv e-prints},
         year = 2013,
        month = aug,
          eid = {arXiv:1308.0847},
        pages = {arXiv:1308.0847},
          doi = {10.48550/arXiv.1308.0847},
archivePrefix = {arXiv},
       eprint = {1308.0847},
 primaryClass = {astro-ph.CO},
       adsurl = {https://ui.adsabs.harvard.edu/abs/2013arXiv1308.0847L}
}

@ARTICLE{2016arXiv161100036D,
       author = {{DESI Collaboration} and {Aghamousa}, Amir and {Aguilar}, Jessica and {Ahlen}, Steve and {Alam}, Shadab and {Allen}, Lori E. and {Allende Prieto}, Carlos and {Annis}, James and {Bailey}, Stephen and {Balland}, Christophe and {Ballester}, Otger and {Baltay}, Charles and {Beaufore}, Lucas and {Bebek}, Chris and {Beers}, Timothy C. and {Bell}, Eric F. and {Bernal}, Jos{\'e} Luis and {Besuner}, Robert and {Beutler}, Florian and {Blake}, Chris and {Bleuler}, Hannes and {Blomqvist}, Michael and {Blum}, Robert and {Bolton}, Adam S. and {Briceno}, Cesar and {Brooks}, David and {Brownstein}, Joel R. and {Buckley-Geer}, Elizabeth and {Burden}, Angela and {Burtin}, Etienne and {Busca}, Nicolas G. and {Cahn}, Robert N. and {Cai}, Yan-Chuan and {Cardiel-Sas}, Laia and {Carlberg}, Raymond G. and {Carton}, Pierre-Henri and {Casas}, Ricard and {Castander}, Francisco J. and {Cervantes-Cota}, Jorge L. and {Claybaugh}, Todd M. and {Close}, Madeline and {Coker}, Carl T. and {Cole}, Shaun and {Comparat}, Johan and {Cooper}, Andrew P. and {Cousinou}, M. -C. and {Crocce}, Martin and {Cuby}, Jean-Gabriel and {Cunningham}, Daniel P. and {Davis}, Tamara M. and {Dawson}, Kyle S. and {de la Macorra}, Axel and {De Vicente}, Juan and {Delubac}, Timoth{\'e}e and {Derwent}, Mark and {Dey}, Arjun and {Dhungana}, Govinda and {Ding}, Zhejie and {Doel}, Peter and {Duan}, Yutong T. and {Ealet}, Anne and {Edelstein}, Jerry and {Eftekharzadeh}, Sarah and {Eisenstein}, Daniel J. and {Elliott}, Ann and {Escoffier}, St{\'e}phanie and {Evatt}, Matthew and {Fagrelius}, Parker and {Fan}, Xiaohui and {Fanning}, Kevin and {Farahi}, Arya and {Farihi}, Jay and {Favole}, Ginevra and {Feng}, Yu and {Fernandez}, Enrique and {Findlay}, Joseph R. and {Finkbeiner}, Douglas P. and {Fitzpatrick}, Michael J. and {Flaugher}, Brenna and {Flender}, Samuel and {Font-Ribera}, Andreu and {Forero-Romero}, Jaime E. and {Fosalba}, Pablo and {Frenk}, Carlos S. and {Fumagalli}, Michele and {Gaensicke}, Boris T. and {Gallo}, Giuseppe and {Garcia-Bellido}, Juan and {Gaztanaga}, Enrique and {Pietro Gentile Fusillo}, Nicola and {Gerard}, Terry and {Gershkovich}, Irena and {Giannantonio}, Tommaso and {Gillet}, Denis and {Gonzalez-de-Rivera}, Guillermo and {Gonzalez-Perez}, Violeta and {Gott}, Shelby and {Graur}, Or and {Gutierrez}, Gaston and {Guy}, Julien and {Habib}, Salman and {Heetderks}, Henry and {Heetderks}, Ian and {Heitmann}, Katrin and {Hellwing}, Wojciech A. and {Herrera}, David A. and {Ho}, Shirley and {Holland}, Stephen and {Honscheid}, Klaus and {Huff}, Eric and {Hutchinson}, Timothy A. and {Huterer}, Dragan and {Hwang}, Ho Seong and {Illa Laguna}, Joseph Maria and {Ishikawa}, Yuzo and {Jacobs}, Dianna and {Jeffrey}, Niall and {Jelinsky}, Patrick and {Jennings}, Elise and {Jiang}, Linhua and {Jimenez}, Jorge and {Johnson}, Jennifer and {Joyce}, Richard and {Jullo}, Eric and {Juneau}, St{\'e}phanie and {Kama}, Sami and {Karcher}, Armin and {Karkar}, Sonia and {Kehoe}, Robert and {Kennamer}, Noble and {Kent}, Stephen and {Kilbinger}, Martin and {Kim}, Alex G. and {Kirkby}, David and {Kisner}, Theodore and {Kitanidis}, Ellie and {Kneib}, Jean-Paul and {Koposov}, Sergey and {Kovacs}, Eve and {Koyama}, Kazuya and {Kremin}, Anthony and {Kron}, Richard and {Kronig}, Luzius and {Kueter-Young}, Andrea and {Lacey}, Cedric G. and {Lafever}, Robin and {Lahav}, Ofer and {Lambert}, Andrew and {Lampton}, Michael and {Landriau}, Martin and {Lang}, Dustin and {Lauer}, Tod R. and {Le Goff}, Jean-Marc and {Le Guillou}, Laurent and {Le Van Suu}, Auguste and {Lee}, Jae Hyeon and {Lee}, Su-Jeong and {Leitner}, Daniela and {Lesser}, Michael and {Levi}, Michael E. and {L'Huillier}, Benjamin and {Li}, Baojiu and {Liang}, Ming and {Lin}, Huan and {Linder}, Eric and {Loebman}, Sarah R. and {Luki{\'c}}, Zarija and {Ma}, Jun and {MacCrann}, Niall and {Magneville}, Christophe and {Makarem}, Laleh and {Manera}, Marc and {Manser}, Christopher J. and {Marshall}, Robert and {Martini}, Paul and {Massey}, Richard and {Matheson}, Thomas and {McCauley}, Jeremy and {McDonald}, Patrick and {McGreer}, Ian D. and {Meisner}, Aaron and {Metcalfe}, Nigel and {Miller}, Timothy N. and {Miquel}, Ramon and {Moustakas}, John and {Myers}, Adam and {Naik}, Milind and {Newman}, Jeffrey A. and {Nichol}, Robert C. and {Nicola}, Andrina and {Nicolati da Costa}, Luiz and {Nie}, Jundan and {Niz}, Gustavo and {Norberg}, Peder and {Nord}, Brian and {Norman}, Dara and {Nugent}, Peter and {O'Brien}, Thomas and {Oh}, Minji and {Olsen}, Knut A.~G.},
        title = "{The DESI Experiment Part I: Science,Targeting, and Survey Design}",
      journal = {arXiv e-prints},
         year = 2016,
        month = oct,
          eid = {arXiv:1611.00036},
        pages = {arXiv:1611.00036},
          doi = {10.48550/arXiv.1611.00036},
archivePrefix = {arXiv},
       eprint = {1611.00036},
 primaryClass = {astro-ph.IM},
       adsurl = {https://ui.adsabs.harvard.edu/abs/2016arXiv161100036D}
}

@ARTICLE{2016arXiv161100037D,
       author = {{DESI Collaboration} and {Aghamousa}, Amir and {Aguilar}, Jessica and {Ahlen}, Steve and {Alam}, Shadab and {Allen}, Lori E. and {Allende Prieto}, Carlos and {Annis}, James and {Bailey}, Stephen and {Balland}, Christophe and {Ballester}, Otger and {Baltay}, Charles and {Beaufore}, Lucas and {Bebek}, Chris and {Beers}, Timothy C. and {Bell}, Eric F. and {Bernal}, Jos{\'e} Luis and {Besuner}, Robert and {Beutler}, Florian and {Blake}, Chris and {Bleuler}, Hannes and {Blomqvist}, Michael and {Blum}, Robert and {Bolton}, Adam S. and {Briceno}, Cesar and {Brooks}, David and {Brownstein}, Joel R. and {Buckley-Geer}, Elizabeth and {Burden}, Angela and {Burtin}, Etienne and {Busca}, Nicolas G. and {Cahn}, Robert N. and {Cai}, Yan-Chuan and {Cardiel-Sas}, Laia and {Carlberg}, Raymond G. and {Carton}, Pierre-Henri and {Casas}, Ricard and {Castander}, Francisco J. and {Cervantes-Cota}, Jorge L. and {Claybaugh}, Todd M. and {Close}, Madeline and {Coker}, Carl T. and {Cole}, Shaun and {Comparat}, Johan and {Cooper}, Andrew P. and {Cousinou}, M. -C. and {Crocce}, Martin and {Cuby}, Jean-Gabriel and {Cunningham}, Daniel P. and {Davis}, Tamara M. and {Dawson}, Kyle S. and {de la Macorra}, Axel and {De Vicente}, Juan and {Delubac}, Timoth{\'e}e and {Derwent}, Mark and {Dey}, Arjun and {Dhungana}, Govinda and {Ding}, Zhejie and {Doel}, Peter and {Duan}, Yutong T. and {Ealet}, Anne and {Edelstein}, Jerry and {Eftekharzadeh}, Sarah and {Eisenstein}, Daniel J. and {Elliott}, Ann and {Escoffier}, St{\'e}phanie and {Evatt}, Matthew and {Fagrelius}, Parker and {Fan}, Xiaohui and {Fanning}, Kevin and {Farahi}, Arya and {Farihi}, Jay and {Favole}, Ginevra and {Feng}, Yu and {Fernandez}, Enrique and {Findlay}, Joseph R. and {Finkbeiner}, Douglas P. and {Fitzpatrick}, Michael J. and {Flaugher}, Brenna and {Flender}, Samuel and {Font-Ribera}, Andreu and {Forero-Romero}, Jaime E. and {Fosalba}, Pablo and {Frenk}, Carlos S. and {Fumagalli}, Michele and {Gaensicke}, Boris T. and {Gallo}, Giuseppe and {Garcia-Bellido}, Juan and {Gaztanaga}, Enrique and {Pietro Gentile Fusillo}, Nicola and {Gerard}, Terry and {Gershkovich}, Irena and {Giannantonio}, Tommaso and {Gillet}, Denis and {Gonzalez-de-Rivera}, Guillermo and {Gonzalez-Perez}, Violeta and {Gott}, Shelby and {Graur}, Or and {Gutierrez}, Gaston and {Guy}, Julien and {Habib}, Salman and {Heetderks}, Henry and {Heetderks}, Ian and {Heitmann}, Katrin and {Hellwing}, Wojciech A. and {Herrera}, David A. and {Ho}, Shirley and {Holland}, Stephen and {Honscheid}, Klaus and {Huff}, Eric and {Hutchinson}, Timothy A. and {Huterer}, Dragan and {Hwang}, Ho Seong and {Illa Laguna}, Joseph Maria and {Ishikawa}, Yuzo and {Jacobs}, Dianna and {Jeffrey}, Niall and {Jelinsky}, Patrick and {Jennings}, Elise and {Jiang}, Linhua and {Jimenez}, Jorge and {Johnson}, Jennifer and {Joyce}, Richard and {Jullo}, Eric and {Juneau}, St{\'e}phanie and {Kama}, Sami and {Karcher}, Armin and {Karkar}, Sonia and {Kehoe}, Robert and {Kennamer}, Noble and {Kent}, Stephen and {Kilbinger}, Martin and {Kim}, Alex G. and {Kirkby}, David and {Kisner}, Theodore and {Kitanidis}, Ellie and {Kneib}, Jean-Paul and {Koposov}, Sergey and {Kovacs}, Eve and {Koyama}, Kazuya and {Kremin}, Anthony and {Kron}, Richard and {Kronig}, Luzius and {Kueter-Young}, Andrea and {Lacey}, Cedric G. and {Lafever}, Robin and {Lahav}, Ofer and {Lambert}, Andrew and {Lampton}, Michael and {Landriau}, Martin and {Lang}, Dustin and {Lauer}, Tod R. and {Le Goff}, Jean-Marc and {Le Guillou}, Laurent and {Le Van Suu}, Auguste and {Lee}, Jae Hyeon and {Lee}, Su-Jeong and {Leitner}, Daniela and {Lesser}, Michael and {Levi}, Michael E. and {L'Huillier}, Benjamin and {Li}, Baojiu and {Liang}, Ming and {Lin}, Huan and {Linder}, Eric and {Loebman}, Sarah R. and {Luki{\'c}}, Zarija and {Ma}, Jun and {MacCrann}, Niall and {Magneville}, Christophe and {Makarem}, Laleh and {Manera}, Marc and {Manser}, Christopher J. and {Marshall}, Robert and {Martini}, Paul and {Massey}, Richard and {Matheson}, Thomas and {McCauley}, Jeremy and {McDonald}, Patrick and {McGreer}, Ian D. and {Meisner}, Aaron and {Metcalfe}, Nigel and {Miller}, Timothy N. and {Miquel}, Ramon and {Moustakas}, John and {Myers}, Adam and {Naik}, Milind and {Newman}, Jeffrey A. and {Nichol}, Robert C. and {Nicola}, Andrina and {Nicolati da Costa}, Luiz and {Nie}, Jundan and {Niz}, Gustavo and {Norberg}, Peder and {Nord}, Brian and {Norman}, Dara and {Nugent}, Peter and {O'Brien}, Thomas and {Oh}, Minji and {Olsen}, Knut A.~G.},
        title = "{The DESI Experiment Part II: Instrument Design}",
      journal = {arXiv e-prints},
         year = 2016,
        month = oct,
          eid = {arXiv:1611.00037},
        pages = {arXiv:1611.00037},
          doi = {10.48550/arXiv.1611.00037},
archivePrefix = {arXiv},
       eprint = {1611.00037},
 primaryClass = {astro-ph.IM},
       adsurl = {https://ui.adsabs.harvard.edu/abs/2016arXiv161100037D}
}

@ARTICLE{2022AJ....164..207D,
       author = {{DESI Collaboration} and {Abareshi}, B. and {Aguilar}, J. and {Ahlen}, S. and {Alam}, Shadab and {Alexander}, David M. and {Alfarsy}, R. and {Allen}, L. and {Allende Prieto}, C. and {Alves}, O. and {Ameel}, J. and {Armengaud}, E. and {Asorey}, J. and {Aviles}, Alejandro and {Bailey}, S. and {Balaguera-Antol{\'\i}nez}, A. and {Ballester}, O. and {Baltay}, C. and {Bault}, A. and {Beltran}, S.~F. and {Benavides}, B. and {BenZvi}, S. and {Berti}, A. and {Besuner}, R. and {Beutler}, Florian and {Bianchi}, D. and {Blake}, C. and {Blanc}, P. and {Blum}, R. and {Bolton}, A. and {Bose}, S. and {Bramall}, D. and {Brieden}, S. and {Brodzeller}, A. and {Brooks}, D. and {Brownewell}, C. and {Buckley-Geer}, E. and {Cahn}, R.~N. and {Cai}, Z. and {Canning}, R. and {Capasso}, R. and {Carnero Rosell}, A. and {Carton}, P. and {Casas}, R. and {Castander}, F.~J. and {Cervantes-Cota}, J.~L. and {Chabanier}, S. and {Chaussidon}, E. and {Chuang}, C. and {Circosta}, C. and {Cole}, S. and {Cooper}, A.~P. and {da Costa}, L. and {Cousinou}, M. -C. and {Cuceu}, A. and {Davis}, T.~M. and {Dawson}, K. and {de la Cruz-Noriega}, R. and {de la Macorra}, A. and {de Mattia}, A. and {Della Costa}, J. and {Demmer}, P. and {Derwent}, M. and {Dey}, A. and {Dey}, B. and {Dhungana}, G. and {Ding}, Z. and {Dobson}, C. and {Doel}, P. and {Donald-McCann}, J. and {Donaldson}, J. and {Douglass}, K. and {Duan}, Y. and {Dunlop}, P. and {Edelstein}, J. and {Eftekharzadeh}, S. and {Eisenstein}, D.~J. and {Enriquez-Vargas}, M. and {Escoffier}, S. and {Evatt}, M. and {Fagrelius}, P. and {Fan}, X. and {Fanning}, K. and {Fawcett}, V.~A. and {Ferraro}, S. and {Ereza}, J. and {Flaugher}, B. and {Font-Ribera}, A. and {Forero-Romero}, J.~E. and {Frenk}, C.~S. and {Fromenteau}, S. and {G{\"a}nsicke}, B.~T. and {Garcia-Quintero}, C. and {Garrison}, L. and {Gazta{\~n}aga}, E. and {Gerardi}, F. and {Gil-Mar{\'\i}n}, H. and {Gontcho A Gontcho}, S. and {Gonzalez-Morales}, Alma X. and {Gonzalez-de-Rivera}, G. and {Gonzalez-Perez}, V. and {Gordon}, C. and {Graur}, O. and {Green}, D. and {Grove}, C. and {Gruen}, D. and {Gutierrez}, G. and {Guy}, J. and {Hahn}, C. and {Harris}, S. and {Herrera}, D. and {Herrera-Alcantar}, Hiram K. and {Honscheid}, K. and {Howlett}, C. and {Huterer}, D. and {Ir{\v{s}}i{\v{c}}}, V. and {Ishak}, M. and {Jelinsky}, P. and {Jiang}, L. and {Jimenez}, J. and {Jing}, Y.~P. and {Joyce}, R. and {Jullo}, E. and {Juneau}, S. and {Kara{\c{c}}ayl{\i}}, N.~G. and {Karamanis}, M. and {Karcher}, A. and {Karim}, T. and {Kehoe}, R. and {Kent}, S. and {Kirkby}, D. and {Kisner}, T. and {Kitaura}, F. and {Koposov}, S.~E. and {Kov{\'a}cs}, A. and {Kremin}, A. and {Krolewski}, Alex and {L'Huillier}, B. and {Lahav}, O. and {Lambert}, A. and {Lamman}, C. and {Lan}, Ting-Wen and {Landriau}, M. and {Lane}, S. and {Lang}, D. and {Lange}, J.~U. and {Lasker}, J. and {Le Guillou}, L. and {Leauthaud}, A. and {Le Van Suu}, A. and {Levi}, Michael E. and {Li}, T.~S. and {Magneville}, C. and {Manera}, M. and {Manser}, Christopher J. and {Marshall}, B. and {Martini}, Paul and {McCollam}, W. and {McDonald}, P. and {Meisner}, Aaron M. and {Mena-Fern{\'a}ndez}, J. and {Meneses-Rizo}, J. and {Mezcua}, M. and {Miller}, T. and {Miquel}, R. and {Montero-Camacho}, P. and {Moon}, J. and {Moustakas}, J. and {Mueller}, E. and {Mu{\~n}oz-Guti{\'e}rrez}, Andrea and {Myers}, Adam D. and {Nadathur}, S. and {Najita}, J. and {Napolitano}, L. and {Neilsen}, E. and {Newman}, Jeffrey A. and {Nie}, J.~D. and {Ning}, Y. and {Niz}, G. and {Norberg}, P. and {Noriega}, Hern{\'a}n E. and {O'Brien}, T. and {Obuljen}, A. and {Palanque-Delabrouille}, N. and {Palmese}, A. and {Zhiwei}, P. and {Pappalardo}, D. and {PENG}, X. and {Percival}, W.~J. and {Perruchot}, S. and {Pogge}, R. and {Poppett}, C. and {Porredon}, A. and {Prada}, F. and {Prochaska}, J. and {Pucha}, R. and {P{\'e}rez-Fern{\'a}ndez}, A. and {P{\'e}rez-R{\`a}fols}, I. and {Rabinowitz}, D. and {Raichoor}, A.},
        title = "{Overview of the Instrumentation for the Dark Energy Spectroscopic Instrument}",
      journal = {\aj},
         year = 2022,
        month = nov,
       volume = {164},
       number = {5},
          eid = {207},
        pages = {207},
          doi = {10.3847/1538-3881/ac882b},
archivePrefix = {arXiv},
       eprint = {2205.10939},
 primaryClass = {astro-ph.IM},
       adsurl = {https://ui.adsabs.harvard.edu/abs/2022AJ....164..207D}
}

@ARTICLE{2023AJ....165....9S,
       author = {{Silber}, Joseph Harry and {Fagrelius}, Parker and {Fanning}, Kevin and {Schubnell}, Michael and {Aguilar}, Jessica Nicole and {Ahlen}, Steven and {Ameel}, Jon and {Ballester}, Otger and {Baltay}, Charles and {Bebek}, Chris and {Benton Beard}, Dominic and {Besuner}, Robert and {Cardiel-Sas}, Laia and {Casas}, Ricard and {Castander}, Francisco Javier and {Claybaugh}, Todd and {Dobson}, Carl and {Duan}, Yutong and {Dunlop}, Patrick and {Edelstein}, Jerry and {Emmet}, William T. and {Elliott}, Ann and {Evatt}, Matthew and {Gershkovich}, Irena and {Guy}, Julien and {Harris}, Stu and {Heetderks}, Henry and {Heetderks}, Ian and {Honscheid}, Klaus and {Illa}, Jose Maria and {Jelinsky}, Patrick and {Jelinsky}, Sharon R. and {Jimenez}, Jorge and {Karcher}, Armin and {Kent}, Stephen and {Kirkby}, David and {Kneib}, Jean-Paul and {Lambert}, Andrew and {Lampton}, Mike and {Leitner}, Daniela and {Levi}, Michael and {McCauley}, Jeremy and {Meisner}, Aaron and {Miller}, Timothy N. and {Miquel}, Ramon and {Mundet}, Juli{\'a} and {Poppett}, Claire and {Rabinowitz}, David and {Reil}, Kevin and {Roman}, David and {Schlegel}, David and {Serrano}, Santiago and {Van Shourt}, William and {Sprayberry}, David and {Tarl{\'e}}, Gregory and {Tie}, Suk Sien and {Weaverdyck}, Curtis and {Zhang}, Kai and {Azzaro}, Marco and {Bailey}, Stephen and {Becerril}, Santiago and {Blackwell}, Tami and {Bouri}, Mohamed and {Brooks}, David and {Buckley-Geer}, Elizabeth and {Castro}, Jose Pe{\~n}ate and {Derwent}, Mark and {Dey}, Arjun and {Dhungana}, Govinda and {Doel}, Peter and {Eisenstein}, Daniel J. and {Fahim}, Nasib and {Garcia-Bellido}, Juan and {Gazta{\~n}aga}, Enrique and {A Gontcho}, Satya Gontcho and {Gutierrez}, Gaston and {H{\"o}rler}, Philipp and {Kehoe}, Robert and {Kisner}, Theodore and {Kremin}, Anthony and {Kronig}, Luzius and {Landriau}, Martin and {Le Guillou}, Laurent and {Martini}, Paul and {Moustakas}, John and {Palanque-Delabrouille}, Nathalie and {Peng}, Xiyan and {Percival}, Will and {Prada}, Francisco and {Allende Prieto}, Carlos and {de Rivera}, Guillermo Gonzalez and {Sanchez}, Eusebio and {Sanchez}, Justo and {Sharples}, Ray and {Soares-Santos}, Marcelle and {Schlafly}, Edward and {Weaver}, Benjamin Alan and {Zhou}, Zhimin and {Zhu}, Yaling and {Zou}, Hu and {DESI Collaboration}},
        title = "{The Robotic Multiobject Focal Plane System of the Dark Energy Spectroscopic Instrument (DESI)}",
      journal = {\aj},
         year = 2023,
        month = jan,
       volume = {165},
       number = {1},
          eid = {9},
        pages = {9},
          doi = {10.3847/1538-3881/ac9ab1},
archivePrefix = {arXiv},
       eprint = {2205.09014},
 primaryClass = {astro-ph.IM},
       adsurl = {https://ui.adsabs.harvard.edu/abs/2023AJ....165....9S}
}

@ARTICLE{2024AJ....168...95M,
       author = {{Miller}, Timothy N. and {Doel}, Peter and {Gutierrez}, Gaston and {Besuner}, Robert and {Brooks}, David and {Gallo}, Giuseppe and {Heetderks}, Henry and {Jelinsky}, Patrick and {Kent}, Stephen M. and {Lampton}, Michael and {Levi}, Michael E. and {Liang}, Ming and {Meisner}, Aaron and {Sholl}, Michael J. and {Silber}, Joseph Harry and {Sprayberry}, David and {Aguilar}, Jessica Nicole and {de la Macorra}, Axel and {Eisenstein}, Daniel and {Fanning}, Kevin and {Font-Ribera}, Andreu and {Gazta{\~n}aga}, Enrique and {Gontcho A Gontcho}, Satya and {Honscheid}, Klaus and {Jimenez}, Jorge and {Joyce}, Dick and {Kehoe}, Robert and {Kisner}, Theodore and {Kremin}, Anthony and {Landriau}, Martin and {Le Guillou}, Laurent and {Magneville}, Christophe and {Martini}, Paul and {Miquel}, Ramon and {Moustakas}, John and {Nie}, Jundan and {Percival}, Will and {Poppett}, Claire and {Prada}, Francisco and {Rossi}, Graziano and {Schlegel}, David and {Schubnell}, Michael and {Seo}, Hee-Jong and {Sharples}, Ray and {Tarl{\'e}}, Gregory and {Vargas-Maga{\~n}a}, Mariana and {Zhou}, Zhimin and {the DESI Collaboration}},
        title = "{The Optical Corrector for the Dark Energy Spectroscopic Instrument}",
      journal = {\aj},
         year = 2024,
        month = aug,
       volume = {168},
       number = {2},
          eid = {95},
        pages = {95},
          doi = {10.3847/1538-3881/ad45fe},
archivePrefix = {arXiv},
       eprint = {2306.06310},
 primaryClass = {astro-ph.IM},
       adsurl = {https://ui.adsabs.harvard.edu/abs/2024AJ....168...95M}
}

@ARTICLE{2024ApJS..270...26G,
       author = {{Guo}, Wei-Jian and {Zou}, Hu and {Fawcett}, Victoria A. and {Canning}, Rebecca and {Juneau}, Stephanie and {Davis}, Tamara M. and {Alexander}, David M. and {Jiang}, Linhua and {Aguilar}, Jessica Nicole and {Ahlen}, Steven and {Brooks}, David and {Claybaugh}, Todd and {de la Macorra}, Axel and {Doel}, Peter and {Fanning}, Kevin and {Forero-Romero}, Jaime E. and {Gontcho A Gontcho}, Satya and {Honscheid}, Klaus and {Kisner}, Theodore and {Kremin}, Anthony and {Landriau}, Martin and {Meisner}, Aaron and {Miquel}, Ramon and {Moustakas}, John and {Nie}, Jundan and {Pan}, Zhiwei and {Poppett}, Claire and {Prada}, Francisco and {Rezaie}, Mehdi and {Rossi}, Graziano and {Siudek}, Ma{\l}gorzata and {Sanchez}, Eusebio and {Schubnell}, Michael and {Seo}, Hee-Jong and {Sui}, Jipeng and {Tarl{\'e}}, Gregory and {Zhou}, Zhimin},
        title = "{Changing-look Active Galactic Nuclei from the Dark Energy Spectroscopic Instrument. I. Sample from the Early Data}",
      journal = {\apjs},
         year = 2024,
        month = feb,
       volume = {270},
       number = {2},
          eid = {26},
        pages = {26},
          doi = {10.3847/1538-4365/ad118a},
archivePrefix = {arXiv},
       eprint = {2307.08289},
 primaryClass = {astro-ph.GA},
       adsurl = {https://ui.adsabs.harvard.edu/abs/2024ApJS..270...26G}
}

@ARTICLE{2023AJ....165..144G,
       author = {{Guy}, J. and {Bailey}, S. and {Kremin}, A. and {Alam}, Shadab and {Alexander}, D.~M. and {Allende Prieto}, C. and {BenZvi}, S. and {Bolton}, A.~S. and {Brooks}, D. and {Chaussidon}, E. and {Cooper}, A.~P. and {Dawson}, K. and {de la Macorra}, A. and {Dey}, A. and {Dey}, Biprateep and {Dhungana}, G. and {Eisenstein}, D.~J. and {Font-Ribera}, A. and {Forero-Romero}, J.~E. and {Gazta{\~n}aga}, E. and {Gontcho A Gontcho}, S. and {Green}, D. and {Honscheid}, K. and {Ishak}, M. and {Kehoe}, R. and {Kirkby}, D. and {Kisner}, T. and {Koposov}, Sergey E. and {Lan}, Ting-Wen and {Landriau}, M. and {Le Guillou}, L. and {Levi}, Michael E. and {Magneville}, C. and {Manser}, Christopher J. and {Martini}, P. and {Meisner}, Aaron M. and {Miquel}, R. and {Moustakas}, J. and {Myers}, Adam D. and {Newman}, Jeffrey A. and {Nie}, Jundan and {Palanque-Delabrouille}, N. and {Percival}, W.~J. and {Poppett}, C. and {Prada}, F. and {Raichoor}, A. and {Ravoux}, C. and {Ross}, A.~J. and {Schlafly}, E.~F. and {Schlegel}, D. and {Schubnell}, M. and {Sharples}, Ray M. and {Tarl{\'e}}, Gregory and {Weaver}, B.~A. and {Y{\'e}che}, Christophe and {Zhou}, Rongpu and {Zhou}, Zhimin and {Zou}, H.},
        title = "{The Spectroscopic Data Processing Pipeline for the Dark Energy Spectroscopic Instrument}",
      journal = {\aj},
         year = 2023,
        month = apr,
       volume = {165},
       number = {4},
          eid = {144},
        pages = {144},
          doi = {10.3847/1538-3881/acb212},
archivePrefix = {arXiv},
       eprint = {2209.14482},
 primaryClass = {astro-ph.IM},
       adsurl = {https://ui.adsabs.harvard.edu/abs/2023AJ....165..144G}
}

@ARTICLE{2023AJ....166..259S,
       author = {{Schlafly}, Edward F. and {Kirkby}, David and {Schlegel}, David J. and {Myers}, Adam D. and {Raichoor}, Anand and {Dawson}, Kyle and {Aguilar}, Jessica and {Allende Prieto}, Carlos and {Bailey}, Stephen and {BenZvi}, Segev and {Bermejo-Climent}, Jose and {Brooks}, David and {de la Macorra}, Axel and {Dey}, Arjun and {Doel}, Peter and {Fanning}, Kevin and {Font-Ribera}, Andreu and {Forero-Romero}, Jaime E. and {Garc{\'\i}a-Bellido}, Juan and {Gontcho A Gontcho}, Satya and {Guy}, Julien and {Hahn}, ChangHoon and {Honscheid}, Klaus and {Ishak}, Mustapha and {Juneau}, St{\'e}phanie and {Kehoe}, Robert and {Kisner}, Theodore and {Kremin}, Anthony and {Landriau}, Martin and {Lang}, Dustin A. and {Lasker}, James and {Levi}, Michael E. and {Magneville}, Christophe and {Manser}, Christopher J. and {Martini}, Paul and {Meisner}, Aaron M. and {Miquel}, Ramon and {Moustakas}, John and {Newman}, Jeffrey A. and {Nie}, Jundan and {Palanque-Delabrouille}, Nathalie. and {Percival}, Will J. and {Poppett}, Claire and {Rockosi}, Constance and {Ross}, Ashley J. and {Rossi}, Graziano and {Tarl{\'e}}, Gregory and {Weaver}, Benjamin A. and {Y{\`e}che}, Christophe and {Zhou}, Rongpu and {DESI Collaboration}},
        title = "{Survey Operations for the Dark Energy Spectroscopic Instrument}",
      journal = {\aj},
         year = 2023,
        month = dec,
       volume = {166},
       number = {6},
          eid = {259},
        pages = {259},
          doi = {10.3847/1538-3881/ad0832},
archivePrefix = {arXiv},
       eprint = {2306.06309},
 primaryClass = {astro-ph.CO},
       adsurl = {https://ui.adsabs.harvard.edu/abs/2023AJ....166..259S}
}

@ARTICLE{2024AJ....168...58D,
       author = {{DESI Collaboration} and {Adame}, A.~G. and {Aguilar}, J. and {Ahlen}, S. and {Alam}, S. and {Aldering}, G. and {Alexander}, D.~M. and {Alfarsy}, R. and {Allende Prieto}, C. and {Alvarez}, M. and {Alves}, O. and {Anand}, A. and {Andrade-Oliveira}, F. and {Armengaud}, E. and {Asorey}, J. and {Avila}, S. and {Aviles}, A. and {Bailey}, S. and {Balaguera-Antol{\'\i}nez}, A. and {Ballester}, O. and {Baltay}, C. and {Bault}, A. and {Bautista}, J. and {Behera}, J. and {Beltran}, S.~F. and {BenZvi}, S. and {Beraldo e Silva}, L. and {Bermejo-Climent}, J.~R. and {Berti}, A. and {Besuner}, R. and {Beutler}, F. and {Bianchi}, D. and {Blake}, C. and {Blum}, R. and {Bolton}, A.~S. and {Brieden}, S. and {Brodzeller}, A. and {Brooks}, D. and {Brown}, Z. and {Buckley-Geer}, E. and {Burtin}, E. and {Cabayol-Garcia}, L. and {Cai}, Z. and {Canning}, R. and {Cardiel-Sas}, L. and {Carnero Rosell}, A. and {Castander}, F.~J. and {Cervantes-Cota}, J.~L. and {Chabanier}, S. and {Chaussidon}, E. and {Chaves-Montero}, J. and {Chen}, S. and {Chen}, X. and {Chuang}, C. and {Claybaugh}, T. and {Cole}, S. and {Cooper}, A.~P. and {Cuceu}, A. and {Davis}, T.~M. and {Dawson}, K. and {de Belsunce}, R. and {de la Cruz}, R. and {de la Macorra}, A. and {Della Costa}, J. and {de Mattia}, A. and {Demina}, R. and {Demirbozan}, U. and {DeRose}, J. and {Dey}, A. and {Dey}, B. and {Dhungana}, G. and {Ding}, J. and {Ding}, Z. and {Doel}, P. and {Doshi}, R. and {Douglass}, K. and {Edge}, A. and {Eftekharzadeh}, S. and {Eisenstein}, D.~J. and {Elliott}, A. and {Ereza}, J. and {Escoffier}, S. and {Fagrelius}, P. and {Fan}, X. and {Fanning}, K. and {Fawcett}, V.~A. and {Ferraro}, S. and {Flaugher}, B. and {Font-Ribera}, A. and {Forero-Romero}, J.~E. and {Forero-S{\'a}nchez}, D. and {Frenk}, C.~S. and {G{\"a}nsicke}, B.~T. and {Garc{\'\i}a}, L. {\'A}. and {Garc{\'\i}a-Bellido}, J. and {Garcia-Quintero}, C. and {Garrison}, L.~H. and {Gil-Mar{\'\i}n}, H. and {Golden-Marx}, J. and {Gontcho A Gontcho}, S. and {Gonzalez-Morales}, A.~X. and {Gonzalez-Perez}, V. and {Gordon}, C. and {Graur}, O. and {Green}, D. and {Gruen}, D. and {Guy}, J. and {Hadzhiyska}, B. and {Hahn}, C. and {Han}, J.~J. and {Hanif}, M.~M.~S. and {Herrera-Alcantar}, H.~K. and {Honscheid}, K. and {Hou}, J. and {Howlett}, C. and {Huterer}, D. and {Ir{\v{s}}i{\v{c}}}, V. and {Ishak}, M. and {Jacques}, A. and {Jana}, A. and {Jiang}, L. and {Jimenez}, J. and {Jing}, Y.~P. and {Joudaki}, S. and {Joyce}, R. and {Jullo}, E. and {Juneau}, S. and {Kara{\c{c}}ayl{\i}}, N.~G. and {Karim}, T. and {Kehoe}, R. and {Kent}, S. and {Khederlarian}, A. and {Kim}, S. and {Kirkby}, D. and {Kisner}, T. and {Kitaura}, F. and {Kizhuprakkat}, N. and {Kneib}, J. and {Koposov}, S.~E. and {Kov{\'a}cs}, A. and {Kremin}, A. and {Krolewski}, A. and {L'Huillier}, B. and {Lahav}, O. and {Lambert}, A. and {Lamman}, C. and {Lan}, T. -W. and {Landriau}, M. and {Lang}, D. and {Lange}, J.~U. and {Lasker}, J. and {Leauthaud}, A. and {Le Guillou}, L. and {Levi}, M.~E. and {Li}, T.~S. and {Linder}, E. and {Lyons}, A. and {Magneville}, C. and {Manera}, M. and {Manser}, C.~J. and {Margala}, D. and {Martini}, P. and {McDonald}, P. and {Medina}, G.~E. and {Medina-Varela}, L. and {Meisner}, A. and {Mena-Fern{\'a}ndez}, J. and {Meneses-Rizo}, J. and {Mezcua}, M. and {Miquel}, R. and {Montero-Camacho}, P. and {Moon}, J. and {Moore}, S. and {Moustakas}, J. and {Mueller}, E. and {Mundet}, J. and {Mu{\~n}oz-Guti{\'e}rrez}, A. and {Myers}, A.~D. and {Nadathur}, S. and {Napolitano}, L. and {Neveux}, R. and {Newman}, J.~A. and {Nie}, J. and {Nikutta}, R. and {Niz}, G. and {Norberg}, P. and {Noriega}, H.~E. and {Paillas}, E. and {Palanque-Delabrouille}, N. and {Palmese}, A. and {Pan}, Z. and {Parkinson}, D. and {Penmetsa}, S. and {Percival}, W.~J. and {P{\'e}rez-Fern{\'a}ndez}, A. and {P{\'e}rez-R{\`a}fols}, I. and {Pieri}, M. and {Poppett}, C. and {Porredon}, A. and {Pothier}, S.},
        title = "{The Early Data Release of the Dark Energy Spectroscopic Instrument}",
      journal = {\aj},
         year = 2024,
        month = aug,
       volume = {168},
       number = {2},
          eid = {58},
        pages = {58},
          doi = {10.3847/1538-3881/ad3217},
archivePrefix = {arXiv},
       eprint = {2306.06308},
 primaryClass = {astro-ph.CO},
       adsurl = {https://ui.adsabs.harvard.edu/abs/2024AJ....168...58D}
}

@ARTICLE{2024AJ....167...62D,
       author = {{DESI Collaboration} and {Adame}, A.~G. and {Aguilar}, J. and {Ahlen}, S. and {Alam}, S. and {Aldering}, G. and {Alexander}, D.~M. and {Alfarsy}, R. and {Allende Prieto}, C. and {Alvarez}, M. and {Alves}, O. and {Anand}, A. and {Andrade-Oliveira}, F. and {Armengaud}, E. and {Asorey}, J. and {Avila}, S. and {Aviles}, A. and {Bailey}, S. and {Balaguera-Antol{\'\i}nez}, A. and {Ballester}, O. and {Baltay}, C. and {Bault}, A. and {Bautista}, J. and {Behera}, J. and {Beltran}, S.~F. and {BenZvi}, S. and {Beraldo e Silva}, L. and {Bermejo-Climent}, J.~R. and {Berti}, A. and {Besuner}, R. and {Beutler}, F. and {Bianchi}, D. and {Blake}, C. and {Blum}, R. and {Bolton}, A.~S. and {Brieden}, S. and {Brodzeller}, A. and {Brooks}, D. and {Brown}, Z. and {Buckley-Geer}, E. and {Burtin}, E. and {Cabayol-Garcia}, L. and {Cai}, Z. and {Canning}, R. and {Cardiel-Sas}, L. and {Carnero Rosell}, A. and {Castander}, F.~J. and {Cervantes-Cota}, J.~L. and {Chabanier}, S. and {Chaussidon}, E. and {Chaves-Montero}, J. and {Chen}, S. and {Chen}, X. and {Chuang}, C. and {Claybaugh}, T. and {Cole}, S. and {Cooper}, A.~P. and {Cuceu}, A. and {Davis}, T.~M. and {Dawson}, K. and {de Belsunce}, R. and {de la Cruz}, R. and {de la Macorra}, A. and {de Mattia}, A. and {Demina}, R. and {Demirbozan}, U. and {DeRose}, J. and {Dey}, A. and {Dey}, B. and {Dhungana}, G. and {Ding}, J. and {Ding}, Z. and {Doel}, P. and {Doshi}, R. and {Douglass}, K. and {Edge}, A. and {Eftekharzadeh}, S. and {Eisenstein}, D.~J. and {Elliott}, A. and {Escoffier}, S. and {Fagrelius}, P. and {Fan}, X. and {Fanning}, K. and {Fawcett}, V.~A. and {Ferraro}, S. and {Ereza}, J. and {Flaugher}, B. and {Font-Ribera}, A. and {Forero-S{\'a}nchez}, D. and {Forero-Romero}, J.~E. and {Frenk}, C.~S. and {G{\"a}nsicke}, B.~T. and {Garc{\'\i}a}, L. {\'A}. and {Garc{\'\i}a-Bellido}, J. and {Garcia-Quintero}, C. and {Garrison}, L.~H. and {Gil-Mar{\'\i}n}, H. and {Golden-Marx}, J. and {Gontcho A Gontcho}, S. and {Gonzalez-Morales}, A.~X. and {Gonzalez-Perez}, V. and {Gordon}, C. and {Graur}, O. and {Green}, D. and {Gruen}, D. and {Guy}, J. and {Hadzhiyska}, B. and {Hahn}, C. and {Han}, J.~J. and {Hanif}, M.~M.~S. and {Herrera-Alcantar}, H.~K. and {Honscheid}, K. and {Hou}, J. and {Howlett}, C. and {Huterer}, D. and {Ir{\v{s}}i{\v{c}}}, V. and {Ishak}, M. and {Jana}, A. and {Jiang}, L. and {Jimenez}, J. and {Jing}, Y.~P. and {Joudaki}, S. and {Jullo}, E. and {Joyce}, R. and {Juneau}, S. and {Kizhuprakkat}, N. and {Kara{\c{c}}ayl{\i}}, N.~G. and {Karim}, T. and {Kehoe}, R. and {Kent}, S. and {Khederlarian}, A. and {Kim}, S. and {Kirkby}, D. and {Kisner}, T. and {Kitaura}, F. and {Kneib}, J. and {Koposov}, S.~E. and {Kov{\'a}cs}, A. and {Kremin}, A. and {Krolewski}, A. and {L'Huillier}, B. and {Lahav}, O. and {Lambert}, A. and {Lamman}, C. and {Lan}, T. -W. and {Landriau}, M. and {Lang}, D. and {Lange}, J.~U. and {Lasker}, J. and {Le Guillou}, L. and {Leauthaud}, A. and {Levi}, M.~E. and {Li}, T.~S. and {Linder}, E. and {Lyons}, A. and {Magneville}, C. and {Manera}, M. and {Manser}, C.~J. and {Margala}, D. and {Martini}, P. and {McDonald}, P. and {Medina}, G.~E. and {Medina-Varela}, L. and {Meisner}, A. and {Mena-Fern{\'a}ndez}, J. and {Meneses-Rizo}, J. and {Mezcua}, M. and {Miquel}, R. and {Montero-Camacho}, P. and {Moon}, J. and {Moore}, S. and {Moustakas}, J. and {Mueller}, E. and {Mundet}, J. and {Mu{\~n}oz-Guti{\'e}rrez}, A. and {Myers}, A.~D. and {Nadathur}, S. and {Napolitano}, L. and {Neveux}, R. and {Newman}, J.~A. and {Nie}, J. and {Niz}, G. and {Norberg}, P. and {Noriega}, H.~E. and {Paillas}, E. and {Palanque-Delabrouille}, N. and {Palmese}, A. and {Zhiwei}, P. and {Parkinson}, D. and {Penmetsa}, S. and {Percival}, W.~J. and {P{\'e}rez-Fern{\'a}ndez}, A. and {P{\'e}rez-R{\`a}fols}, I. and {Pieri}, M. and {Poppett}, C. and {Porredon}, A. and {Prada}, F. and {Pucha}, R. and {Raichoor}, A. and {Ram{\'\i}rez-P{\'e}rez}, C.},
        title = "{Validation of the Scientific Program for the Dark Energy Spectroscopic Instrument}",
      journal = {\aj},
         year = 2024,
        month = feb,
       volume = {167},
       number = {2},
          eid = {62},
        pages = {62},
          doi = {10.3847/1538-3881/ad0b08},
archivePrefix = {arXiv},
       eprint = {2306.06307},
 primaryClass = {astro-ph.CO},
       adsurl = {https://ui.adsabs.harvard.edu/abs/2024AJ....167...62D}
}

@ARTICLE{2006AJ....131.2332G,
       author = {{Gunn}, James E. and {Siegmund}, Walter A. and {Mannery}, Edward J. and {Owen}, Russell E. and {Hull}, Charles L. and {Leger}, R. French and {Carey}, Larry N. and {Knapp}, Gillian R. and {York}, Donald G. and {Boroski}, William N. and {Kent}, Stephen M. and {Lupton}, Robert H. and {Rockosi}, Constance M. and {Evans}, Michael L. and {Waddell}, Patrick and {Anderson}, John E. and {Annis}, James and {Barentine}, John C. and {Bartoszek}, Larry M. and {Bastian}, Steven and {Bracker}, Stephen B. and {Brewington}, Howard J. and {Briegel}, Charles I. and {Brinkmann}, Jon and {Brown}, Yorke J. and {Carr}, Michael A. and {Czarapata}, Paul C. and {Drennan}, Craig C. and {Dombeck}, Thomas and {Federwitz}, Glenn R. and {Gillespie}, Bruce A. and {Gonzales}, Carlos and {Hansen}, Sten U. and {Harvanek}, Michael and {Hayes}, Jeffrey and {Jordan}, Wendell and {Kinney}, Ellyne and {Klaene}, Mark and {Kleinman}, S.~J. and {Kron}, Richard G. and {Kresinski}, Jurek and {Lee}, Glenn and {Limmongkol}, Siriluk and {Lindenmeyer}, Carl W. and {Long}, Daniel C. and {Loomis}, Craig L. and {McGehee}, Peregrine M. and {Mantsch}, Paul M. and {Neilsen}, Jr., Eric H. and {Neswold}, Richard M. and {Newman}, Peter R. and {Nitta}, Atsuko and {Peoples}, Jr., John and {Pier}, Jeffrey R. and {Prieto}, Peter S. and {Prosapio}, Angela and {Rivetta}, Claudio and {Schneider}, Donald P. and {Snedden}, Stephanie and {Wang}, Shu-i.},
        title = "{The 2.5 m Telescope of the Sloan Digital Sky Survey}",
      journal = {\aj},
         year = 2006,
        month = apr,
       volume = {131},
       number = {4},
        pages = {2332-2359},
          doi = {10.1086/500975},
archivePrefix = {arXiv},
       eprint = {astro-ph/0602326},
 primaryClass = {astro-ph},
       adsurl = {https://ui.adsabs.harvard.edu/abs/2006AJ....131.2332G}
}

@ARTICLE{2023ApJS..267...44A,
       author = {{Almeida}, Andr{\'e}s and {Anderson}, Scott F. and {Argudo-Fern{\'a}ndez}, Maria and {Badenes}, Carles and {Barger}, Kat and {Barrera-Ballesteros}, Jorge K. and {Bender}, Chad F. and {Benitez}, Erika and {Besser}, Felipe and {Bird}, Jonathan C. and {Bizyaev}, Dmitry and {Blanton}, Michael R. and {Bochanski}, John and {Bovy}, Jo and {Brandt}, William Nielsen and {Brownstein}, Joel R. and {Buchner}, Johannes and {Bulbul}, Esra and {Burchett}, Joseph N. and {Cano D{\'\i}az}, Mariana and {Carlberg}, Joleen K. and {Casey}, Andrew R. and {Chandra}, Vedant and {Cherinka}, Brian and {Chiappini}, Cristina and {Coker}, Abigail A. and {Comparat}, Johan and {Conroy}, Charlie and {Contardo}, Gabriella and {Cortes}, Arlin and {Covey}, Kevin and {Crane}, Jeffrey D. and {Cunha}, Katia and {Dabbieri}, Collin and {Davidson}, James W. and {Davis}, Megan C. and {de Andrade Queiroz}, Anna Barbara and {De Lee}, Nathan and {M{\'e}ndez Delgado}, Jos{\'e} Eduardo and {Demasi}, Sebastian and {Di Mille}, Francesco and {Donor}, John and {Dow}, Peter and {Dwelly}, Tom and {Eracleous}, Mike and {Eriksen}, Jamey and {Fan}, Xiaohui and {Farr}, Emily and {Frederick}, Sara and {Fries}, Logan and {Frinchaboy}, Peter and {G{\"a}nsicke}, Boris T. and {Ge}, Junqiang and {Gonz{\'a}lez {\'A}vila}, Consuelo and {Grabowski}, Katie and {Grier}, Catherine and {Guiglion}, Guillaume and {Gupta}, Pramod and {Hall}, Patrick and {Hawkins}, Keith and {Hayes}, Christian R. and {Hermes}, J.~J. and {Hern{\'a}ndez-Garc{\'\i}a}, Lorena and {Hogg}, David W. and {Holtzman}, Jon A. and {Ibarra-Medel}, Hector Javier and {Ji}, Alexander and {Jofre}, Paula and {Johnson}, Jennifer A. and {Jones}, Amy M. and {Kinemuchi}, Karen and {Kluge}, Matthias and {Koekemoer}, Anton and {Kollmeier}, Juna A. and {Kounkel}, Marina and {Krishnarao}, Dhanesh and {Krumpe}, Mirko and {Lacerna}, Ivan and {Lago}, Paulo Jakson Assuncao and {Laporte}, Chervin and {Liu}, Chao and {Liu}, Ang and {Liu}, Xin and {Lopes}, Alexandre Roman and {Macktoobian}, Matin and {Majewski}, Steven R. and {Malanushenko}, Viktor and {Maoz}, Dan and {Masseron}, Thomas and {Masters}, Karen L. and {Matijevic}, Gal and {McBride}, Aidan and {Medan}, Ilija and {Merloni}, Andrea and {Morrison}, Sean and {Myers}, Natalie and {M{\'e}sz{\'a}ros}, Szabolcs and {Negrete}, C. Alenka and {Nidever}, David L. and {Nitschelm}, Christian and {Oravetz}, Daniel and {Oravetz}, Audrey and {Pan}, Kaike and {Peng}, Yingjie and {Pinsonneault}, Marc H. and {Pogge}, Rick and {Qiu}, Dan and {Ramirez}, Solange V. and {Rix}, Hans-Walter and {Fern{\'a}ndez Rosso}, Daniela and {Runnoe}, Jessie and {Salvato}, Mara and {Sanchez}, Sebastian F. and {Santana}, Felipe A. and {Saydjari}, Andrew and {Sayres}, Conor and {Schlaufman}, Kevin C. and {Schneider}, Donald P. and {Schwope}, Axel and {Serna}, Javier and {Shen}, Yue and {Sobeck}, Jennifer and {Song}, Ying-Yi and {Souto}, Diogo and {Spoo}, Taylor and {Stassun}, Keivan G. and {Steinmetz}, Matthias and {Straumit}, Ilya and {Stringfellow}, Guy and {S{\'a}nchez-Gallego}, Jos{\'e} and {Taghizadeh-Popp}, Manuchehr and {Tayar}, Jamie and {Thakar}, Ani and {Tissera}, Patricia B. and {Tkachenko}, Andrew and {Hernandez Toledo}, Hector and {Trakhtenbrot}, Benny and {Fern{\'a}ndez-Trincado}, Jos{\'e} G. and {Troup}, Nicholas and {Trump}, Jonathan R. and {Tuttle}, Sarah and {Ulloa}, Natalie and {Vazquez-Mata}, Jose Antonio and {Vera Alfaro}, Pablo and {Villanova}, Sandro and {Wachter}, Stefanie and {Weijmans}, Anne-Marie and {Wheeler}, Adam and {Wilson}, John and {Wojno}, Leigh and {Wolf}, Julien and {Xue}, Xiang-Xiang and {Ybarra}, Jason E. and {Zari}, Eleonora and {Zasowski}, Gail},
        title = "{The Eighteenth Data Release of the Sloan Digital Sky Surveys: Targeting and First Spectra from SDSS-V}",
      journal = {\apjs},
         year = 2023,
        month = aug,
       volume = {267},
       number = {2},
          eid = {44},
        pages = {44},
          doi = {10.3847/1538-4365/acda98},
archivePrefix = {arXiv},
       eprint = {2301.07688},
 primaryClass = {astro-ph.GA},
       adsurl = {https://ui.adsabs.harvard.edu/abs/2023ApJS..267...44A}
}

@ARTICLE{2009ApJS..182..543A,
       author = {{Abazajian}, Kevork N. and {Adelman-McCarthy}, Jennifer K. and {Ag{\"u}eros}, Marcel A. and {Allam}, Sahar S. and {Allende Prieto}, Carlos and {An}, Deokkeun and {Anderson}, Kurt S.~J. and {Anderson}, Scott F. and {Annis}, James and {Bahcall}, Neta A. and {Bailer-Jones}, C.~A.~L. and {Barentine}, J.~C. and {Bassett}, Bruce A. and {Becker}, Andrew C. and {Beers}, Timothy C. and {Bell}, Eric F. and {Belokurov}, Vasily and {Berlind}, Andreas A. and {Berman}, Eileen F. and {Bernardi}, Mariangela and {Bickerton}, Steven J. and {Bizyaev}, Dmitry and {Blakeslee}, John P. and {Blanton}, Michael R. and {Bochanski}, John J. and {Boroski}, William N. and {Brewington}, Howard J. and {Brinchmann}, Jarle and {Brinkmann}, J. and {Brunner}, Robert J. and {Budav{\'a}ri}, Tam{\'a}s and {Carey}, Larry N. and {Carliles}, Samuel and {Carr}, Michael A. and {Castander}, Francisco J. and {Cinabro}, David and {Connolly}, A.~J. and {Csabai}, Istv{\'a}n and {Cunha}, Carlos E. and {Czarapata}, Paul C. and {Davenport}, James R.~A. and {de Haas}, Ernst and {Dilday}, Ben and {Doi}, Mamoru and {Eisenstein}, Daniel J. and {Evans}, Michael L. and {Evans}, N.~W. and {Fan}, Xiaohui and {Friedman}, Scott D. and {Frieman}, Joshua A. and {Fukugita}, Masataka and {G{\"a}nsicke}, Boris T. and {Gates}, Evalyn and {Gillespie}, Bruce and {Gilmore}, G. and {Gonzalez}, Belinda and {Gonzalez}, Carlos F. and {Grebel}, Eva K. and {Gunn}, James E. and {Gy{\"o}ry}, Zsuzsanna and {Hall}, Patrick B. and {Harding}, Paul and {Harris}, Frederick H. and {Harvanek}, Michael and {Hawley}, Suzanne L. and {Hayes}, Jeffrey J.~E. and {Heckman}, Timothy M. and {Hendry}, John S. and {Hennessy}, Gregory S. and {Hindsley}, Robert B. and {Hoblitt}, J. and {Hogan}, Craig J. and {Hogg}, David W. and {Holtzman}, Jon A. and {Hyde}, Joseph B. and {Ichikawa}, Shin-ichi and {Ichikawa}, Takashi and {Im}, Myungshin and {Ivezi{\'c}}, {\v{Z}}eljko and {Jester}, Sebastian and {Jiang}, Linhua and {Johnson}, Jennifer A. and {Jorgensen}, Anders M. and {Juri{\'c}}, Mario and {Kent}, Stephen M. and {Kessler}, R. and {Kleinman}, S.~J. and {Knapp}, G.~R. and {Konishi}, Kohki and {Kron}, Richard G. and {Krzesinski}, Jurek and {Kuropatkin}, Nikolay and {Lampeitl}, Hubert and {Lebedeva}, Svetlana and {Lee}, Myung Gyoon and {Lee}, Young Sun and {French Leger}, R. and {L{\'e}pine}, S{\'e}bastien and {Li}, Nolan and {Lima}, Marcos and {Lin}, Huan and {Long}, Daniel C. and {Loomis}, Craig P. and {Loveday}, Jon and {Lupton}, Robert H. and {Magnier}, Eugene and {Malanushenko}, Olena and {Malanushenko}, Viktor and {Mandelbaum}, Rachel and {Margon}, Bruce and {Marriner}, John P. and {Mart{\'\i}nez-Delgado}, David and {Matsubara}, Takahiko and {McGehee}, Peregrine M. and {McKay}, Timothy A. and {Meiksin}, Avery and {Morrison}, Heather L. and {Mullally}, Fergal and {Munn}, Jeffrey A. and {Murphy}, Tara and {Nash}, Thomas and {Nebot}, Ada and {Neilsen}, Jr., Eric H. and {Newberg}, Heidi Jo and {Newman}, Peter R. and {Nichol}, Robert C. and {Nicinski}, Tom and {Nieto-Santisteban}, Maria and {Nitta}, Atsuko and {Okamura}, Sadanori and {Oravetz}, Daniel J. and {Ostriker}, Jeremiah P. and {Owen}, Russell and {Padmanabhan}, Nikhil and {Pan}, Kaike and {Park}, Changbom and {Pauls}, George and {Peoples}, Jr., John and {Percival}, Will J. and {Pier}, Jeffrey R. and {Pope}, Adrian C. and {Pourbaix}, Dimitri and {Price}, Paul A. and {Purger}, Norbert and {Quinn}, Thomas and {Raddick}, M. Jordan and {Re Fiorentin}, Paola and {Richards}, Gordon T. and {Richmond}, Michael W. and {Riess}, Adam G. and {Rix}, Hans-Walter and {Rockosi}, Constance M. and {Sako}, Masao and {Schlegel}, David J. and {Schneider}, Donald P. and {Scholz}, Ralf-Dieter and {Schreiber}, Matthias R. and {Schwope}, Axel D. and {Seljak}, Uro{\v{s}} and {Sesar}, Branimir and {Sheldon}, Erin and {Shimasaku}, Kazu and {Sibley}, Valena C. and {Simmons}, A.~E. and {Sivarani}, Thirupathi and {Allyn Smith}, J. and {Smith}, Martin C. and {Smol{\v{c}}i{\'c}}, Vernesa and {Snedden}, Stephanie A. and {Stebbins}, Albert and {Steinmetz}, Matthias and {Stoughton}, Chris and {Strauss}, Michael A. and {SubbaRao}, Mark and {Suto}, Yasushi and {Szalay}, Alexander S. and {Szapudi}, Istv{\'a}n and {Szkody}, Paula and {Tanaka}, Masayuki and {Tegmark}, Max and {Teodoro}, Luis F.~A. and {Thakar}, Aniruddha R. and {Tremonti}, Christy A. and {Tucker}, Douglas L. and {Uomoto}, Alan and {Vanden Berk}, Daniel E. and {Vandenberg}, Jan and {Vidrih}, S. and {Vogeley}, Michael S. and {Voges}, Wolfgang and {Vogt}, Nicole P. and {Wadadekar}, Yogesh and {Watters}, Shannon and {Weinberg}, David H. and {West}, Andrew A. and {White}, Simon D.~M. and {Wilhite}, Brian C. and {Wonders}, Alainna C. and {Yanny}, Brian and {Yocum}, D.~R.},
        title = "{The Seventh Data Release of the Sloan Digital Sky Survey}",
      journal = {\apjs},
         year = 2009,
        month = jun,
       volume = {182},
       number = {2},
        pages = {543-558},
          doi = {10.1088/0067-0049/182/2/543},
archivePrefix = {arXiv},
       eprint = {0812.0649},
 primaryClass = {astro-ph},
       adsurl = {https://ui.adsabs.harvard.edu/abs/2009ApJS..182..543A}
}

@ARTICLE{2011AJ....142...72E,
       author = {{Eisenstein}, Daniel J. and {Weinberg}, David H. and {Agol}, Eric and {Aihara}, Hiroaki and {Allende Prieto}, Carlos and {Anderson}, Scott F. and {Arns}, James A. and {Aubourg}, {\'E}ric and {Bailey}, Stephen and {Balbinot}, Eduardo and {Barkhouser}, Robert and {Beers}, Timothy C. and {Berlind}, Andreas A. and {Bickerton}, Steven J. and {Bizyaev}, Dmitry and {Blanton}, Michael R. and {Bochanski}, John J. and {Bolton}, Adam S. and {Bosman}, Casey T. and {Bovy}, Jo and {Brandt}, W.~N. and {Breslauer}, Ben and {Brewington}, Howard J. and {Brinkmann}, J. and {Brown}, Peter J. and {Brownstein}, Joel R. and {Burger}, Dan and {Busca}, Nicolas G. and {Campbell}, Heather and {Cargile}, Phillip A. and {Carithers}, William C. and {Carlberg}, Joleen K. and {Carr}, Michael A. and {Chang}, Liang and {Chen}, Yanmei and {Chiappini}, Cristina and {Comparat}, Johan and {Connolly}, Natalia and {Cortes}, Marina and {Croft}, Rupert A.~C. and {Cunha}, Katia and {da Costa}, Luiz N. and {Davenport}, James R.~A. and {Dawson}, Kyle and {De Lee}, Nathan and {Porto de Mello}, Gustavo F. and {de Simoni}, Fernando and {Dean}, Janice and {Dhital}, Saurav and {Ealet}, Anne and {Ebelke}, Garrett L. and {Edmondson}, Edward M. and {Eiting}, Jacob M. and {Escoffier}, Stephanie and {Esposito}, Massimiliano and {Evans}, Michael L. and {Fan}, Xiaohui and {Femen{\'\i}a Castell{\'a}}, Bruno and {Dutra Ferreira}, Leticia and {Fitzgerald}, Greg and {Fleming}, Scott W. and {Font-Ribera}, Andreu and {Ford}, Eric B. and {Frinchaboy}, Peter M. and {Garc{\'\i}a P{\'e}rez}, Ana Elia and {Gaudi}, B. Scott and {Ge}, Jian and {Ghezzi}, Luan and {Gillespie}, Bruce A. and {Gilmore}, G. and {Girardi}, L{\'e}o and {Gott}, J. Richard and {Gould}, Andrew and {Grebel}, Eva K. and {Gunn}, James E. and {Hamilton}, Jean-Christophe and {Harding}, Paul and {Harris}, David W. and {Hawley}, Suzanne L. and {Hearty}, Frederick R. and {Hennawi}, Joseph F. and {Gonz{\'a}lez Hern{\'a}ndez}, Jonay I. and {Ho}, Shirley and {Hogg}, David W. and {Holtzman}, Jon A. and {Honscheid}, Klaus and {Inada}, Naohisa and {Ivans}, Inese I. and {Jiang}, Linhua and {Jiang}, Peng and {Johnson}, Jennifer A. and {Jordan}, Cathy and {Jordan}, Wendell P. and {Kauffmann}, Guinevere and {Kazin}, Eyal and {Kirkby}, David and {Klaene}, Mark A. and {Knapp}, G.~R. and {Kneib}, Jean-Paul and {Kochanek}, C.~S. and {Koesterke}, Lars and {Kollmeier}, Juna A. and {Kron}, Richard G. and {Lampeitl}, Hubert and {Lang}, Dustin and {Lawler}, James E. and {Le Goff}, Jean-Marc and {Lee}, Brian L. and {Lee}, Young Sun and {Leisenring}, Jarron M. and {Lin}, Yen-Ting and {Liu}, Jian and {Long}, Daniel C. and {Loomis}, Craig P. and {Lucatello}, Sara and {Lundgren}, Britt and {Lupton}, Robert H. and {Ma}, Bo and {Ma}, Zhibo and {MacDonald}, Nicholas and {Mack}, Claude and {Mahadevan}, Suvrath and {Maia}, Marcio A.~G. and {Majewski}, Steven R. and {Makler}, Martin and {Malanushenko}, Elena and {Malanushenko}, Viktor and {Mandelbaum}, Rachel and {Maraston}, Claudia and {Margala}, Daniel and {Maseman}, Paul and {Masters}, Karen L. and {McBride}, Cameron K. and {McDonald}, Patrick and {McGreer}, Ian D. and {McMahon}, Richard G. and {Mena Requejo}, Olga and {M{\'e}nard}, Brice and {Miralda-Escud{\'e}}, Jordi and {Morrison}, Heather L. and {Mullally}, Fergal and {Muna}, Demitri and {Murayama}, Hitoshi and {Myers}, Adam D. and {Naugle}, Tracy and {Neto}, Angelo Fausti and {Nguyen}, Duy Cuong and {Nichol}, Robert C. and {Nidever}, David L. and {O'Connell}, Robert W. and {Ogando}, Ricardo L.~C. and {Olmstead}, Matthew D. and {Oravetz}, Daniel J. and {Padmanabhan}, Nikhil and {Paegert}, Martin and {Palanque-Delabrouille}, Nathalie and {Pan}, Kaike and {Pandey}, Parul and {Parejko}, John K. and {P{\^a}ris}, Isabelle and {Pellegrini}, Paulo and {Pepper}, Joshua and {Percival}, Will J. and {Petitjean}, Patrick and {Pfaffenberger}, Robert and {Pforr}, Janine and {Phleps}, Stefanie and {Pichon}, Christophe and {Pieri}, Matthew M. and {Prada}, Francisco and {Price-Whelan}, Adrian M. and {Raddick}, M. Jordan and {Ramos}, Beatriz H.~F. and {Reid}, I. Neill and {Reyle}, Celine and {Rich}, James and {Richards}, Gordon T. and {Rieke}, George H. and {Rieke}, Marcia J. and {Rix}, Hans-Walter and {Robin}, Annie C. and {Rocha-Pinto}, Helio J. and {Rockosi}, Constance M. and {Roe}, Natalie A. and {Rollinde}, Emmanuel and {Ross}, Ashley J. and {Ross}, Nicholas P. and {Rossetto}, Bruno and {S{\'a}nchez}, Ariel G. and {Santiago}, Basilio and {Sayres}, Conor and {Schiavon}, Ricardo and {Schlegel}, David J. and {Schlesinger}, Katharine J. and {Schmidt}, Sarah J. and {Schneider}, Donald P. and {Sellgren}, Kris and {Shelden}, Alaina and {Sheldon}, Erin and {Shetrone}, Matthew},
        title = "{SDSS-III: Massive Spectroscopic Surveys of the Distant Universe, the Milky Way, and Extra-Solar Planetary Systems}",
      journal = {\aj},
         year = 2011,
        month = sep,
       volume = {142},
       number = {3},
          eid = {72},
        pages = {72},
          doi = {10.1088/0004-6256/142/3/72},
archivePrefix = {arXiv},
       eprint = {1101.1529},
 primaryClass = {astro-ph.IM},
       adsurl = {https://ui.adsabs.harvard.edu/abs/2011AJ....142...72E}
}

@ARTICLE{2013AJ....146...32S,
       author = {{Smee}, Stephen A. and {Gunn}, James E. and {Uomoto}, Alan and {Roe}, Natalie and {Schlegel}, David and {Rockosi}, Constance M. and {Carr}, Michael A. and {Leger}, French and {Dawson}, Kyle S. and {Olmstead}, Matthew D. and {Brinkmann}, Jon and {Owen}, Russell and {Barkhouser}, Robert H. and {Honscheid}, Klaus and {Harding}, Paul and {Long}, Dan and {Lupton}, Robert H. and {Loomis}, Craig and {Anderson}, Lauren and {Annis}, James and {Bernardi}, Mariangela and {Bhardwaj}, Vaishali and {Bizyaev}, Dmitry and {Bolton}, Adam S. and {Brewington}, Howard and {Briggs}, John W. and {Burles}, Scott and {Burns}, James G. and {Castander}, Francisco Javier and {Connolly}, Andrew and {Davenport}, James R.~A. and {Ebelke}, Garrett and {Epps}, Harland and {Feldman}, Paul D. and {Friedman}, Scott D. and {Frieman}, Joshua and {Heckman}, Timothy and {Hull}, Charles L. and {Knapp}, Gillian R. and {Lawrence}, David M. and {Loveday}, Jon and {Mannery}, Edward J. and {Malanushenko}, Elena and {Malanushenko}, Viktor and {Merrelli}, Aronne James and {Muna}, Demitri and {Newman}, Peter R. and {Nichol}, Robert C. and {Oravetz}, Daniel and {Pan}, Kaike and {Pope}, Adrian C. and {Ricketts}, Paul G. and {Shelden}, Alaina and {Sandford}, Dale and {Siegmund}, Walter and {Simmons}, Audrey and {Smith}, D. Shane and {Snedden}, Stephanie and {Schneider}, Donald P. and {SubbaRao}, Mark and {Tremonti}, Christy and {Waddell}, Patrick and {York}, Donald G.},
        title = "{The Multi-object, Fiber-fed Spectrographs for the Sloan Digital Sky Survey and the Baryon Oscillation Spectroscopic Survey}",
      journal = {\aj},
         year = 2013,
        month = aug,
       volume = {146},
       number = {2},
          eid = {32},
        pages = {32},
          doi = {10.1088/0004-6256/146/2/32},
archivePrefix = {arXiv},
       eprint = {1208.2233},
 primaryClass = {astro-ph.IM},
       adsurl = {https://ui.adsabs.harvard.edu/abs/2013AJ....146...32S}
}

@ARTICLE{2004AJ....128..502A,
       author = {{Abazajian}, Kevork and {Adelman-McCarthy}, Jennifer K. and {Ag{\"u}eros}, Marcel A. and {Allam}, Sahar S. and {Anderson}, Kurt and {Anderson}, Scott F. and {Annis}, James and {Bahcall}, Neta A. and {Baldry}, Ivan K. and {Bastian}, Steven and {Berlind}, Andreas and {Bernardi}, Mariangela and {Blanton}, Michael R. and {Bochanski}, Jr., John J. and {Boroski}, William N. and {Briggs}, John W. and {Brinkmann}, J. and {Brunner}, Robert J. and {Budav{\'a}ri}, Tam{\'a}s and {Carey}, Larry N. and {Carliles}, Samuel and {Castander}, Francisco J. and {Connolly}, A.~J. and {Csabai}, Istv{\'a}n and {Doi}, Mamoru and {Dong}, Feng and {Eisenstein}, Daniel J. and {Evans}, Michael L. and {Fan}, Xiaohui and {Finkbeiner}, Douglas P. and {Friedman}, Scott D. and {Frieman}, Joshua A. and {Fukugita}, Masataka and {Gal}, Roy R. and {Gillespie}, Bruce and {Glazebrook}, Karl and {Gray}, Jim and {Grebel}, Eva K. and {Gunn}, James E. and {Gurbani}, Vijay K. and {Hall}, Patrick B. and {Hamabe}, Masaru and {Harris}, Frederick H. and {Harris}, Hugh C. and {Harvanek}, Michael and {Heckman}, Timothy M. and {Hendry}, John S. and {Hennessy}, Gregory S. and {Hindsley}, Robert B. and {Hogan}, Craig J. and {Hogg}, David W. and {Holmgren}, Donald J. and {Ichikawa}, Shin-ichi and {Ichikawa}, Takashi and {Ivezi{\'c}}, {\v{Z}}eljko and {Jester}, Sebastian and {Johnston}, David E. and {Jorgensen}, Anders M. and {Kent}, Stephen M. and {Kleinman}, S.~J. and {Knapp}, G.~R. and {Kniazev}, Alexei Yu. and {Kron}, Richard G. and {Krzesinski}, Jurek and {Kunszt}, Peter Z. and {Kuropatkin}, Nickolai and {Lamb}, Donald Q. and {Lampeitl}, Hubert and {Lee}, Brian C. and {Leger}, R. French and {Li}, Nolan and {Lin}, Huan and {Loh}, Yeong-Shang and {Long}, Daniel C. and {Loveday}, Jon and {Lupton}, Robert H. and {Malik}, Tanu and {Margon}, Bruce and {Matsubara}, Takahiko and {McGehee}, Peregrine M. and {McKay}, Timothy A. and {Meiksin}, Avery and {Munn}, Jeffrey A. and {Nakajima}, Reiko and {Nash}, Thomas and {Neilsen}, Jr., Eric H. and {Newberg}, Heidi Jo and {Newman}, Peter R. and {Nichol}, Robert C. and {Nicinski}, Tom and {Nieto-Santisteban}, Maria and {Nitta}, Atsuko and {Okamura}, Sadanori and {O'Mullane}, William and {Ostriker}, Jeremiah P. and {Owen}, Russell and {Padmanabhan}, Nikhil and {Peoples}, John and {Pier}, Jeffrey R. and {Pope}, Adrian C. and {Quinn}, Thomas R. and {Richards}, Gordon T. and {Richmond}, Michael W. and {Rix}, Hans-Walter and {Rockosi}, Constance M. and {Schlegel}, David J. and {Schneider}, Donald P. and {Scranton}, Ryan and {Sekiguchi}, Maki and {Seljak}, Uros and {Sergey}, Gary and {Sesar}, Branimir and {Sheldon}, Erin and {Shimasaku}, Kazu and {Siegmund}, Walter A. and {Silvestri}, Nicole M. and {Smith}, J. Allyn and {Smol{\v{c}}i{\'c}}, Vernesa and {Snedden}, Stephanie A. and {Stebbins}, Albert and {Stoughton}, Chris and {Strauss}, Michael A. and {SubbaRao}, Mark and {Szalay}, Alexander S. and {Szapudi}, Istv{\'a}n and {Szkody}, Paula and {Szokoly}, Gyula P. and {Tegmark}, Max and {Teodoro}, Luis and {Thakar}, Aniruddha R. and {Tremonti}, Christy and {Tucker}, Douglas L. and {Uomoto}, Alan and {Vanden Berk}, Daniel E. and {Vandenberg}, Jan and {Vogeley}, Michael S. and {Voges}, Wolfgang and {Vogt}, Nicole P. and {Walkowicz}, Lucianne M. and {Wang}, Shu-i. and {Weinberg}, David H. and {West}, Andrew A. and {White}, Simon D.~M. and {Wilhite}, Brian C. and {Xu}, Yongzhong and {Yanny}, Brian and {Yasuda}, Naoki and {Yip}, Ching-Wa and {Yocum}, D.~R. and {York}, Donald G. and {Zehavi}, Idit and {Zibetti}, Stefano and {Zucker}, Daniel B.},
        title = "{The Second Data Release of the Sloan Digital Sky Survey}",
      journal = {\aj},
         year = 2004,
        month = jul,
       volume = {128},
       number = {1},
        pages = {502-512},
          doi = {10.1086/421365},
archivePrefix = {arXiv},
       eprint = {astro-ph/0403325},
 primaryClass = {astro-ph},
       adsurl = {https://ui.adsabs.harvard.edu/abs/2004AJ....128..502A}
}

@ARTICLE{2012AJ....144..144B,
       author = {{Bolton}, Adam S. and {Schlegel}, David J. and {Aubourg}, {\'E}ric and {Bailey}, Stephen and {Bhardwaj}, Vaishali and {Brownstein}, Joel R. and {Burles}, Scott and {Chen}, Yan-Mei and {Dawson}, Kyle and {Eisenstein}, Daniel J. and {Gunn}, James E. and {Knapp}, G.~R. and {Loomis}, Craig P. and {Lupton}, Robert H. and {Maraston}, Claudia and {Muna}, Demitri and {Myers}, Adam D. and {Olmstead}, Matthew D. and {Padmanabhan}, Nikhil and {P{\^a}ris}, Isabelle and {Percival}, Will J. and {Petitjean}, Patrick and {Rockosi}, Constance M. and {Ross}, Nicholas P. and {Schneider}, Donald P. and {Shu}, Yiping and {Strauss}, Michael A. and {Thomas}, Daniel and {Tremonti}, Christy A. and {Wake}, David A. and {Weaver}, Benjamin A. and {Wood-Vasey}, W. Michael},
        title = "{Spectral Classification and Redshift Measurement for the SDSS-III Baryon Oscillation Spectroscopic Survey}",
      journal = {\aj},
         year = 2012,
        month = nov,
       volume = {144},
       number = {5},
          eid = {144},
        pages = {144},
          doi = {10.1088/0004-6256/144/5/144},
archivePrefix = {arXiv},
       eprint = {1207.7326},
 primaryClass = {astro-ph.CO},
       adsurl = {https://ui.adsabs.harvard.edu/abs/2012AJ....144..144B}
}

@ARTICLE{2016AJ....152..205H,
       author = {{Hutchinson}, Timothy A. and {Bolton}, Adam S. and {Dawson}, Kyle S. and {Allende Prieto}, Carlos and {Bailey}, Stephen and {Bautista}, Julian E. and {Brownstein}, Joel R. and {Conroy}, Charlie and {Guy}, Julien and {Myers}, Adam D. and {Newman}, Jeffrey A. and {Prakash}, Abhishek and {Carnero-Rosell}, Aurelio and {Seo}, Hee-Jong and {Tojeiro}, Rita and {Vivek}, M. and {Ben Zhu}, Guangtun},
        title = "{Redshift Measurement and Spectral Classification for eBOSS Galaxies with the redmonster Software}",
      journal = {\aj},
         year = 2016,
        month = dec,
       volume = {152},
       number = {6},
          eid = {205},
        pages = {205},
          doi = {10.3847/0004-6256/152/6/205},
archivePrefix = {arXiv},
       eprint = {1607.02432},
 primaryClass = {astro-ph.IM},
       adsurl = {https://ui.adsabs.harvard.edu/abs/2016AJ....152..205H}
}

@ARTICLE{2009ApJ...696..870D,
       author = {{Drake}, A.~J. and {Djorgovski}, S.~G. and {Mahabal}, A. and {Beshore}, E. and {Larson}, S. and {Graham}, M.~J. and {Williams}, R. and {Christensen}, E. and {Catelan}, M. and {Boattini}, A. and {Gibbs}, A. and {Hill}, R. and {Kowalski}, R.},
        title = "{First Results from the Catalina Real-Time Transient Survey}",
      journal = {\apj},
         year = 2009,
        month = may,
       volume = {696},
       number = {1},
        pages = {870-884},
          doi = {10.1088/0004-637X/696/1/870},
archivePrefix = {arXiv},
       eprint = {0809.1394},
 primaryClass = {astro-ph},
       adsurl = {https://ui.adsabs.harvard.edu/abs/2009ApJ...696..870D}
}

@INPROCEEDINGS{2012IAUS..285..306D,
       author = {{Drake}, A.~J. and {Djorgovski}, S.~G. and {Mahabal}, A. and {Prieto}, J.~L. and {Beshore}, E. and {Graham}, M.~J. and {Catalan}, M. and {Larson}, S. and {Christensen}, E. and {Donalek}, C. and {Williams}, R.},
        title = "{The Catalina Real-time Transient Survey}",
    booktitle = {New Horizons in Time Domain Astronomy},
         year = 2012,
       editor = {{Griffin}, Elizabeth and {Hanisch}, Robert and {Seaman}, Rob},
       series = {IAU Symposium},
       volume = {285},
        month = apr,
        pages = {306-308},
          doi = {10.1017/S1743921312000889},
archivePrefix = {arXiv},
       eprint = {1111.2566},
 primaryClass = {astro-ph.CO},
       adsurl = {https://ui.adsabs.harvard.edu/abs/2012IAUS..285..306D}
}

@ARTICLE{2016arXiv161205560C,
       author = {{Chambers}, K.~C. and {Magnier}, E.~A. and {Metcalfe}, N. and {Flewelling}, H.~A. and {Huber}, M.~E. and {Waters}, C.~Z. and {Denneau}, L. and {Draper}, P.~W. and {Farrow}, D. and {Finkbeiner}, D.~P. and {Holmberg}, C. and {Koppenhoefer}, J. and {Price}, P.~A. and {Rest}, A. and {Saglia}, R.~P. and {Schlafly}, E.~F. and {Smartt}, S.~J. and {Sweeney}, W. and {Wainscoat}, R.~J. and {Burgett}, W.~S. and {Chastel}, S. and {Grav}, T. and {Heasley}, J.~N. and {Hodapp}, K.~W. and {Jedicke}, R. and {Kaiser}, N. and {Kudritzki}, R. -P. and {Luppino}, G.~A. and {Lupton}, R.~H. and {Monet}, D.~G. and {Morgan}, J.~S. and {Onaka}, P.~M. and {Shiao}, B. and {Stubbs}, C.~W. and {Tonry}, J.~L. and {White}, R. and {Ba{\~n}ados}, E. and {Bell}, E.~F. and {Bender}, R. and {Bernard}, E.~J. and {Boegner}, M. and {Boffi}, F. and {Botticella}, M.~T. and {Calamida}, A. and {Casertano}, S. and {Chen}, W. -P. and {Chen}, X. and {Cole}, S. and {Deacon}, N. and {Frenk}, C. and {Fitzsimmons}, A. and {Gezari}, S. and {Gibbs}, V. and {Goessl}, C. and {Goggia}, T. and {Gourgue}, R. and {Goldman}, B. and {Grant}, P. and {Grebel}, E.~K. and {Hambly}, N.~C. and {Hasinger}, G. and {Heavens}, A.~F. and {Heckman}, T.~M. and {Henderson}, R. and {Henning}, T. and {Holman}, M. and {Hopp}, U. and {Ip}, W. -H. and {Isani}, S. and {Jackson}, M. and {Keyes}, C.~D. and {Koekemoer}, A.~M. and {Kotak}, R. and {Le}, D. and {Liska}, D. and {Long}, K.~S. and {Lucey}, J.~R. and {Liu}, M. and {Martin}, N.~F. and {Masci}, G. and {McLean}, B. and {Mindel}, E. and {Misra}, P. and {Morganson}, E. and {Murphy}, D.~N.~A. and {Obaika}, A. and {Narayan}, G. and {Nieto-Santisteban}, M.~A. and {Norberg}, P. and {Peacock}, J.~A. and {Pier}, E.~A. and {Postman}, M. and {Primak}, N. and {Rae}, C. and {Rai}, A. and {Riess}, A. and {Riffeser}, A. and {Rix}, H.~W. and {R{\"o}ser}, S. and {Russel}, R. and {Rutz}, L. and {Schilbach}, E. and {Schultz}, A.~S.~B. and {Scolnic}, D. and {Strolger}, L. and {Szalay}, A. and {Seitz}, S. and {Small}, E. and {Smith}, K.~W. and {Soderblom}, D.~R. and {Taylor}, P. and {Thomson}, R. and {Taylor}, A.~N. and {Thakar}, A.~R. and {Thiel}, J. and {Thilker}, D. and {Unger}, D. and {Urata}, Y. and {Valenti}, J. and {Wagner}, J. and {Walder}, T. and {Walter}, F. and {Watters}, S.~P. and {Werner}, S. and {Wood-Vasey}, W.~M. and {Wyse}, R.},
        title = "{The Pan-STARRS1 Surveys}",
      journal = {arXiv e-prints},
         year = 2016,
        month = dec,
          eid = {arXiv:1612.05560},
        pages = {arXiv:1612.05560},
          doi = {10.48550/arXiv.1612.05560},
archivePrefix = {arXiv},
       eprint = {1612.05560},
 primaryClass = {astro-ph.IM},
       adsurl = {https://ui.adsabs.harvard.edu/abs/2016arXiv161205560C}
}

@ARTICLE{2019PASP..131a8002B,
       author = {{Bellm}, Eric C. and {Kulkarni}, Shrinivas R. and {Graham}, Matthew J. and {Dekany}, Richard and {Smith}, Roger M. and {Riddle}, Reed and {Masci}, Frank J. and {Helou}, George and {Prince}, Thomas A. and {Adams}, Scott M. and {Barbarino}, C. and {Barlow}, Tom and {Bauer}, James and {Beck}, Ron and {Belicki}, Justin and {Biswas}, Rahul and {Blagorodnova}, Nadejda and {Bodewits}, Dennis and {Bolin}, Bryce and {Brinnel}, Valery and {Brooke}, Tim and {Bue}, Brian and {Bulla}, Mattia and {Burruss}, Rick and {Cenko}, S. Bradley and {Chang}, Chan-Kao and {Connolly}, Andrew and {Coughlin}, Michael and {Cromer}, John and {Cunningham}, Virginia and {De}, Kishalay and {Delacroix}, Alex and {Desai}, Vandana and {Duev}, Dmitry A. and {Eadie}, Gwendolyn and {Farnham}, Tony L. and {Feeney}, Michael and {Feindt}, Ulrich and {Flynn}, David and {Franckowiak}, Anna and {Frederick}, S. and {Fremling}, C. and {Gal-Yam}, Avishay and {Gezari}, Suvi and {Giomi}, Matteo and {Goldstein}, Daniel A. and {Golkhou}, V. Zach and {Goobar}, Ariel and {Groom}, Steven and {Hacopians}, Eugean and {Hale}, David and {Henning}, John and {Ho}, Anna Y.~Q. and {Hover}, David and {Howell}, Justin and {Hung}, Tiara and {Huppenkothen}, Daniela and {Imel}, David and {Ip}, Wing-Huen and {Ivezi{\'c}}, {\v{Z}}eljko and {Jackson}, Edward and {Jones}, Lynne and {Juric}, Mario and {Kasliwal}, Mansi M. and {Kaspi}, S. and {Kaye}, Stephen and {Kelley}, Michael S.~P. and {Kowalski}, Marek and {Kramer}, Emily and {Kupfer}, Thomas and {Landry}, Walter and {Laher}, Russ R. and {Lee}, Chien-De and {Lin}, Hsing Wen and {Lin}, Zhong-Yi and {Lunnan}, Ragnhild and {Giomi}, Matteo and {Mahabal}, Ashish and {Mao}, Peter and {Miller}, Adam A. and {Monkewitz}, Serge and {Murphy}, Patrick and {Ngeow}, Chow-Choong and {Nordin}, Jakob and {Nugent}, Peter and {Ofek}, Eran and {Patterson}, Maria T. and {Penprase}, Bryan and {Porter}, Michael and {Rauch}, Ludwig and {Rebbapragada}, Umaa and {Reiley}, Dan and {Rigault}, Mickael and {Rodriguez}, Hector and {van Roestel}, Jan and {Rusholme}, Ben and {van Santen}, Jakob and {Schulze}, S. and {Shupe}, David L. and {Singer}, Leo P. and {Soumagnac}, Maayane T. and {Stein}, Robert and {Surace}, Jason and {Sollerman}, Jesper and {Szkody}, Paula and {Taddia}, F. and {Terek}, Scott and {Van Sistine}, Angela and {van Velzen}, Sjoert and {Vestrand}, W. Thomas and {Walters}, Richard and {Ward}, Charlotte and {Ye}, Quan-Zhi and {Yu}, Po-Chieh and {Yan}, Lin and {Zolkower}, Jeffry},
        title = "{The Zwicky Transient Facility: System Overview, Performance, and First Results}",
      journal = {\pasp},
         year = 2019,
        month = jan,
       volume = {131},
       number = {995},
        pages = {018002},
          doi = {10.1088/1538-3873/aaecbe},
archivePrefix = {arXiv},
       eprint = {1902.01932},
 primaryClass = {astro-ph.IM},
       adsurl = {https://ui.adsabs.harvard.edu/abs/2019PASP..131a8002B}
}

@ARTICLE{2019PASP..131a8003M,
       author = {{Masci}, Frank J. and {Laher}, Russ R. and {Rusholme}, Ben and {Shupe}, David L. and {Groom}, Steven and {Surace}, Jason and {Jackson}, Edward and {Monkewitz}, Serge and {Beck}, Ron and {Flynn}, David and {Terek}, Scott and {Landry}, Walter and {Hacopians}, Eugean and {Desai}, Vandana and {Howell}, Justin and {Brooke}, Tim and {Imel}, David and {Wachter}, Stefanie and {Ye}, Quan-Zhi and {Lin}, Hsing-Wen and {Cenko}, S. Bradley and {Cunningham}, Virginia and {Rebbapragada}, Umaa and {Bue}, Brian and {Miller}, Adam A. and {Mahabal}, Ashish and {Bellm}, Eric C. and {Patterson}, Maria T. and {Juri{\'c}}, Mario and {Golkhou}, V. Zach and {Ofek}, Eran O. and {Walters}, Richard and {Graham}, Matthew and {Kasliwal}, Mansi M. and {Dekany}, Richard G. and {Kupfer}, Thomas and {Burdge}, Kevin and {Cannella}, Christopher B. and {Barlow}, Tom and {Van Sistine}, Angela and {Giomi}, Matteo and {Fremling}, Christoffer and {Blagorodnova}, Nadejda and {Levitan}, David and {Riddle}, Reed and {Smith}, Roger M. and {Helou}, George and {Prince}, Thomas A. and {Kulkarni}, Shrinivas R.},
        title = "{The Zwicky Transient Facility: Data Processing, Products, and Archive}",
      journal = {\pasp},
         year = 2019,
        month = jan,
       volume = {131},
       number = {995},
        pages = {018003},
          doi = {10.1088/1538-3873/aae8ac},
archivePrefix = {arXiv},
       eprint = {1902.01872},
 primaryClass = {astro-ph.IM},
       adsurl = {https://ui.adsabs.harvard.edu/abs/2019PASP..131a8003M}
}

@ARTICLE{1999PASP..111...63F,
       author = {{Fitzpatrick}, Edward L.},
        title = "{Correcting for the Effects of Interstellar Extinction}",
      journal = {\pasp},
         year = 1999,
        month = jan,
       volume = {111},
       number = {755},
        pages = {63-75},
          doi = {10.1086/316293},
archivePrefix = {arXiv},
       eprint = {astro-ph/9809387},
 primaryClass = {astro-ph},
       adsurl = {https://ui.adsabs.harvard.edu/abs/1999PASP..111...63F}
}

@ARTICLE{2022ApJ...933..180G,
       author = {{Green}, Paul J. and {Pulgarin-Duque}, Lina and {Anderson}, Scott F. and {MacLeod}, Chelsea L. and {Eracleous}, Michael and {Ruan}, John J. and {Runnoe}, Jessie and {Graham}, Matthew and {Roulston}, Benjamin R. and {Schneider}, Donald P. and {Ahlf}, Austin and {Bizyaev}, Dmitry and {Brownstein}, Joel R. and {del Casal}, Sonia Joesephine and {Dodd}, Sierra A. and {Hoover}, Daniel and {Matt}, Cayenne and {Merloni}, Andrea and {Pan}, Kaike and {Ramirez}, Arnulfo and {Ridder}, Margaret and {Moseley}, Serena},
        title = "{The Time Domain Spectroscopic Survey: Changing-look Quasar Candidates from Multi-epoch Spectroscopy in SDSS-IV}",
      journal = {\apj},
         year = 2022,
        month = jul,
       volume = {933},
       number = {2},
          eid = {180},
        pages = {180},
          doi = {10.3847/1538-4357/ac743f},
archivePrefix = {arXiv},
       eprint = {2201.09123},
 primaryClass = {astro-ph.GA},
       adsurl = {https://ui.adsabs.harvard.edu/abs/2022ApJ...933..180G}
}

@ARTICLE{2018ApJ...862..109Y,
       author = {{Yang}, Qian and {Wu}, Xue-Bing and {Fan}, Xiaohui and {Jiang}, Linhua and {McGreer}, Ian and {Shangguan}, Jinyi and {Yao}, Su and {Wang}, Bingquan and {Joshi}, Ravi and {Green}, Richard and {Wang}, Feige and {Feng}, Xiaotong and {Fu}, Yuming and {Yang}, Jinyi and {Liu}, Yuanqi},
        title = "{Discovery of 21 New Changing-look AGNs in the Northern Sky}",
      journal = {\apj},
         year = 2018,
        month = aug,
       volume = {862},
       number = {2},
          eid = {109},
        pages = {109},
          doi = {10.3847/1538-4357/aaca3a},
archivePrefix = {arXiv},
       eprint = {1711.08122},
 primaryClass = {astro-ph.GA},
       adsurl = {https://ui.adsabs.harvard.edu/abs/2018ApJ...862..109Y}
}

@ARTICLE{2016MNRAS.457..389M,
       author = {{MacLeod}, Chelsea L. and {Ross}, Nicholas P. and {Lawrence}, Andy and {Goad}, Mike and {Horne}, Keith and {Burgett}, William and {Chambers}, Ken C. and {Flewelling}, Heather and {Hodapp}, Klaus and {Kaiser}, Nick and {Magnier}, Eugene and {Wainscoat}, Richard and {Waters}, Christopher},
        title = "{A systematic search for changing-look quasars in SDSS}",
      journal = {\mnras},
         year = 2016,
        month = mar,
       volume = {457},
       number = {1},
        pages = {389-404},
          doi = {10.1093/mnras/stv2997},
archivePrefix = {arXiv},
       eprint = {1509.08393},
 primaryClass = {astro-ph.GA},
       adsurl = {https://ui.adsabs.harvard.edu/abs/2016MNRAS.457..389M}
}

@ARTICLE{2019ApJ...874....8M,
       author = {{MacLeod}, Chelsea L. and {Green}, Paul J. and {Anderson}, Scott F. and {Bruce}, Alastair and {Eracleous}, Michael and {Graham}, Matthew and {Homan}, David and {Lawrence}, Andy and {LeBleu}, Amy and {Ross}, Nicholas P. and {Ruan}, John J. and {Runnoe}, Jessie and {Stern}, Daniel and {Burgett}, William and {Chambers}, Kenneth C. and {Kaiser}, Nick and {Magnier}, Eugene and {Metcalfe}, Nigel},
        title = "{Changing-look Quasar Candidates: First Results from Follow-up Spectroscopy of Highly Optically Variable Quasars}",
      journal = {\apj},
         year = 2019,
        month = mar,
       volume = {874},
       number = {1},
          eid = {8},
        pages = {8},
          doi = {10.3847/1538-4357/ab05e2},
archivePrefix = {arXiv},
       eprint = {1810.00087},
 primaryClass = {astro-ph.GA},
       adsurl = {https://ui.adsabs.harvard.edu/abs/2019ApJ...874....8M}
}

@ARTICLE{2019ApJ...883L..44G,
       author = {{Guo}, Hengxiao and {Sun}, Mouyuan and {Liu}, Xin and {Wang}, Tinggui and {Kong}, Minzhi and {Wang}, Shu and {Sheng}, Zhenfeng and {He}, Zhicheng},
        title = "{Discovery of an Mg II Changing-look Active Galactic Nucleus and Its Implications for a Unification Sequence of Changing-look Active Galactic Nuclei}",
      journal = {\apjl},
         year = 2019,
        month = oct,
       volume = {883},
       number = {2},
          eid = {L44},
        pages = {L44},
          doi = {10.3847/2041-8213/ab4138},
archivePrefix = {arXiv},
       eprint = {1908.00072},
 primaryClass = {astro-ph.GA},
       adsurl = {https://ui.adsabs.harvard.edu/abs/2019ApJ...883L..44G}
}

@ARTICLE{2020ApJ...888...58G,
       author = {{Guo}, Hengxiao and {Shen}, Yue and {He}, Zhicheng and {Wang}, Tinggui and {Liu}, Xin and {Wang}, Shu and {Sun}, Mouyuan and {Yang}, Qian and {Kong}, Minzhi and {Sheng}, Zhenfeng},
        title = "{Understanding Broad Mg II Variability in Quasars with Photoionization: Implications for Reverberation Mapping and Changing-look Quasars}",
      journal = {\apj},
         year = 2020,
        month = jan,
       volume = {888},
       number = {2},
          eid = {58},
        pages = {58},
          doi = {10.3847/1538-4357/ab5db0},
archivePrefix = {arXiv},
       eprint = {1907.06669},
 primaryClass = {astro-ph.GA},
       adsurl = {https://ui.adsabs.harvard.edu/abs/2020ApJ...888...58G}
}

@ARTICLE{2024ApJS..272...13P,
       author = {{Panda}, Swayamtrupta and {{\'S}niegowska}, Marzena},
        title = "{Changing-look Active Galactic Nuclei. I. Tracking the Transition on the Main Sequence of Quasars}",
      journal = {\apjs},
         year = 2024,
        month = may,
       volume = {272},
       number = {1},
          eid = {13},
        pages = {13},
          doi = {10.3847/1538-4365/ad344f},
archivePrefix = {arXiv},
       eprint = {2206.10056},
 primaryClass = {astro-ph.HE},
       adsurl = {https://ui.adsabs.harvard.edu/abs/2024ApJS..272...13P}
}

@ARTICLE{2024ApJ...966..128W,
       author = {{Wang}, Shu and {Woo}, Jong-Hak and {Gallo}, Elena and {Guo}, Hengxiao and {Son}, Donghoon and {Kong}, Minzhi and {Mandal}, Amit Kumar and {Cho}, Hojin and {Kim}, Changseok and {Shin}, Jaejin},
        title = "{Identifying Changing-look AGNs Using Variability Characteristics}",
      journal = {\apj},
         year = 2024,
        month = may,
       volume = {966},
       number = {1},
          eid = {128},
        pages = {128},
          doi = {10.3847/1538-4357/ad3049},
archivePrefix = {arXiv},
       eprint = {2402.18131},
 primaryClass = {astro-ph.GA},
       adsurl = {https://ui.adsabs.harvard.edu/abs/2024ApJ...966..128W}
}

@ARTICLE{2024arXiv241015587W,
       author = {{Wang}, Shu and {Woo}, Jong-Hak and {Gallo}, Elena and {Son}, Donghoon and {Yang}, Qian and {Jin}, Junjie and {Guo}, Hengxiao and {Kong}, Minzhi},
        title = "{Dormancy and Reawakening Over Years: Eight New Recurrent Changing-Look AGNs}",
      journal = {arXiv e-prints},
         year = 2024,
        month = oct,
          eid = {arXiv:2410.15587},
        pages = {arXiv:2410.15587},
          doi = {10.48550/arXiv.2410.15587},
archivePrefix = {arXiv},
       eprint = {2410.15587},
 primaryClass = {astro-ph.GA},
       adsurl = {https://ui.adsabs.harvard.edu/abs/2024arXiv241015587W}
}

@ARTICLE{2015ApJ...800..144L,
       author = {{LaMassa}, Stephanie M. and {Cales}, Sabrina and {Moran}, Edward C. and {Myers}, Adam D. and {Richards}, Gordon T. and {Eracleous}, Michael and {Heckman}, Timothy M. and {Gallo}, Luigi and {Urry}, C. Megan},
        title = "{The Discovery of the First {\textquotedblleft}Changing Look{\textquotedblright} Quasar: New Insights Into the Physics and Phenomenology of Active Galactic Nucleus}",
      journal = {\apj},
         year = 2015,
        month = feb,
       volume = {800},
       number = {2},
          eid = {144},
        pages = {144},
          doi = {10.1088/0004-637X/800/2/144},
archivePrefix = {arXiv},
       eprint = {1412.2136},
 primaryClass = {astro-ph.GA},
       adsurl = {https://ui.adsabs.harvard.edu/abs/2015ApJ...800..144L}
}

@ARTICLE{1986ApJS...61..249W,
       author = {{Wolfe}, A.~M. and {Turnshek}, D.~A. and {Smith}, H.~E. and {Cohen}, R.~D.},
        title = "{Damped Lyman-Alpha Absorption by Disk Galaxies with Large Redshifts. I. The Lick Survey}",
      journal = {\apjs},
         year = 1986,
        month = jun,
       volume = {61},
        pages = {249},
          doi = {10.1086/191114},
       adsurl = {https://ui.adsabs.harvard.edu/abs/1986ApJS...61..249W}
}
\bibliographystyle{aasjournal}

\clearpage
\appendix

\section{Fiber-drop Case}\label{fiber_drop}

\begin{figure*}
    \centering
    \includegraphics[width=0.85\textwidth]{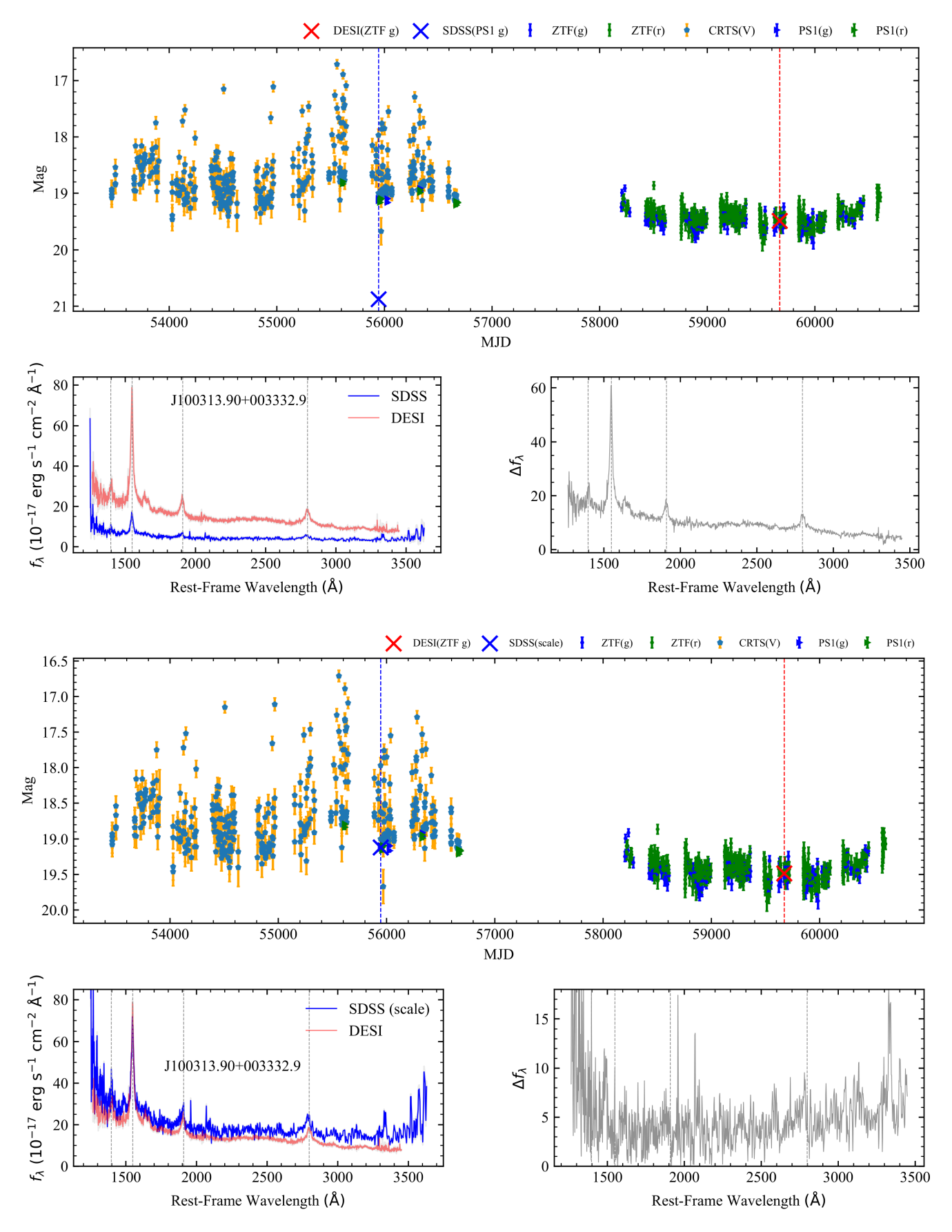}
    \caption{An example of a false CL quasar caused by fiber drop issues. The top panel shows the nearly 20-year light curve compiled from CRTS $V$-band (dark blue pentagons), PS1 $g$-band (blue triangles), $r$-band (green triangles), ZTF $g$-band (blue circles), and $r$-band (green circles). The blue and red dashed vertical lines indicate the epochs of SDSS and DESI spectroscopic observations, respectively. The corresponding ``X'' markers represent the pseudo-magnitudes derived by convolving the spectra with filter response curves. The lower-left panel displays the original spectra from SDSS (blue) and DESI (red). The lower-right panel on the left side shows the difference in flux between the two spectra. The bottom panels are identical in layout, except that the SDSS pseudo magnitudes and spectra have been flux-scaled to match the actual photometric level (PS1 $g$-band).}
    \label{fig: flux_calibratio}
\end{figure*}

Fiber drop refers to the failure to obtain a valid spectrum due to issues such as fiber misalignment, fiber collision, or mechanical malfunction, resulting in problematic or unusable flux data (\citealt{2015ApJS..216....4S, 2015ApJ...811...42S, 2020ApJ...888...58G}). This phenomenon becomes more prominent at higher redshifts, where targets are typically fainter and more susceptible to observational limitations. For studies aiming to identify CL quasars, a rare class of quasars exhibiting dramatic variability in emission line fluxes on timescales of months to years, fiber drop poses a significant challenge, as it can mimic or obscure the spectral transitions that are critical for CL quasar classification. As shown in Figure \ref{fig: flux_calibratio}, the spectra from SDSS and DESI exhibit a striking discrepancy in flux levels. In the SDSS spectrum, the $\rm Mg\, \textsc{ii}$, $\rm C\,\textsc{iii}]$, and $\rm Si\, \textsc{iv}$ emission lines are nearly buried in noise, while in the DESI spectrum, these lines are clearly visible, so much so that the object was initially misidentified as a CL quasar. However, we found that the pseudo-magnitude derived from the SDSS spectrum is significantly fainter than the contemporaneous PS1 photometry. After the SDSS flux was adjusted to match the actual photometric brightness, the spectral difference between SDSS and DESI largely disappeared. Identifying and removing such false CL quasar candidates caused by fiber drop is, therefore, a necessary and critical step in our analysis.

\section{High-redshift CL quasar Candidate Catalog \label{appendix b}}

\begin{deluxetable*}{lcccccccl}[!ht]
    \setlength{\tabcolsep}{8pt}
    \renewcommand{\arraystretch}{1}
    \tablenum{6}
    \centering
    \tablecaption{CL quasar candidates identified in this work \label{tab:candidates}}
    \tablehead{
        \colhead{SDSS\_NAME} & \colhead{R.A.} & \colhead{Dec.} & \colhead{Redshift} & \colhead{MJD\_SDSS} & \colhead{MJD\_DESI} & \colhead{Transition}\\ 
        \colhead{(1)} & \colhead{(2)} & \colhead{(3)} & \colhead{(4)} & \colhead{(5)} & \colhead{(6)} & \colhead{(7)}
    }
    \startdata
    \hline
    J000908.69+303733.6 & 2.2862   & 30.626  & 2.3471   & 56566     & 59475.43985 & on    \\
J002245.89+061729.0 & 5.6912   & 6.2914  & 1.4869   & 58396     & 59525.21465 & on    \\
J002829.91-005055.4 & 7.1246   & -0.8487 & 2.361    & 55447     & 59492.30014 & off   \\
J003111.32-001121.2 & 7.7972   & -0.1892 & 1.5139   & 58045     & 59494.29318 & on    \\
J003646.67+070100.0 & 9.1944   & 7.0167  & 1.5224   & 58398     & 59498.29944 & on    \\
J003827.88+314744.2 & 9.6162   & 31.7956 & 1.7597   & 57364     & 59531.15362 & off   \\
J004108.69+223401.7 & 10.2862  & 22.5672 & 1.0473   & 56949     & 59554.14933 & on    \\
J004537.42+061652.6 & 11.4059  & 6.2813  & 2.465    & 55888     & 59532.54183 & on    \\
J004830.19+010622.6 & 12.1258  & 1.1063  & 1.6745   & 57278     & 59484.34243 & off   \\
J010345.18-000711.7 & 15.9382  & -0.1199 & 2.3171   & 55475     & 59493.08173 & on    \\
    \hline
    \enddata
    \tablecomments{The table is in its entirety in the machine-readable format.}
\end{deluxetable*}

\end{document}